\documentclass[aps,preprint,prd,nofootinbib]{revtex4}
\usepackage{graphicx}
\usepackage{psfrag}

\usepackage{subfigure}
\usepackage{array}
\usepackage{graphicx}
\usepackage{amsmath}
\usepackage{amssymb}
\usepackage{mathrsfs}
\usepackage{bm}
\usepackage{slashed}
\usepackage{threeparttable}
\usepackage{multirow}
\usepackage[colorlinks, linkcolor=black, anchorcolor=black, citecolor=black]{hyperref}
\usepackage[amsmath,thmmarks,hyperref]{ntheorem}
\usepackage{soul}
\usepackage{ulem}

\newcommand{\ud}{\mathrm d}

\newcommand{\cancel}{\slashed}
\allowdisplaybreaks[4]

\begin{document}

\title{Isgur-Wise function in $B_c$ decays to charmonium with the Bethe-Salpeter method}
\author{Zi-Kan Geng, Yue Jiang\footnote{Joint first author, Corresponding author: jiangure@hit.edu.cn}, Tianhong Wang, Hui-Wen Zheng, Guo-Li Wang\\}
\address{Department of Physics, Harbin Institute of Technology, Harbin, 150001, China}

\baselineskip=20pt

\begin{abstract}
The Isgur-Wise function vastly reduces the weak-decay form factors of hadrons containing one heavy quark. In this paper, we extract the Isgur-Wise functions from the instantaneous Bethe-Salpeter method, and give the numerical results for the $B_c$ decays to charmonium where the final states include $1S$, $1P$, $2S$ and $2P$. The overlapping integral of the wave functions for the initial and final states is the Isgur-Wise function, as the heavy quark effective theory does. In the case of accurate calculation,  describing form factors need to introduce more relativistic corrections which are the overlapping integrals with the relative momentum between the quark and antiquark to Isgur-Wise function. The relativistic corrections to Isgur-Wise function provide greater contributions especially involving the excited state, and therefore are necessary to be adopted.
 
\end{abstract}

\maketitle

\section{Introduction}
Under the heavy quark effective theory (HQET), a semileptonic decay process can be related to a rotation of the heavy quark flavor or spin \cite{Neubert:1991xw,Neubert:1996qg}. In the limit $m_Q\to\infty$ ($Q$ denotes the heavy quark or anti-quark), this rotation is a symmetry transformation. The form factors depend only on the Lorentz boost $\gamma=v\cdot v'$ which connects the rest frames of the initial state and final state. The transition can be described by a dimensionless function $\xi(v\cdot v')$. Heavy-quark symmetry reduces the weak-decay form factors of heavy hadrons to this universal function. These relations were derived by Isgur and Wise firstly \cite{Isgur:1988gb,Isgur:1989ed}, so called Isgur-Wise function (IWF). 

HQET vastly simplifies the calculations, and plays a crucial role in extracting the values of $|V_{cb}|$ and $|V_{ub}|$. For example, the differential semileptonic decay rate for $B\to D$ in the heavy-quark limit can be  model-independent described by \cite{Neubert:1996qg}
\begin{equation}
\frac{\ud\Gamma(\bar B\to D\ell\bar\nu)}{\ud\omega}=\frac{G_F^2}{48\pi^3}|V_{cb}|^2(m_B+m_D)^2m_D^3(\omega^2-1)^{3/2}\xi^2(\omega).
\end{equation}
The decay rate depends on only two quantities, $|V_{cb}|$ and $\xi(\omega)$. If the differential semileptonic decay rate is measured by experiments, one can obtain the value of $|V_{cb}|\xi(\omega)$. Exploiting the normalization of Isgur-Wise function $\xi(1)=1$, the value of $|V_{cb}|$ can be extracted. Conversely exploiting the given value of $|V_{cb}|$, the differential semileptonic decay rate can be obtained by calculating the Isgur-Wise function. But Isgur-Wise function can not be calculated by perturbation theory in principle, and can only be abtained by various phenomenological models. Obviously the latter is model-dependent. Then a great deal of efforts were directed to study the Isgur-Wise function and its applications in different frameworks. Kiselev et al. exploited the Isgur-Wise function in a covariant potential model to calculate the semileptonic $B\to D^{(*)}\ell\nu$ Decays \cite{Kiselev:1994ay}. Olsson et al. explored  the Isgur-Wise Function in the relativistic flux tube model \cite{Olsson:1994us}. Huang et al. calculated the leading Isgur-Wise function parametrizing the semileptonic decays $\Lambda_b\to\Lambda_{c1}^{1/2,3/2}$  by using the QCD sum rules \cite{Huang:2000xw}. Lee et al. calculated the subleading Isgur-Wise function for the same processes by using the QCD sum rules \cite{Lee:2002ka}. Hassanabadi and Rahmani et al. investigated the Isgur-Wise Function for semileptonic decays of bottom and charmed hadrons in a series of models, including a Cornell potential combined a improved variational approach \cite{Rahmani:2014aia}, the baryon hyperspherical approach \cite{Hassanabadi:2014kka}, a combination of linear confining and Hulth\'en potentials \cite{Hassanabadi:2014isa}, the three-boy Schr\"odinger equation\cite{Hassanabadi:2015jya}, a combination of Deng-Fan-type and harmonic potentials \cite{Hassanabadi:2015scl}, a quantum isotonic nonlinear oscillator potential model \cite{Rahmani:2017vbg} and so on. Choudhury et al. had studied the renormalization scale dependence of IW function by using a wave function with linear part of the Cornell potential as perturbation \cite{Choudhury:2014cia}. Yazarloo et al. employed the nonrelativistic Schr\"odinger equation with a new potential model to calculate the semileptonic decay width and branching ratio of the $B\to D^{*}\ell\nu$ process \cite{Yazarloo:2016luc}.  In a potential model, the Isgur-Wise function usually corresponds to the overlapping integral of the wave functions obtained by this model.

However the lowest order result is not accurate enough due to the heavy quark approximation. The symmetry-breaking corrections are needed when the study becomes more precise, since the masses of the heavy quarks or anti-quarks are not infinite actually. In addition the radiation correction can not be ignored. The HQET provides a systematic framework to analyze these corrections. For example, Luke analyzed the $1/m_Q$ corrections for the more complicated case of weak decay form factors \cite{Luke:1990eg}. Falk et al. analyzed the structure of $1/m^2_Q$ corrections for both meson and baryon weak decay form factors, and calculated the leading QCD radiative corrections \cite{Falk:1992fm,Falk:1990yz}. Neubert et al. developed a method to express the universal functions, that appear in leading and next-to-leading order in the $1/m_Q$ expansion of current matrix elements, in terms of hadronic form factors evaluated at maximum recoil \cite{Neubert:1991xw}. Leibovich et al. investigated the $1/m_Q$ corrections for semileptonic $\Lambda_b$ decays to excited $\Lambda_c$ at zero recoil \cite{Leibovich:1997az}. Cardarelli presented an investigation of the Isgur-Wise form factor relevant for the semileptonic decay $\Lambda_b\to\Lambda_c\ell^-\nu_\ell$ performed within a light-front constituent quark model, including both radiative effects and first-order power corrections in the inverse heavy-quark mass \cite{Cardarelli:1999mi}. Coleman et al. investigated a new effective IW function with a shift in slope, and all of $1/m_Q$ corrections and QCD radiative corrections are distilled \cite{PhysRevD.63.032006}.  Other efforts of complements are too many to be list here.

The flavor-spin symmetry can be used not only in weak decays of ground hardons containing one heavy quark, but also in the processes involving orbitally and radially excited states containing one heavy quark. Isgur and Wise exploited the flavor-spin symmetry to obtain model-independent predictions in weak decays from pseudoscalar meson of a heavy quark $Q_i$ to P wave excited states of another heavy quark $Q_j$ in term of two Isgur-Wise functions \cite{Isgur:1990jf}. Gan et al. employed the QCD sum rule approach to estimate the Isgur-Wise functions for the semileptonic decays of the ground statemeson $B_s$ into orbitally excited D-wave $\bar cs$ mesons \cite{Gan:2014jxa}. Yaouanc calculated the Isgur-Wise functions that involves only heavy mesons with light cloud $j^P = 1/2^-$ and their radial excitations by Bakamjian-Thomas relativistic quark model \cite{Yaouanc:2015xfa}. Be\u cirevi\'c et al. have studied the $1/m_Q$ corrections of form factors for semi-leptonic transitions between $B$ mesons and radially excited $D$ mesons in the Bakamjian-Thomas relativistic quark model \cite{Becirevic:2017vxe,Becirevic:2017njz}.

However, the validity of HQET is suspectable, when the systems contain two or more heavy degrees of freedom. Some work have explored the application of the heavy quark symmetry to describe the weak decays of hadrons containing two heavy quarks. Sanchis-Lozano estimated that the flavor symmetry losts, while the spin symmetry holds for the double-heavy meson \cite{SanchisLozano:1994vh}. As for the baryon containing two heavy quarks, two heavy quarks constitute a bosonic diquark whose mass could be regarded as infinite, and therefore HQET still holds true for the diquark-light quark system. But HQET is not suitable for dealing with the diquark subsystem. Ebert et al. exploited the relativistic quark model to obtain the diquark wave functions, and then the transition amplitudes of heavy diquarks $bb$ and $bc$ going respectively to $bc$ and $cc$ are expressed through the overlap integrals of corresponding diquark wave functions \cite{Ebert:2004ck}. Pathak et al. treated $B_c$ meson as a typical heavy-light meson like $B$ or $D$ within a QCD potential model, and the semileptonic decay rates of $B_c$ meson into $\eta_c$, $J/\psi$ are exploited \cite{Pathak:2013dra}. Das et al. computed the slopes and curvatures of the form factors of semileptonic decays of heavy-light mesons included $B_c$ \cite{Das:2016hmv}. The error of the result for $B_c$ is very large. Wang at al. obtained form factors for the $B_c$ into S-wave and P-wave charmonium by three universal Isgur-Wise functions \cite{Wang:2018duy}. Almost all of these studies just use the formulas in HQET directly, but the usefulness of the heavy quark symmetry to describe double heavy hadrons has been little reported.

The aim of this paper is to investigate the heavy quark symmetry in the double-heavy mesons without the use of heavy quark limit. The heavy quark symmetry, if exists, must be reflected in the results obtained by a suitable phenomenological model without the use of heavy quark limit. The instantaneous Bethe-Salpeter (BS) equation is just a very effective method to deal with double-heavy mesons. This method has a comparatively solid foundation because both the BS equation and the Mandelstam formula are established on relativistic quantum field theory. Meanwhile the instantaneous approximation is reasonable, since the quark and antiquark in double-heavy mesons are both heavy. This method can give an analytic expression, so the symmetry can be found intuitively. However, the accuracy may not be as good as Lattice QCD. We choose the semileptonic $B_c$ decays to charmonium , and the final mesons involve the orbitally and radially excited states. We conclude that the flavor symmetry breaks, while the spin symmetry holds for the double-heavy meson, as Sanchis-Lozano estimated \cite{SanchisLozano:1994vh}. Further, we investigate the behaviors of the Isugr-Wise function and other high-order relativistic corrections to Isgur-Wise function.

Actually, some works have been done on IWF with BS equation. Kugo et al. expanded BS equation in orders of the inverse heavy quark mass and defined the leading term in the expansion of the first form factor as IWF \cite{Kugo:1993rd}. El-Hady et al. pointed out that the IWF can be related to the overlap integral of normalized meson wave functions in the infinite momentum frame and it should be possible to calculate the form factors directly without using the heavy quark limit \cite{AbdElHady:1994dr}. Zoller et al. calculated the numerical IWF by multiplying quark masses with a large factor directly \cite{Zoller:1994eh}. Chang et al. obtained two universal functions in $B_c\to h_c,\chi_{c}$ with the instantaneous BS method, but the wave functions they used are nonrelativistic \cite{Chang:1992pt,Chang:2001pm}. Nowadays the instantaneous BS method has developed to be quite covariant, and the full Salpeter equations are solved for different $J^{P(C)}$ states \cite{Cvetic:2004qg,Chang:2004im,Wang:2007av,Wang:2009er}. So the relativistic correction which equate to symmetry-breaking correction has been taken into account. In this paper we shall not derive these corrections from HQET, but attempt to extract the IWF from the solutions of the instantaneous BS method. Note that we do not use the heavy quark limit. The results show that only the lowest order correction to IWF is not enough, and higher-order relativistic corrections need to be introduced for more accurate results.

The paper is organized as follows. In section \ref{sec:formula}, we give the useful formulas for the $B_c$ decays to charmonium. In section \ref{sec:wavefunction}, we give the relativistic wave function for $0^-$ state in the instantaneous BS method. In section \ref{sec:IWF}, we extract the IWF and give the analytical results. In section \ref{sec:results}, we give the numerical results and discussions. We summarize and conclude in section \ref{sec:conclusion}, and put the Salpeter equation and some wave functions in the appendix \ref{appendix}.

\section{\label{sec:formula} Form factors and semileptonic decay width }

\begin{figure}[tbp]
\centering
\includegraphics[width=0.45\textwidth]{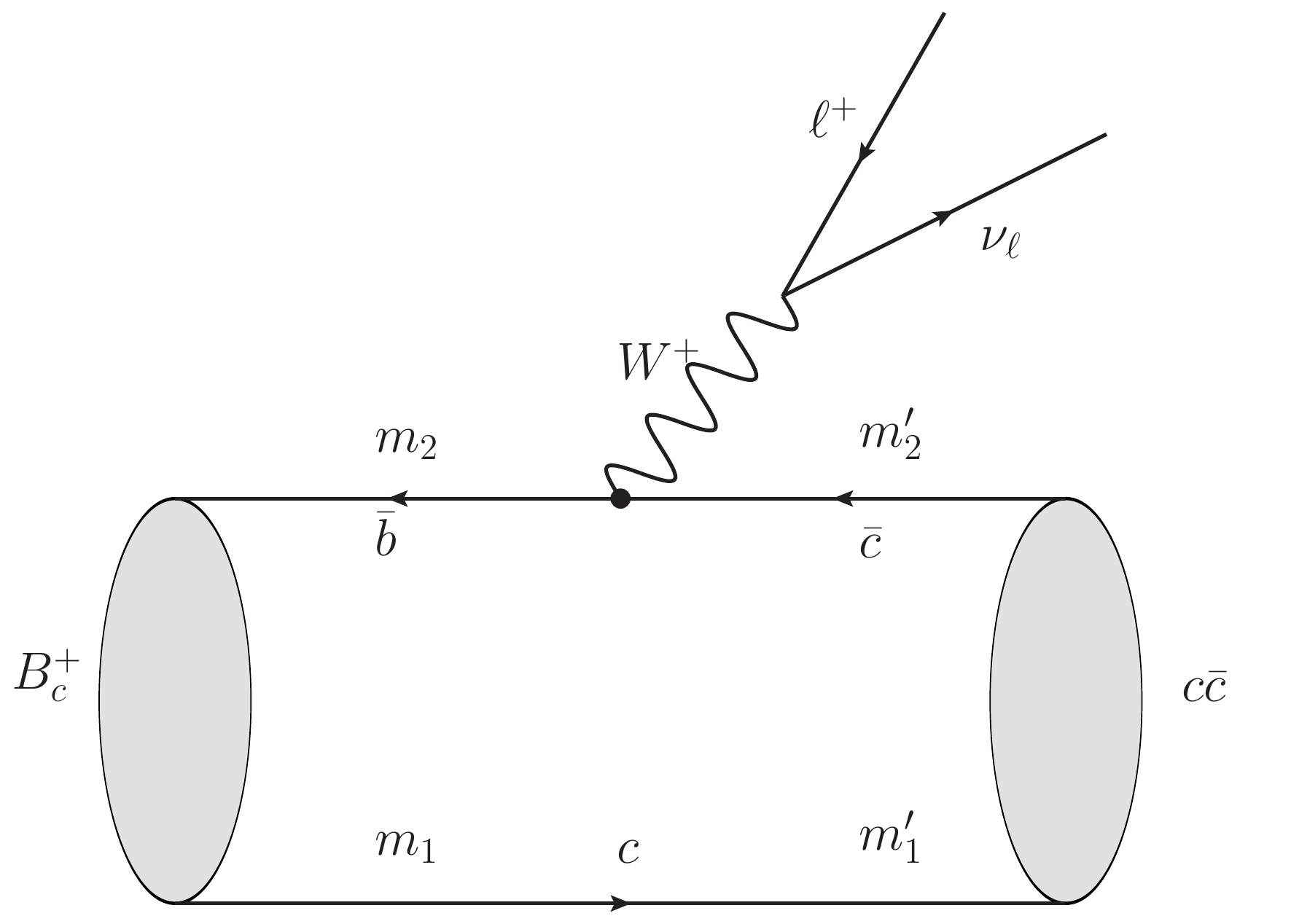}
\caption{Feynman diagram corresponding to the semileptonic decays $B_c^+\to (c\bar c)\ell^+\nu_\ell$.}\label{fig:feymanw}
\end{figure}

For the $B_c^+\to (c\bar c)\ell^+\nu_\ell$ processes shown in figure~\ref{fig:feymanw}, the transition amplitude element  reads
\begin{equation}
\begin{aligned}
T=\frac{G_F}{\sqrt 2}V_{cb}\bar{u}_{\nu_\ell}\gamma^\mu(1-\gamma_5)v_\ell\left\langle (c\bar c)(P_f)|J_\mu|B_c^+(P)\right\rangle,
\end{aligned}
\end{equation}
where $(c\bar c)$ denotes charmonium; $V_{cb}$ is the Cabibbo-Kobayashi-Maskawa (CKM) matrix element; $J_\mu\equiv V_\mu-A_\mu$ is the charged current responsible for the decays; $P$ and $P_f$ are the momenta of the initial $B_c^+$ and the final charmonium, respectively.

The hadronic transition element can be written as the overlapping integral over the initial and final relativistic BS wave functions within Mandelstam formalism. We would not solve the full BS equation, but the instantaneous one, namely, the full Salpeter equation. We perform the instantaneous approximation to the transition element \cite{Chang:2006tc} and write it as
\begin{equation}
\langle(c\bar c)| \bar b\gamma^\mu(1-\gamma^5)c|B_c^+\rangle = \int\frac{\ud\vec q}{(2\pi)^3} {\rm Tr}\bigg[\overline\varphi_{P_f}^{++}(\vec q\:')\frac{\slashed P}{M}\varphi_P^{++}(\vec q\:)\gamma^\mu(1-\gamma^5)\bigg],\label{eq:mandelstam}
\end{equation}
where $\varphi_P^{++}$ denotes the positive energy component of the instantaneous BS wave function of the initial state; $\overline\varphi_{P_f}^{++}\equiv\gamma^0(\varphi_{P_f}^{++})^\dagger\gamma^0$ is the Dirac conjugate of the positive energy component of the final state; $m'_1$ and $m'_2$ are the masses of quark and antiquark in the final state, respectively, and $\vec q\:'=\vec q-\frac{m'_1}{m'_1+m'_2} \vec P_f$ is the relative momentum between them. In this paper, we keep only the positive energy component $\varphi^{++}$ of the relativistic wave functions, because the contributions from other components are much smaller than $1\%$ in transition of $B_c \to (c\bar c)$ \cite{Wang:2016enc}. This matrix element can also be written in the framework in which the momentum $\vec q\:'$ is the integral argument by means of a suitable Jacobi transformation,
\begin{equation}\langle(c\bar c)| \bar b\gamma^\mu(1-\gamma^5)c|B_c^+\rangle = \int\frac{\vec q\:'^2\ud|\vec q\:'|\ud(\cos\theta)}{(2\pi)^2} {\rm Tr}\bigg[\overline\varphi_{P_f}^{++}(\vec q\:')\frac{\slashed P}{M}\varphi_P^{++}(\vec q\:)\gamma^\mu(1-\gamma^5)\bigg],\label{eq:mandelstam2}
\end{equation}
where $\vec q=\vec q\:'+\alpha \vec P_f, \alpha=\frac{m'_1}{m'_1+m'_2}$ and $\theta$ is the angle between $\vec q\:'$ and $\vec P_f$. The Eq.~(\ref{eq:mandelstam2}) is more convenient, because some matrix elements we calculate in this paper involve a $P$-wave final state \cite{Chang:2001pm}.

For $B_c^+\to P\ell^+\nu_\ell$ (here $P$ denotes $\eta_c$ or $\chi_{c0}$), the hadronic matrix element can be written as
\begin{equation}
\left\langle P|\bar b\gamma^\mu(1-\gamma^5)c|B_c^+\right\rangle=S_+(P+P_f)^\mu+S_-(P-P_f)^\mu\label{eq:0-ff},
\end{equation}
where $S_{+}$ and $S_{-}$ are the form factors.
For $B_c^+\to V\ell^+\nu_\ell$ (here $V$ denotes $J/\psi$, $h_c$ or $\chi_{c1}$), the hadronic matrix element can be written as
\begin{equation}
\langle V|\bar b\gamma^\mu(1-\gamma^5)c|B_c^+\rangle=(t_1 P^\mu+ t_2 P_f^\mu)\frac{\epsilon\cdot P}{M}+t_3(M+M_f)\epsilon^{\mu}+\frac{2t_4}{M+M_f}\mathrm i\varepsilon^{\mu\nu\sigma\delta}\epsilon_{\nu} P_\sigma P_{f\delta},
\end{equation}
where $\epsilon^\mu$ is the polarization vector of the final vector meson; $t_1$, $t_2$, $t_3$ and $t_4$ are the form factors.
For $B_c^+\to T\ell^+\nu_\ell$ (here $T$ denotes $\chi_{c2}$), the hadronic matrix element can be written as
\begin{equation}
\begin{aligned}
\langle T|\bar b\gamma^\mu(1-\gamma^5)c|B_c^+\rangle&=(t_1 P^\mu+ t_2 P_f^\mu)\epsilon_{\alpha\beta}\frac{P^\alpha P^\beta}{M^2}+t_3(M+M_f)\epsilon^{\mu\alpha}\frac{P_\alpha}{M}\\
&+\frac{2t_4}{M+M_f}\mathrm i\varepsilon^{\mu\beta\sigma\delta}\epsilon_{\alpha\beta}\frac{P^\alpha}{M} P_\sigma P_{f\delta},
\end{aligned}
\end{equation}
where $\epsilon_{\alpha\beta}$ is the polarization tensor of the final tensor meson; $t_1$, $t_2$, $t_3$ and $t_4$ are the form factors.

The summation formulas for polarization of the final vector meson are
\begin{equation}
\begin{aligned}
\epsilon_\mu^{(\lambda)}(P_f)P_f^\mu&=0,\\
\sum_{\lambda=0,\pm}\epsilon_\mu^{(\lambda)}(P_f)\epsilon_\nu^{\dagger(\lambda)}(P_f)&=-g_{\mu\nu}+\frac{P_{f\mu}P_{f\nu}}{M_f^2}.
\end{aligned}
\end{equation}
The summation formulas for polarization of the final tensor meson are
\begin{equation}
\begin{aligned}
\epsilon_{\alpha\beta}^{(\lambda)}(P_f)P_f^\alpha&=0,\\
\sum_{\lambda=0,\pm 1,\pm 2}\epsilon_{\mu\nu}^{(\lambda)}(P_f)\epsilon_{\alpha\beta}^{\dagger(\lambda)}(P_f)&=\frac{1}{2}(S_{\mu\alpha}S_{\nu\beta}+S_{\mu\beta}S_{\nu\alpha})-\frac{1}{3}S_{\mu\nu}S_{\alpha\beta},
\end{aligned}
\end{equation}
Where $S_{\mu\nu}=-g_{\mu\nu}+\frac{P_{f\mu}P_{f\nu}}{M_f^2}$.
Finally, the semileptonic decay width can be expressed as
\begin{equation}
\Gamma=\frac{1}{8M(2\pi)^3}\int\frac{|\vec P_\ell|}{E_\ell}\ud |\vec P_\ell|\int\frac{|\vec P_f|}{E_f}\ud |\vec P_f|\sum_{\lambda}|T|^2\label{eq:width},
\end{equation}
where $\vec P_\ell$ is the three-dimensional momentum of the final lepton, and $\vec P_f$ is the three-dimensional momentum of the final meson. In this paper, we only calculate the form factors but no longer calculate the decay widths.

\section{\label{sec:wavefunction} Relativistic wave function}

Usually, the nonrelativistic wave function for a pseudoscalar is written as \cite{Chang:1992pt}
\begin{equation}
\Psi_P(\vec{q}\:)=(\slashed P+M)\gamma_5 f(\vec{q}\:)\label{eq:schrodinger},
\end{equation}
where $M$ and $P$ are the mass and momentum of the meson, respectively; $\vec{q}$ is the relative momentum between the quark and antiquark in the meson, and the radial wave function $f(\vec{q})$ can be obtained numerically by solving the Schrodinger equation.

But in our method, we solve the full Salpeter equation. The form of wave function is relativistic and depends on the $J^{P(C)}$ quantum number of the corresponding meson. For a pseudoscalar, the relativistic wave function can be written as the four items constructed by $P$, $q_{\perp}$ and $\gamma$-matrices \cite{Kim:2003ny}
\begin{equation}
\begin{aligned}
\varphi_{0^-}(q_\perp)&=M\left[\frac{\slashed{P}}{M}f_1(q_\perp)
+f_2(q_\perp)+\frac{\slashed{q}_\perp}{M}f_3(q_\perp)+\frac{\slashed{P}\slashed{q}_\perp}{M^2}f_4(q_\perp)\right]\gamma_5,
\end{aligned}\label{eq:BS wave function}
\end{equation}
where $q=p_1-\alpha_1 P=\alpha_2 P-p_2$ is the relative momentum between quark (with momentum $p_1$ and mass $m_1$) and antiquark (momentum $p_2$ and mass $m_2$), $\alpha_1=\frac{m_1}{m_1+m_2}$, $\alpha_2=\frac{m_2}{m_1+m_2}$;  $q_{\perp}=q-\frac{P\cdot q}{M^2}P$, in the rest frame of the meson, $q_{\perp}=(0,\vec{q})$.

All the items in the wave function Eq. (\ref{eq:BS wave function}) have the quantum number of $0^-$. This wave function is a general relativistic form for a pseudoscalar with the instantaneous approximation. If we set the items with $f_3$ and $f_4$ to zero, and set $f_1=f_2$, the relativistic wave function is reduced to the  Schrodinger wave function Eq.~(\ref{eq:schrodinger}).

Taking into account the last two equations in Eq. (\ref{eq:phi}), we obtain the relations
\begin{equation}
\begin{aligned}
f_3(q_\perp)&=\frac{M(\omega_2-\omega_1)}{(m_1\omega_2+m_2\omega_1)}f_1, \\
f_4(q_\perp)&=-\frac{M(\omega_1+\omega_2)}{(m_1\omega_2+m_2\omega_1)}f_2,
\end{aligned}
\end{equation}
where the quark energy $\omega_i=\sqrt{m^2_i-q_{\perp}^2}=\sqrt{m^2_i+\vec{q}^{\:2}}$ ($i=1,2$).
The wave function corresponding to the positive energy projection has the form
\begin{equation}
\begin{aligned}
\varphi_{0^-}^{++}(q_{\perp})=\left[A_1(q_{\perp})+\frac{\slashed P}{M}A_2(q_{\perp})+\frac{\slashed q_{\perp}}{M}A_3(q_{\perp})+\frac{\slashed P\slashed q_{\perp}}{M^2}A_4(q_{\perp})\right]\gamma^5\label{eq:++wave},
\end{aligned}
\end{equation}
where
\begin{equation}
\begin{aligned}
A_1&=\frac{M}{2}\left[\frac{\omega_1+\omega_2}{m_1+m_2}f_1+f_2\right],\qquad A_3=-\frac{M(\omega_1-\omega_2)}{m_1\omega_2+m_2\omega_1}A_1,\\
A_2&=\frac{M}{2}\left[f_1+\frac{m_1+m_2}{\omega_1+\omega_2}f_2\right],\qquad A_4=-\frac{M(m_1+m_2)}{m_1\omega_2+m_2\omega_1}A_1\label{eq:A}.
\end{aligned}
\end{equation}

The normalization condition reads
\begin{equation}
\int\frac{\ud\vec q}{(2\pi)^3}4f_1f_2M^2\left\{\frac{m_1+m_2}{\omega_1+\omega_2}+\frac{\omega_1+\omega_2}{m_1+m_2}+\frac{2\vec q^{\:2}(m_1\omega_1+m_2\omega_2)}{(m_2\omega_1+m_1\omega_2)^2}\right\}=2M.\label{eq:norma}
\end{equation}

By solving the full Salpeter equation, the numerical values of wave functions $f_1$ and $f_2$ are obtained. The positive energy component Eq.~(\ref{eq:++wave}) is brought into the Mandelstam formula Eq.~(\ref{eq:mandelstam}). After the trace and integral are finished, the form factors $S_+$ and $S_-$ can be calculated numerically. In this paper, besides the wave function for $0^-$ state, we also need the wave functions for the states of $1^{--}$ ($J/\psi$), $1^{+-}$ ($h_c$), $0^{++}$ ($\chi_{c0}$), $1^{++}$ ($\chi_{c1}$) and $2^{++}$ ($\chi_{c2}$). We put the $2^{++}$ state wave function in the appendix \ref{appendix} and the others can be referred to \cite{Geng:2018qrl}.

\section{\label{sec:IWF}Isgur-Wise function}

Since $B_c$ and charmonium are the weak-binding states, the approximation 
\begin{equation}
\omega_i\equiv\sqrt{m_i^2+\vec{q}\:^2}\approx m_i+\frac{\vec{q}\:^2}{2m_i}\label{eq:weakbind}
\end{equation}
is taken in this paper. This approximation requires the three-dimensional relative momentum $|\vec q|$ between quarks much less than the masses of quarks, but its contribution will be suppressed by the wave function $f_i(\vec q)$ in the large $|\vec q|$ interval. After performing this approximation and the trace on the matrix element Eq.~(\ref{eq:mandelstam2}), the dependence of all the form factors on the overlapping integrals of the wave functions for the initial state and the final state becomes transparent. For instance, one type of overlapping integrals are
\begin{equation}
\begin{split}
&\int\frac{\ud \vec q\:'}{(2\pi)^3}f_1(|\vec q|)f'_1(|\vec q\:'|),~~\int\frac{\ud \vec q\:'}{(2\pi)^3}f_1(|\vec q|)f'_2(|\vec q\:'|),\\
&\int\frac{\ud \vec q\:'}{(2\pi)^3}f_2(|\vec q|)f'_1(|\vec q\:'|),~~\int\frac{\ud \vec q\:'}{(2\pi)^3}f_2(|\vec q|)f'_2(|\vec q\:'|),
\end{split}\label{eq:overlap}
\end{equation}
where $f_i$ denotes the wave function of initial state, and $f'_i$ denotes the wave function of final state. Two wave functions from the same meson is very close numerically, i.e., $f_1\approx f_2$ and $f'_1\approx f'_2$. So the four overlapping integrals in Eq.~(\ref{eq:overlap}) are approximately equal, and for convenience they are replaced by their average which is denoted as
\begin{equation}
\xi_{00}(v\cdot v')=C\int\frac{\ud \vec q\:'}{(2\pi)^3}\overline{ff'},\label{eq:iw00}
\end{equation}
where $C$ is the normalized coefficient; $v,v'$ are the four dimensional velocities of the initial state and final state respectively, and $\overline{ff'}=\frac{f_1f'_1+f_1f'_2+f_2f'_1+f_2f'_2}{4}$. There are other overlapping integrals with the relative momentum $\vec q\:'$ being inserted. They may be the relativistic corrections to the function $\xi_{00}$. We denote them as $\xi_{qx}$, where subscript $q$ denotes the power of the relative momentum $\vec q\:'$, subscript $x$ denotes the power of $\cos\theta$, and $\theta$ is the angle between $\vec q\:'$ and $\vec P_f$, i.e.,
\begin{equation}
\begin{split}
\xi_{11}&=C\int\frac{\ud \vec q\:'}{(2\pi)^3}\overline{ff'}\frac{|\vec q\:'|\cos\theta}{\sqrt{MM'}},\quad\xi_{20}=C\int\frac{\ud \vec q\:'}{(2\pi)^3}\overline{ff'}\frac{\vec q\:'^2}{MM'},\\
\xi_{22}&=C\int\frac{\ud \vec q\:'}{(2\pi)^3}\overline{ff'}\frac{\vec q\:'^2\cos^2\theta}{MM'},\quad\xi_{31}=C\int\frac{\ud \vec q\:'}{(2\pi)^3}\overline{ff'}\frac{|\vec q\:'|^3\cos\theta}{\sqrt{(MM')^3}},\\
\xi_{33}&=C\int\frac{\ud \vec q\:'}{(2\pi)^3}\overline{ff'}\frac{|\vec q\:'|^3\cos^3\theta}{\sqrt{(MM')^3}},\quad\xi_{40}=C\int\frac{\ud \vec q\:'}{(2\pi)^3}\overline{ff'}\frac{\vec q\:'^4}{(MM')^2},\\
\xi_{42}&=C\int\frac{\ud \vec q\:'}{(2\pi)^3}\overline{ff'}\frac{\vec q\:'^4\cos^2\theta}{(MM')^2},\label{eq:iwother}
\end{split}
\end{equation}
and so on. When the final state is S-wave meson, we keep the first six functions and abandon the higher order $\mathcal O(q^4)$; When the final state is P-wave meson whose wave function includes a $\vec q$, $\xi_{00}$ disappears, thus we reserve the first eight functions and abandon the higher order $\mathcal O(q^5)$. The normalized coefficients based on the normalized formulas are shown in Table.~\ref{tab:coeff} for each process. Taking the process $B_c\to\eta_c$ as an example, the initial and final states are both $0^-$ state. With the approximations $\vec q=0$, $f_1=f_2$ and $\omega_i=m_i$, Eq~.(\ref{eq:norma}) is deduced as
\begin{equation}
\int\frac{\ud\vec q}{(2\pi)^3}4Mf^2=1.\label{eq:norma2}
\end{equation}
So the normalized wave function of $0^-$ state is $2\sqrt{M}f$, and the normalized coefficient is $4\sqrt{MM'}$ for the process $B_c\to\eta_c$. 

\begin{table}[!hbp]
\caption{ The normalization coefficients of different processes.}
\vspace{0.2cm}
\setlength{\tabcolsep}{0.2cm}
\centering
\begin{tabular*}{\textwidth}{@{}@{\extracolsep{\fill}}ccccccccc}
\hline\hline
 & final state & $\eta_c$ & $J/\psi$ & $h_c$ & $\chi_{c0}$ & $\chi_{c1}$ & $\chi_{c2}$ &\\
\hline 
& $C$ & $4\sqrt{MM'}$  & $4\sqrt{MM'}$  & $\frac{4M}{\sqrt{3}} $ & $ 4M $ & $4\sqrt{\frac{2}{3}}M$ & $\frac{4MM'}{\sqrt{3}}$ &\\
\hline\hline
\end{tabular*}
\label{tab:coeff}
\end{table}

The form factors of semileptonic decay $B_c\to\eta_c\ell\nu_\ell$ can be written as
\begin{equation}
\begin{split}
S_+&=-\frac{M+M'}{2\sqrt{MM'}}\xi_{00}+\frac{1}{4\sqrt{MM'}}\left[b_1\frac{1}{m_1}+b_2\frac{1}{m_2}\right]\alpha\xi_{00}\\
&+\frac{1}{4P'}\left[-b_1\frac{1}{m_1}-b_2\frac{1}{m_2}+a_1\frac{1}{m'_1}+a_2\frac{1}{m'_2}\right]\xi_{11}+\frac{(M-M')P'^2}{8\sqrt{MM'}}\frac{1}{m_1m_2}\alpha^2\xi_{00}\\
&+\frac{P'}{8}\left[(E'+M')\left(\frac{1}{m_1m'_1}+\frac{1}{m_2m'_2}\right)+(E'-M')\left(\frac{1}{m_1m'_2}+\frac{1}{m_2m'_1}\right)-(M-M')\frac{1}{m_1m_2}\right]\alpha\xi_{11}\\
&+\frac{\sqrt{MM'}}{8}\left[(M-M')\left(\frac{1}{m_1m_2}-\frac{1}{m_1m'_2}-\frac{1}{m_2m'_1}+\frac{1}{m'_1m'_2}\right)-(M+M')\left(\frac{1}{m_1m'_1}+\frac{1}{m_2m'_2}\right)\right]\xi_{20}\\
&+\frac{\sqrt{MM'}}{8}(M-E')\left[\frac{1}{m_1m'_1}+\frac{1}{m_1m'_2}+\frac{1}{m_2m'_1}+\frac{1}{m_2m'_2}\right]\xi_{22}\\
&-\frac{P'^2}{16\sqrt{MM'}}\left[b_1\frac{1}{m_1^3}+b_2\frac{1}{m_2^3}\right]\alpha^3\xi_{00}+\frac{P'}{16}\left[b_1\frac{3}{m_1^3}+b_2\frac{3}{m_2^3}+a_1\frac{1}{m_1m_2m'_2}+a_2\frac{1}{m_1m_2m'_1}\right]\alpha^2\xi_{11}\\
&-\frac{\sqrt{MM'}}{16}\left[b_1\left(\frac{1}{m_1^3}-\frac{1}{m_2m'_1m'_2}\right)+b_2\left(\frac{1}{m_2^3}-\frac{1}{m_1m'_1m'_2}\right)\right]\alpha\xi_{20}\\
&-\frac{\sqrt{MM'}}{8}\left[b_1\frac{1}{m_1^3}+b_2\frac{1}{m_2^3}+a_1\frac{1}{m_1m_2m'_2}+a_2\frac{1}{m_1m_2m'_1}\right]\alpha\xi_{22}\\
&+\frac{MM'}{16P'}\left[b_1\left(\frac{1}{m_1^3}-\frac{1}{m_2m'_1m'_2}\right)+b_2\left(\frac{1}{m_2^3}-\frac{1}{m_1m'_1m'_2}\right)\right.\\
&\left.-a_1\left(\frac{1}{m_1^{'3}}-\frac{1}{m_1m_2m'_2}\right)-a_2\left(\frac{1}{m_2^{'3}}-\frac{1}{m_1m_2m'_1}\right)\right]\xi_{31}\\
\end{split}\label{eq:ffs+}
\end{equation}
where $a_1=E'^2-E'M+E'M'-MM'=M'(E'-M)(\omega+1)$,~$a_2=E'^2-E'M-E'M'+MM'=M'(E'-M)(\omega-1)$,~$b_1=MM'-E'M'+E'M-M'^2=M'(M-M')(1+\omega)$,~$b_2=MM'-E'M'-E'M+M'^2=M'(M+M')(1-\omega)$,~$b_1=\vec P_f^2-a_1$,~$b_2=a_2-\vec P_f^2$.

\begin{equation}
\begin{split}
S_-&=\frac{M-M'}{2\sqrt{MM'}}\xi_{00}+\frac{1}{4\sqrt{MM'}}\left[-c_1\frac{1}{m_1}+c_2\frac{1}{m_2}\right]\alpha\xi_{00}\\
&+\frac{1}{4P'}\left[c_1\frac{1}{m_1}-c_2\frac{1}{m_2}+d_1\frac{1}{m'_1}+d_2\frac{1}{m'_2}\right]\xi_{11}-\frac{(M+M')P'^2}{8\sqrt{MM'}}\frac{1}{m_1m_2}\alpha^2\xi_{00}\\
&+\frac{P'}{8}\left[(M'+E')\left(\frac{1}{m_1m'_1}+\frac{1}{m_2m'_2}\right)+(E'-M')\left(\frac{1}{m_1m'_2}+\frac{1}{m_2m'_1}\right)+2(M+M')\frac{1}{m_1m_2}\right]\alpha\xi_{11}\\
&+\frac{\sqrt{MM'}}{8}\left[(M+M')\left(-\frac{1}{m_1m_2}+\frac{1}{m_1m'_2}+\frac{1}{m_2m'_1}-\frac{1}{m'_1m'_2}\right)+(M-M')\left(\frac{1}{m_1m'_1}+\frac{1}{m_2m'_2}\right)\right]\xi_{20}\\
&-\frac{\sqrt{MM'}}{8}(M+E')\left[\frac{1}{m_1m'_1}+\frac{1}{m_1m'_2}+\frac{1}{m_2m'_1}+\frac{1}{m_2m'_2}\right]\xi_{22}\\
&+\frac{P'^2}{16\sqrt{MM'}}\left[c_1\frac{1}{m_1^3}-c_2\frac{1}{m_2^3}\right]\alpha^3\xi_{00}+\frac{P'}{16}\left[-c_1\frac{3}{m_1^3}+c_2\frac{3}{m_2^3}+d_1\frac{1}{m_1m_2m'_2}+d_2\frac{1}{m_1m_2m'_1}\right]\alpha^2\xi_{11}\\
&+\frac{\sqrt{MM'}}{16}\left[c_1\left(\frac{1}{m_1^3}-\frac{1}{m_2m'_1m'_2}\right)-c_2\left(\frac{1}{m_2^3}-\frac{1}{m_1m'_1m'_2}\right)\right]\alpha\xi_{20}\\
&-\frac{\sqrt{MM'}}{8}\left[c_1\frac{1}{m_1^3}-c_2\frac{1}{m_2^3}-d_1\frac{1}{m_1m_2m'_2}-d_2\frac{1}{m_1m_2m'_1}\right]\alpha\xi_{22}\\
&+\frac{MM'}{16P'}\left[-c_1\left(\frac{1}{m_1^3}-\frac{1}{m_2m'_1m'_2}\right)+c_2\left(\frac{1}{m_2^3}-\frac{1}{m_1m'_1m'_2}\right)\right.\\
&\left.-d_1\left(\frac{1}{m_1^{'3}}-\frac{1}{m_1m_2m'_2}\right)-d_2\left(\frac{1}{m_2^{'3}}-\frac{1}{m_1m_2m'_1}\right)\right]\xi_{31}\\
\end{split}\label{eq:ffs-}
\end{equation}
where $c_1=E'M+E'M'+MM'+M'^2$,~$c_2=E'M-E'M'-MM'+M'^2$,~$d_1=E'^2+E'M+E'M'+MM'$,~$d_2=E'^2+E'M-E'M'-MM'$,~$c_1=d_1-\vec P_f^2$,~$c_2=d_2-\vec P_f^2$.

The function $\xi_{00}$ may be directly related to the Isgur-Wise function appearing in HQET for $0^-\to 0^-$ decays. Because the form factors in this process will degenerate into those in the nonrelativistic limit if only the function $\xi_{00}$ is considered \cite{Neubert:1996qg}, 
\begin{equation}
\begin{split}
\langle \eta_c|\bar b\gamma^\mu(1-\gamma^5)c|B_c^+\rangle&=-\sqrt{MM_f}\left[v^\mu+v_f^\mu\right]\xi_{00},\\
S_{\pm}&=\mp\frac{M\pm M'}{2\sqrt{MM'}}\xi_{00}.\label{eq:limit0-}
\end{split}
\end{equation}
The other functions are the relativistic corrections ($1/m_i$ corrections) to the leading order IWF $\xi_{00}$, where $i$ denote a quark or anti-quark in the initial and final mesons. The number of $\vec q\:'$ contained in the function $\xi_{qx}$ (subscript $q$) corresponds to the order of the correction. Note that there should have been another type of overlapping integrals with the relative momentum $\vec q$ in the initial state. For example,
\begin{equation}
C\int\frac{\ud \vec q\:'}{(2\pi)^3}\overline{ff'}\frac{|\vec q\:|\cos\beta}{\sqrt{MM'}},\label{eq:qoverlap}
\end{equation}
where $\beta$ is the angle between $\vec q$ and $\vec P_f$. Due to the relation $\vec q=\vec q\:'+\alpha \vec P_f, \alpha=\frac{m'_1}{m'_1+m'_2}$, this overlapping integral Eq.~(\ref{eq:qoverlap}) is decomposed into $\xi_{11}+\alpha|\vec P_f\:|\xi_{00}$. So the item involving $\alpha\xi_{00}$ should be considered as relativistic correction of the same order as $\xi_{11}$. Generally, the item involving $\alpha^n\xi_{qx}$ is the $q+n$ order relativistic correction ($1/m_i^{q+n}$ correction) which can be confirmed in Eq.~(\ref{eq:ffs+}) and (\ref{eq:ffs-}). The process $0^-\to 1^{--}$ is the same as above case. The leading order result is agree with HQET \cite{Neubert:1996qg}, i.e.,
\begin{equation}
\langle J/\psi|\bar b\gamma^\mu(1-\gamma^5)c|B_c^+\rangle=\sqrt{MM_f}\left[\epsilon\cdot v v_f^\mu-(v\cdot v_f+1)\epsilon^{\mu}+\mathrm i\varepsilon^{\mu\nu\sigma\delta}\epsilon_{\nu} v_\sigma v_{f\delta}\right]\xi_{00}.\label{eq:limit1--}
\end{equation}
It is very natural that the leading order results in this paper are entirely consistent with HQET for $0^-\to 0^-$ or $1^{--}$ processes. Because for the leading order results, the terms involving $\slashed q$ disappear and $\omega_i=m_i$, so that the BS wave functions degenerate into the nonrelativistic case, i.e., 
\begin{equation}
0^-:\frac{M+\slashed P}{2\sqrt M}\gamma^5\Psi,\quad 1^{--}:\frac{M+\slashed P}{2\sqrt M}\slashed\epsilon \Psi.\\\label{eq:nrwf1}
\end{equation}
A pseudoscalar meson and its corresponding vector have the same radial wave function $\Psi$ in the nonrelativistic limit. But in this paper, the radial wave functions are obtained by solving BS equation, and the $\Psi$ in Eq.~(\ref{eq:nrwf1}) corresponds to the normalized wave function $2\sqrt{M}f_i$. They are not exactly the same numerically. And then, the numerical results of IWF is not close to HQET and contains part of the relativistic correction.

For P-wave meson as the final state, the nonrelativistic wave functions are usually written as
\begin{equation}
\begin{split}
0^{++}&:\frac{\slashed q_{\perp}}{|\vec q~|}\frac{M+\slashed P}{2\sqrt M}\Phi,\quad 1^{++}:\mathrm i\varepsilon_{\mu\nu\alpha\beta}\sqrt\frac{3}{2}\frac{P^\nu}{M}\frac{q_\perp^\alpha}{|\vec q~|}\epsilon^\beta\frac{M+\slashed P}{2\sqrt M}\gamma^\mu\Phi,\\
2^{++}&:\sqrt{3}\epsilon_{\mu\nu}\gamma^\mu\frac{q_\perp^\nu}{|\vec q~|}\frac{M+\slashed P}{2\sqrt M}\Phi,\quad 1^{+-}:\sqrt 3\frac{q_{\perp}\cdot\epsilon}{|\vec q~|}\frac{M+\slashed P}{2\sqrt M}\gamma^5\Phi.\label{eq:nrwf2}
\end{split}
\end{equation}
and these states have the same radial wave function $\Phi$. Similarly in this paper, the radial wave functions are obtained by solving BS equation, and the $\Phi$ in Eq.~(\ref{eq:nrwf2}) corresponds to the normalized wave function. In these cases, $\xi_{00}$ disappears and $\xi_{11}$ is IWF. Due to the change of orbital angular momentum, this weak decay process can not correspond to a scattering process simply. We give the leading order results in the case that only the function $\xi_{11}$ is considered,
\begin{equation}
\begin{split}
\langle h_c|\bar b\gamma^\mu(1-\gamma^5)c|B_c^+\rangle&=\sqrt{3MM_f}\frac{v\cdot v_f}{|\vec v_f|}(\epsilon\cdot v)\left[v^\mu+v_f^\mu\right]\xi_{11},\\
\langle \chi_{c0}|\bar b\gamma^\mu(1-\gamma^5)c|B_c^+\rangle&=-\sqrt{MM_f}\frac{v\cdot v_f+1}{|\vec v_f|}\left[v\cdot v_fv^\mu-v_f^\mu\right]\xi_{11},\\
\langle \chi_{c1}|\bar b\gamma^\mu(1-\gamma^5)c|B_c^+\rangle&=\sqrt{\frac{3MM_f}{2}}\frac{v\cdot v_f}{|\vec v_f|}\left[\epsilon\cdot v(v^\mu-v\cdot v_fv_f^\mu)\right.\\
&\left.+\vec v_f^2\epsilon^{\mu}+\mathrm i(v\cdot v_f+1)\varepsilon^{\mu\nu\sigma\delta}\epsilon_{\nu} v_\sigma v_{f\delta}\right]\xi_{11},\\
\langle\chi_{c2}|\bar b\gamma^\mu(1-\gamma^5)c|B_c^+\rangle&=-\sqrt{3MM_f}\frac{v\cdot v_f}{|\vec v_f|}\left[\epsilon_{\alpha\beta}v^\alpha v^\beta v_f^\mu-(v\cdot v_f+1)\epsilon^{\mu\alpha}v_\alpha\right.\\
&\left.+\mathrm i\epsilon_{\alpha\beta}v^\alpha\varepsilon^{\mu\beta\sigma\delta}v_\sigma v_{f\delta}\right]\xi_{11}.\label{eq:heavylimit}
\end{split}
\end{equation}
These results are not agree with Ref.~\cite{Wang:2018duy}. The latter analyzes the reduction of form factors in the heavy quark limit, and there are two IWFs $\xi_E,\xi_Fv_\alpha$ for $B_c$ to P-wave charmonium. While we only need IWF $\xi_{11}$ for these processes in the leading order. Ref.~\cite{Wang:2018duy} does not further describe the used IWFs. The difference needs further examination. Note that the above results are not confined to the processes of $B_c$ to charmonium, but hold true for any possible process where the initial and final mesons are corresponding $J^{PC}$ states. In the next section, we will give the numerical results and discussions on the specific processes.

\section{\label{sec:results}Results and Discussions}

\begin{figure}[!hbp]
\centering
\subfigure[$B_c$ 介子]{\label{fig:wfbc}
			      \includegraphics[width=0.3\textwidth]{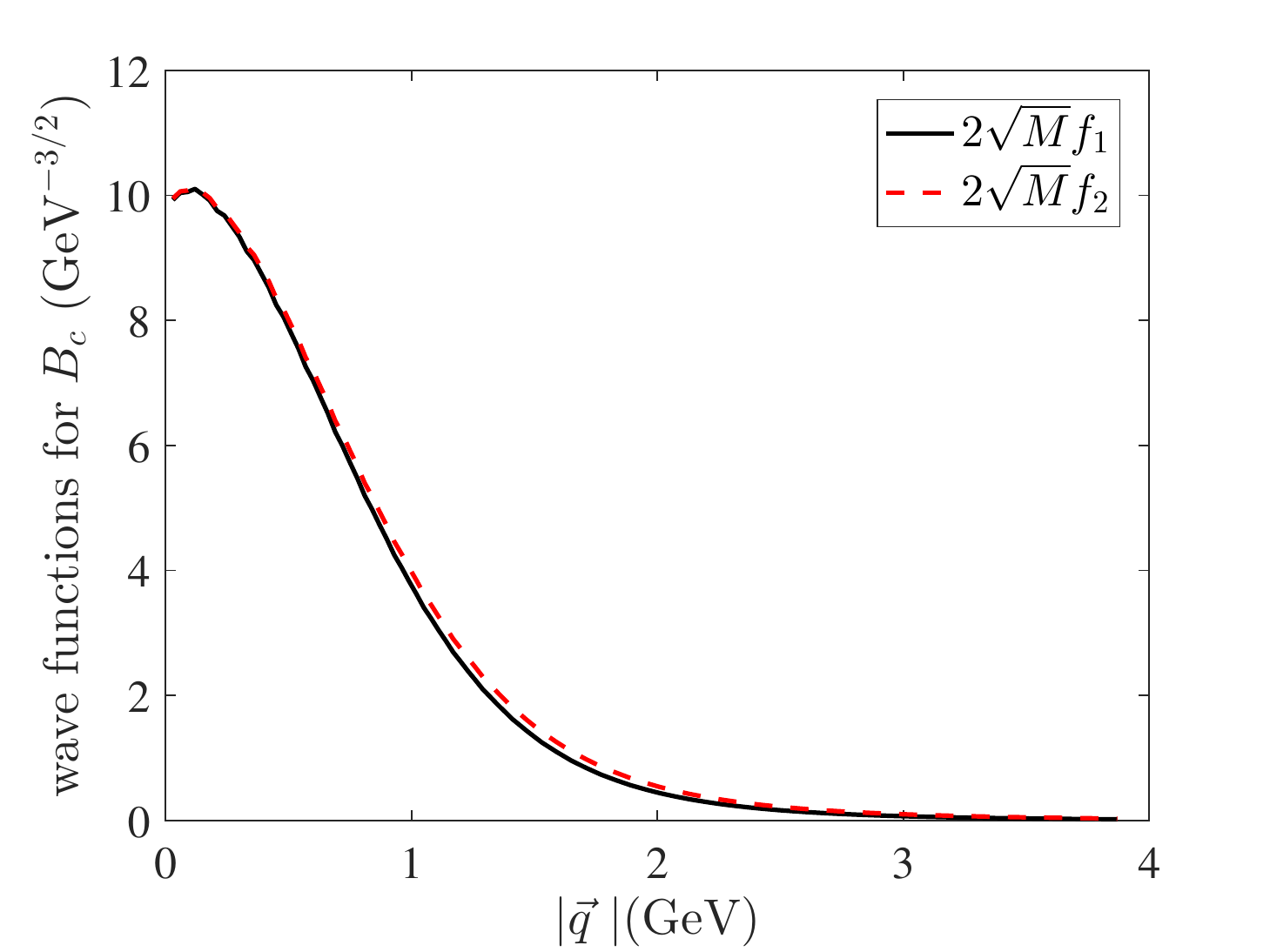}}
\subfigure[$\eta_c$ 介子]{\label{fig:wfetac}
			      \includegraphics[width=0.3\textwidth]{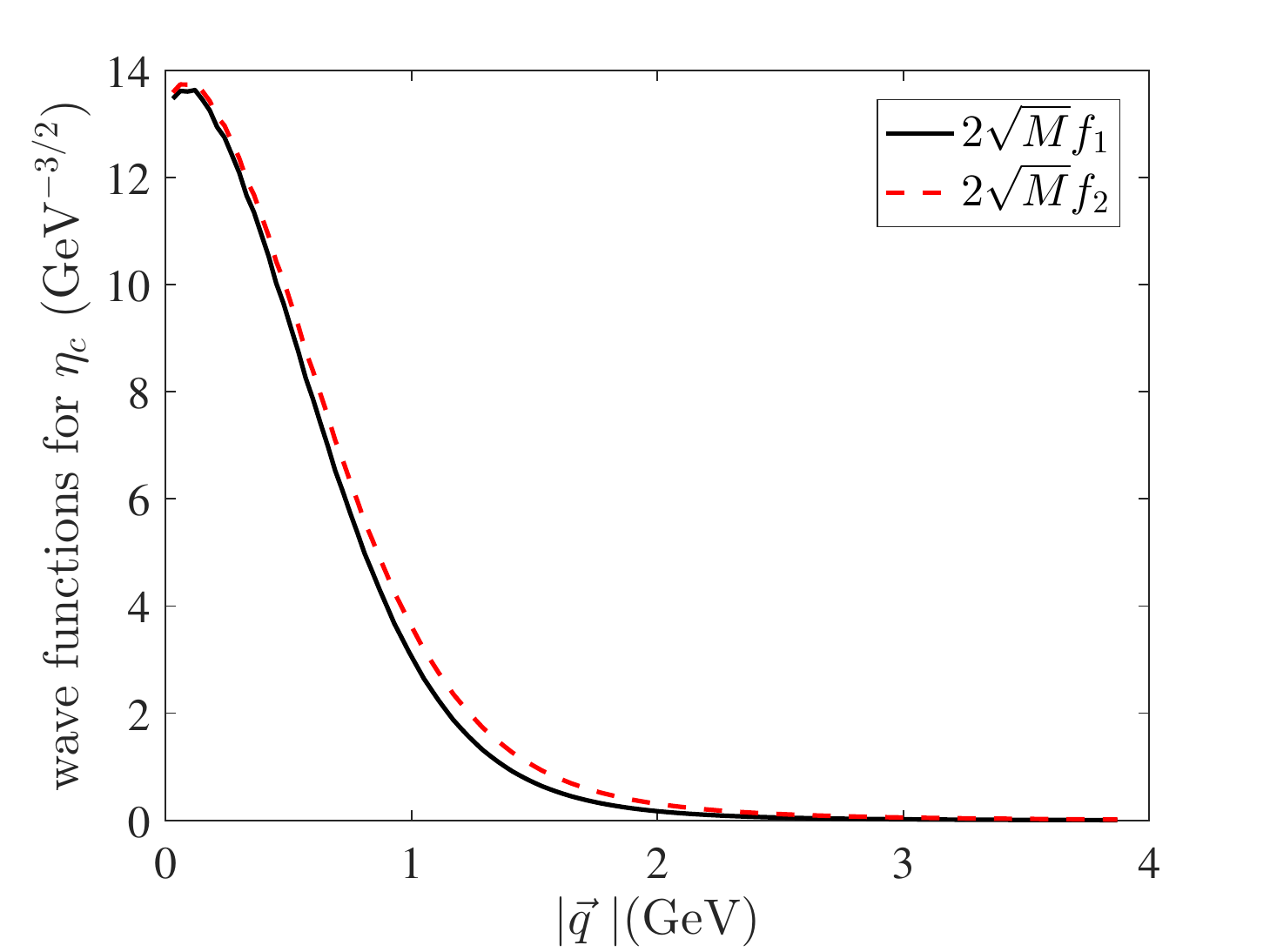}}
\subfigure[$J/\psi$ 介子]{\label{fig:wfjpsi}
			      \includegraphics[width=0.3\textwidth]{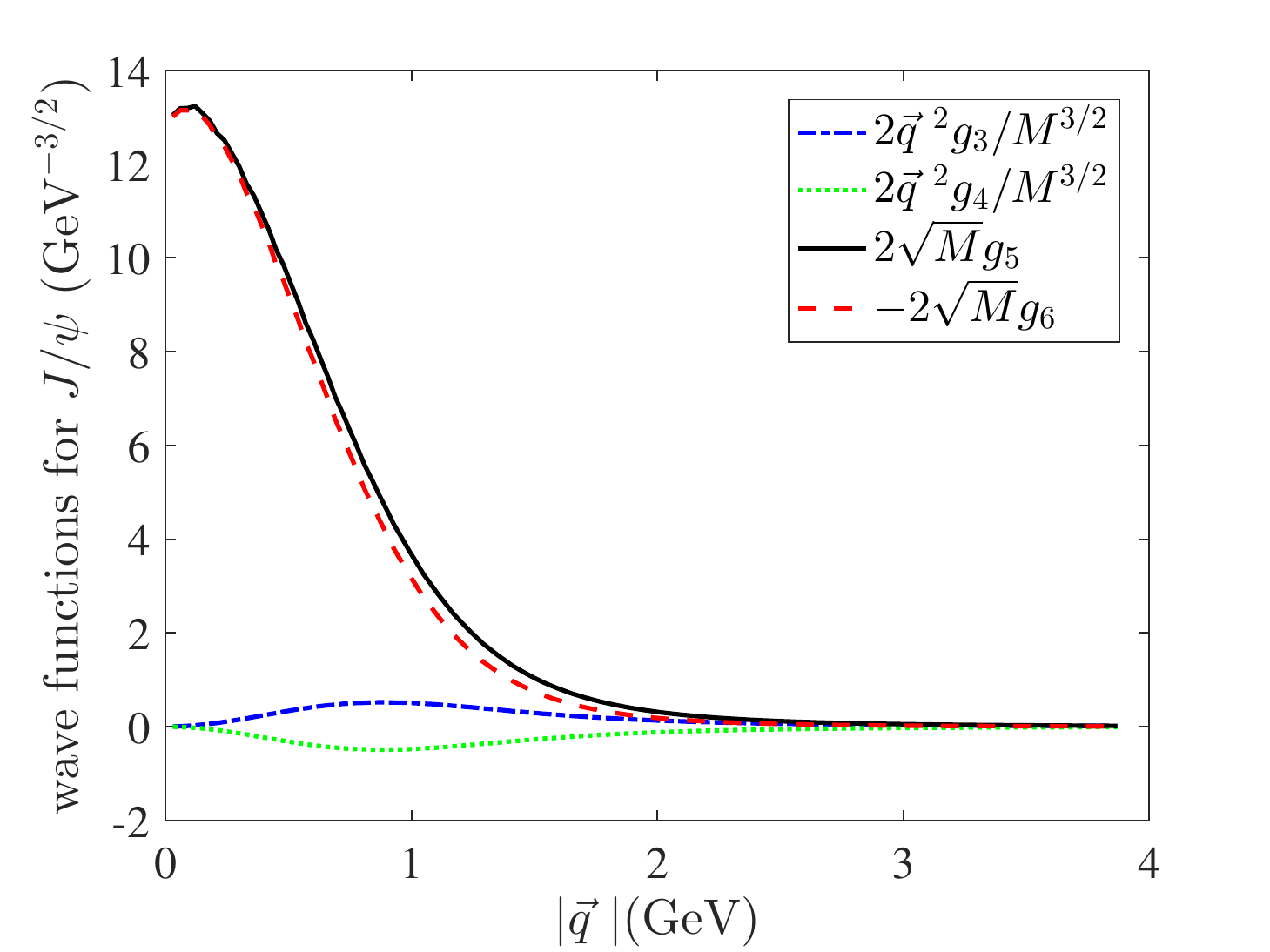}}
\subfigure[$h_c$ 介子]{\label{fig:wfhc}
			      \includegraphics[width=0.3\textwidth]{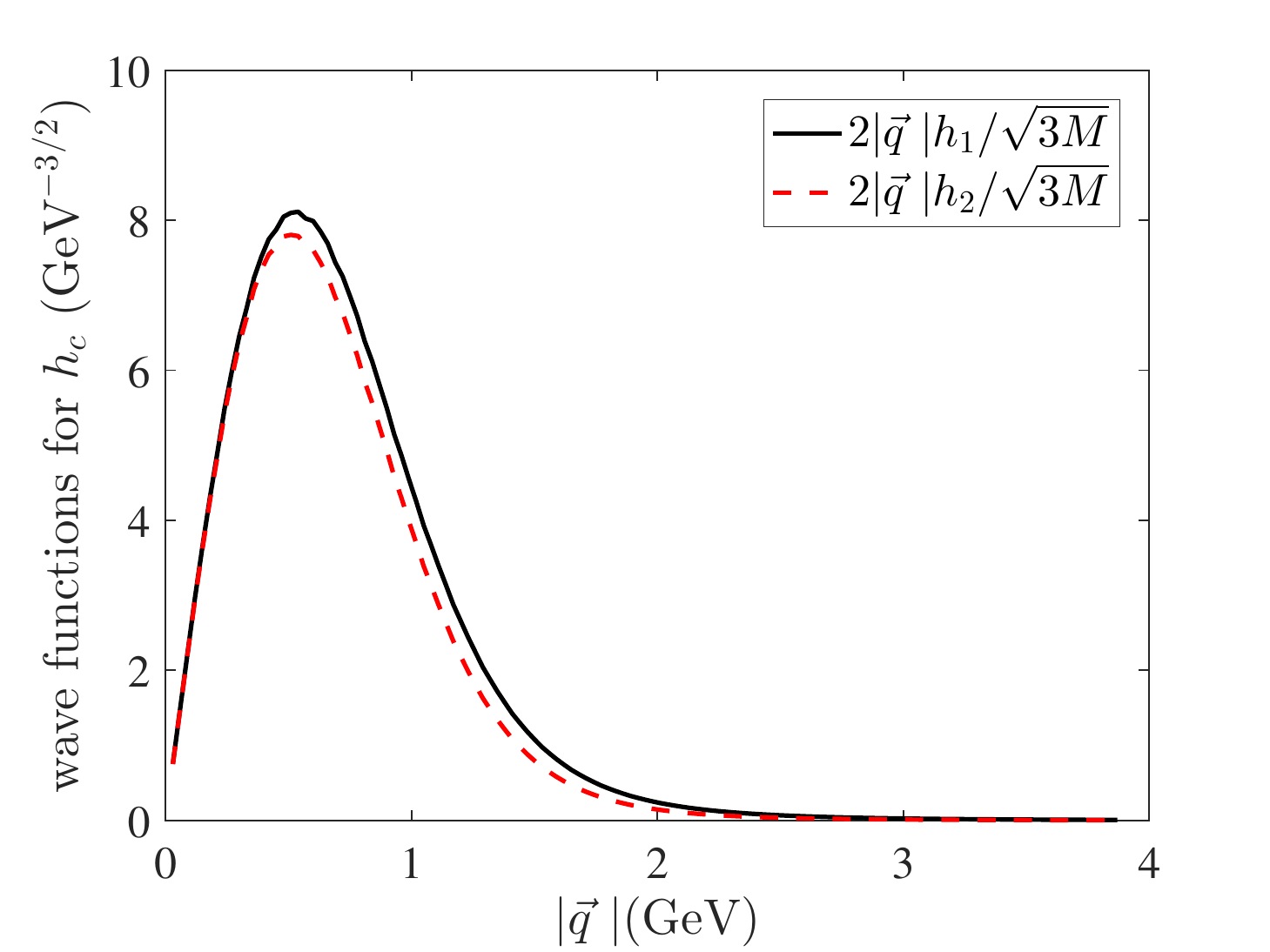}}
\subfigure[$\chi_{c0}$ 介子]{\label{fig:wfxc0}
			      \includegraphics[width=0.3\textwidth]{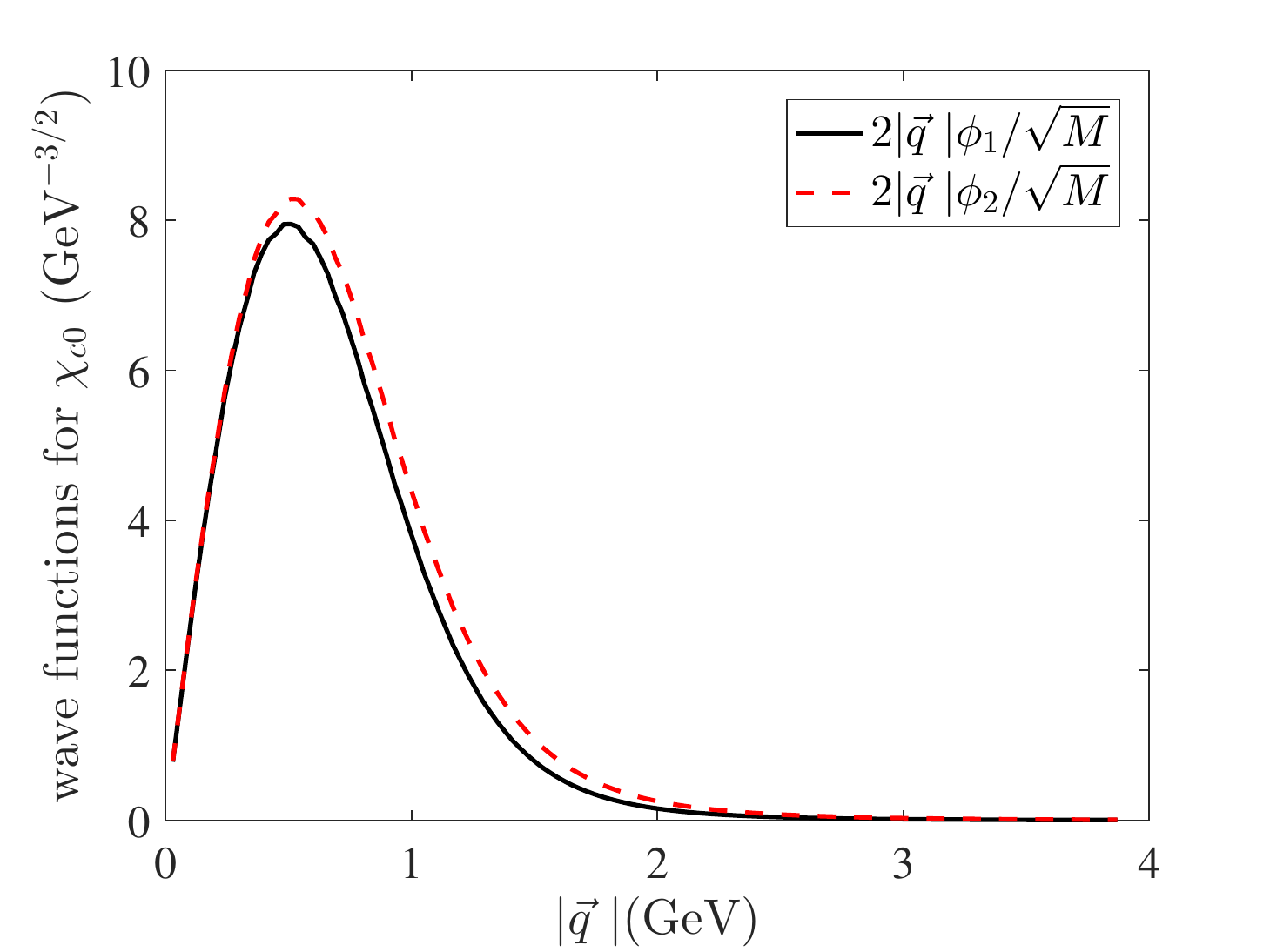}}
\subfigure[$\chi_{c1}$ 介子]{\label{fig:wfxc1}
			      \includegraphics[width=0.3\textwidth]{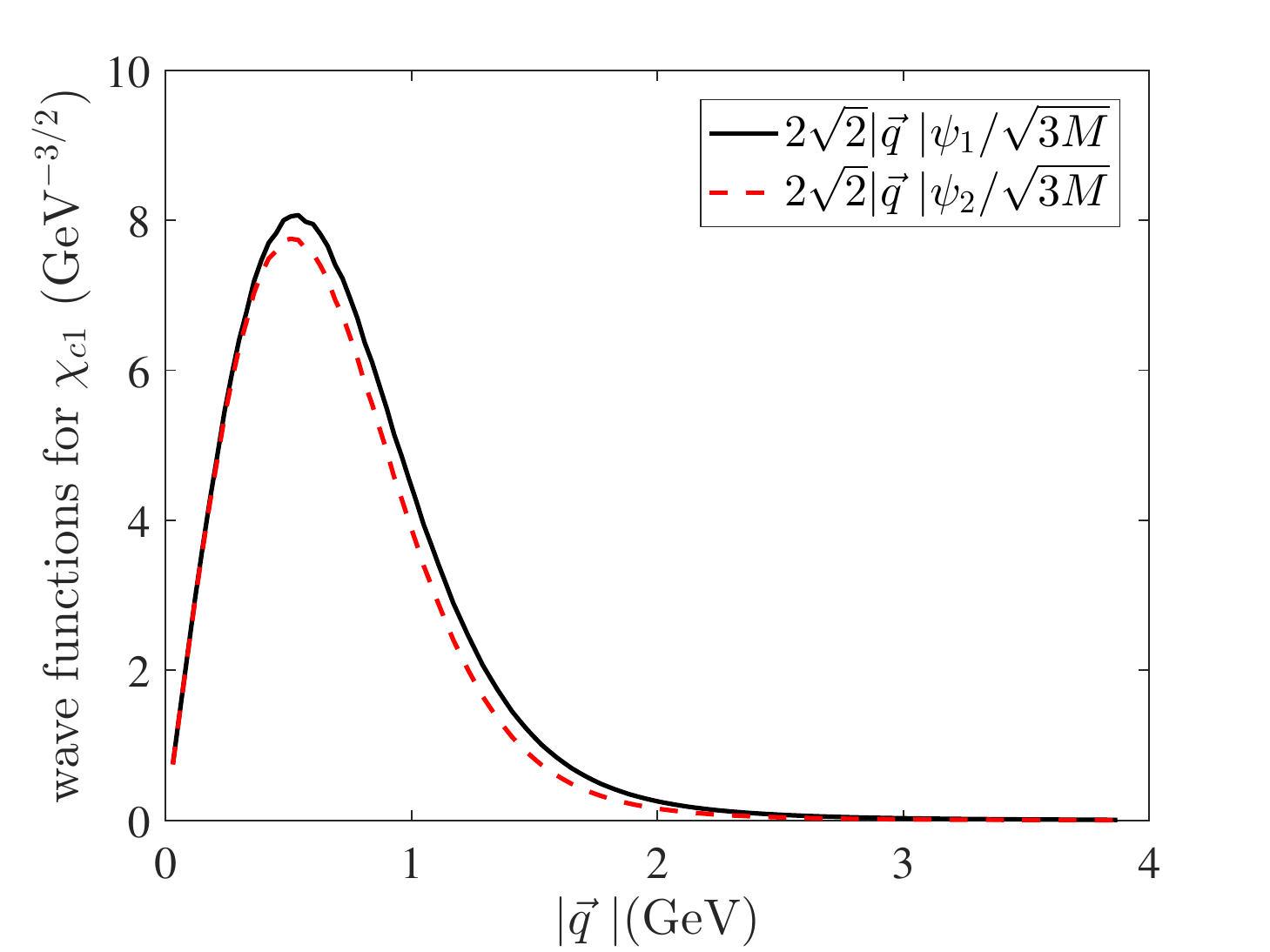}}
\subfigure[$\chi_{c2}$ 介子]{\label{fig:wfxc2}
		      	\includegraphics[width=0.3\textwidth]{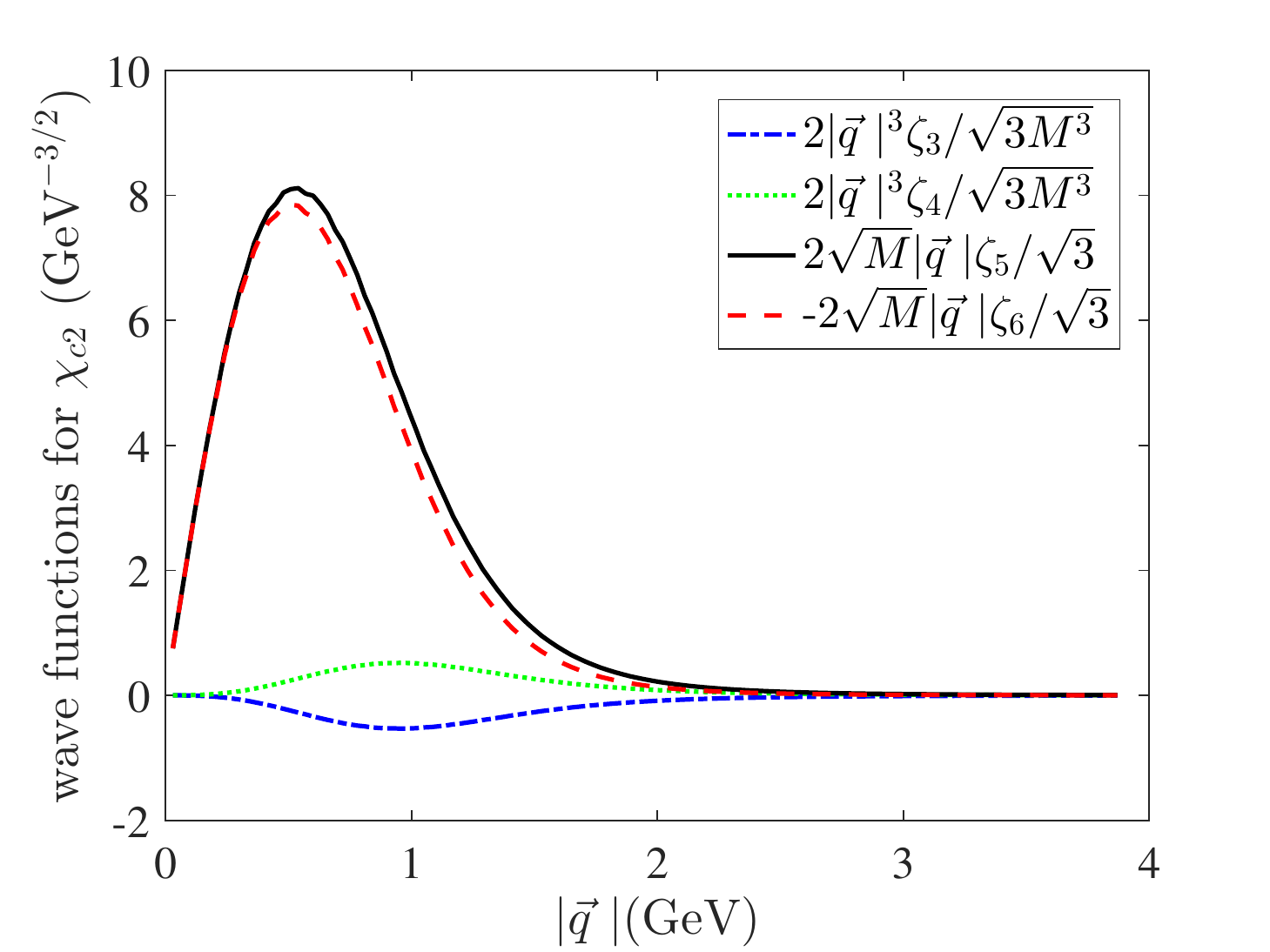}}
\caption{The normalized radial wave functions of $B_c$ and charmonium ($n=1$).}\label{fig:wfs}
\end{figure}

\begin{figure}[!hbp]
\centering
\subfigure[$\eta_c(2S)$ 介子]{\label{fig:wfetac2s}
			      \includegraphics[width=0.4\textwidth]{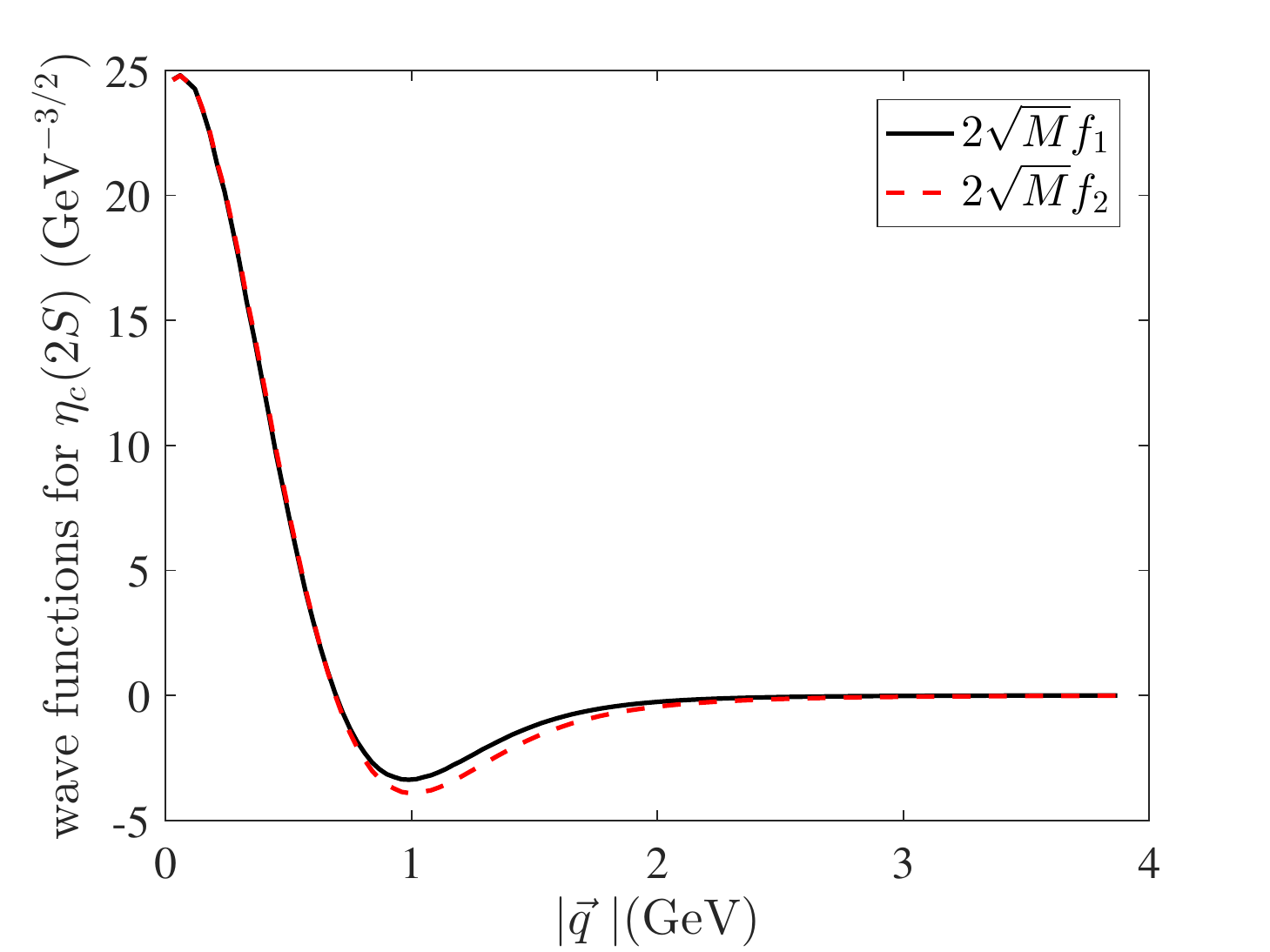}}
\subfigure[$\psi(2S)$ 介子]{\label{fig:wfpsi2s}
			      \includegraphics[width=0.4\textwidth]{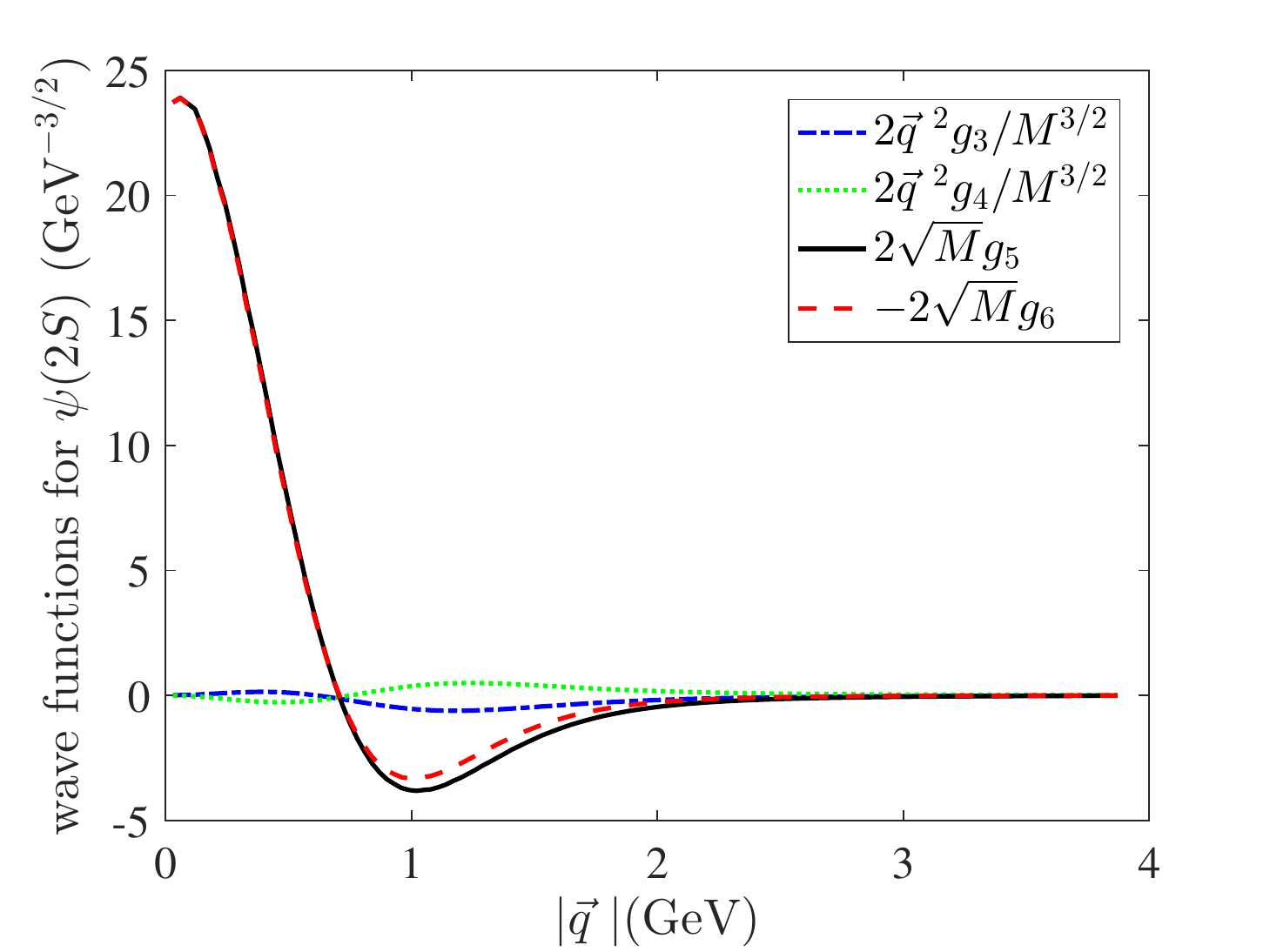}}
\subfigure[$h_c(2P)$ 介子]{\label{fig:wfhcwp}
			      \includegraphics[width=0.4\textwidth]{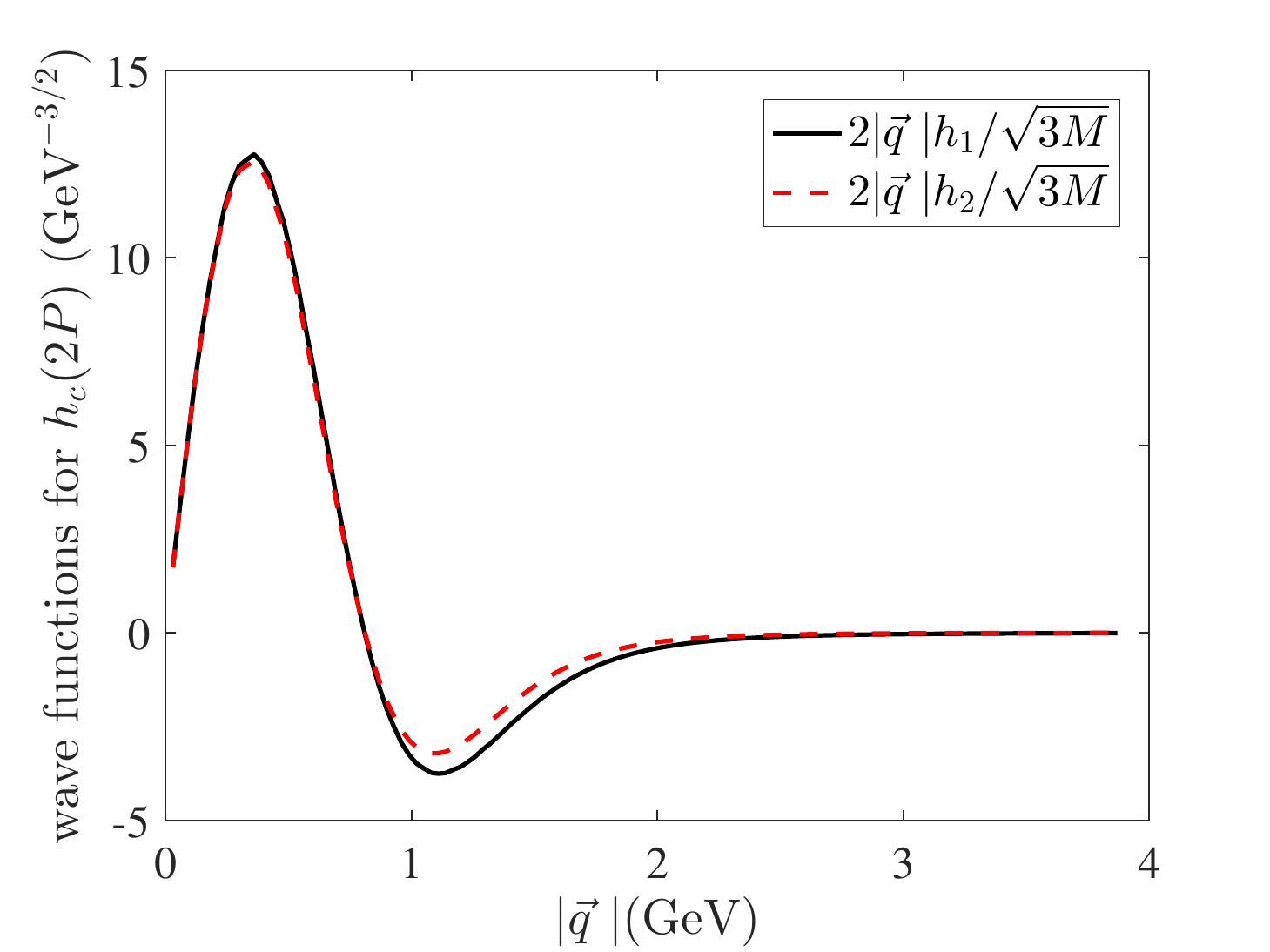}}
\subfigure[$\chi_{c0}(2P)$ 介子]{\label{fig:wfxc02p}
			      \includegraphics[width=0.4\textwidth]{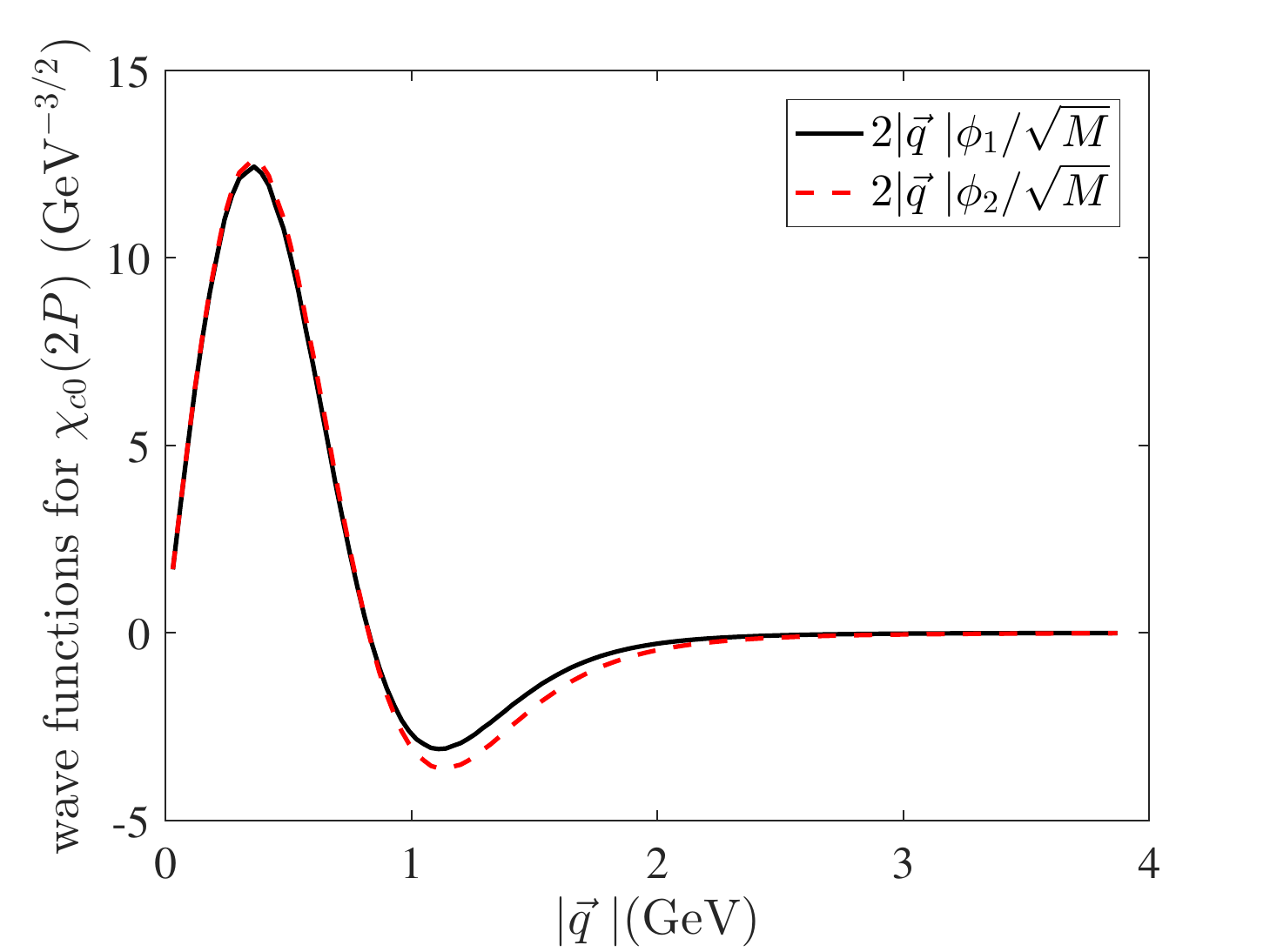}}
\subfigure[$\chi_{c1}(2P)$ 介子]{\label{fig:wfxc12p}
			      \includegraphics[width=0.4\textwidth]{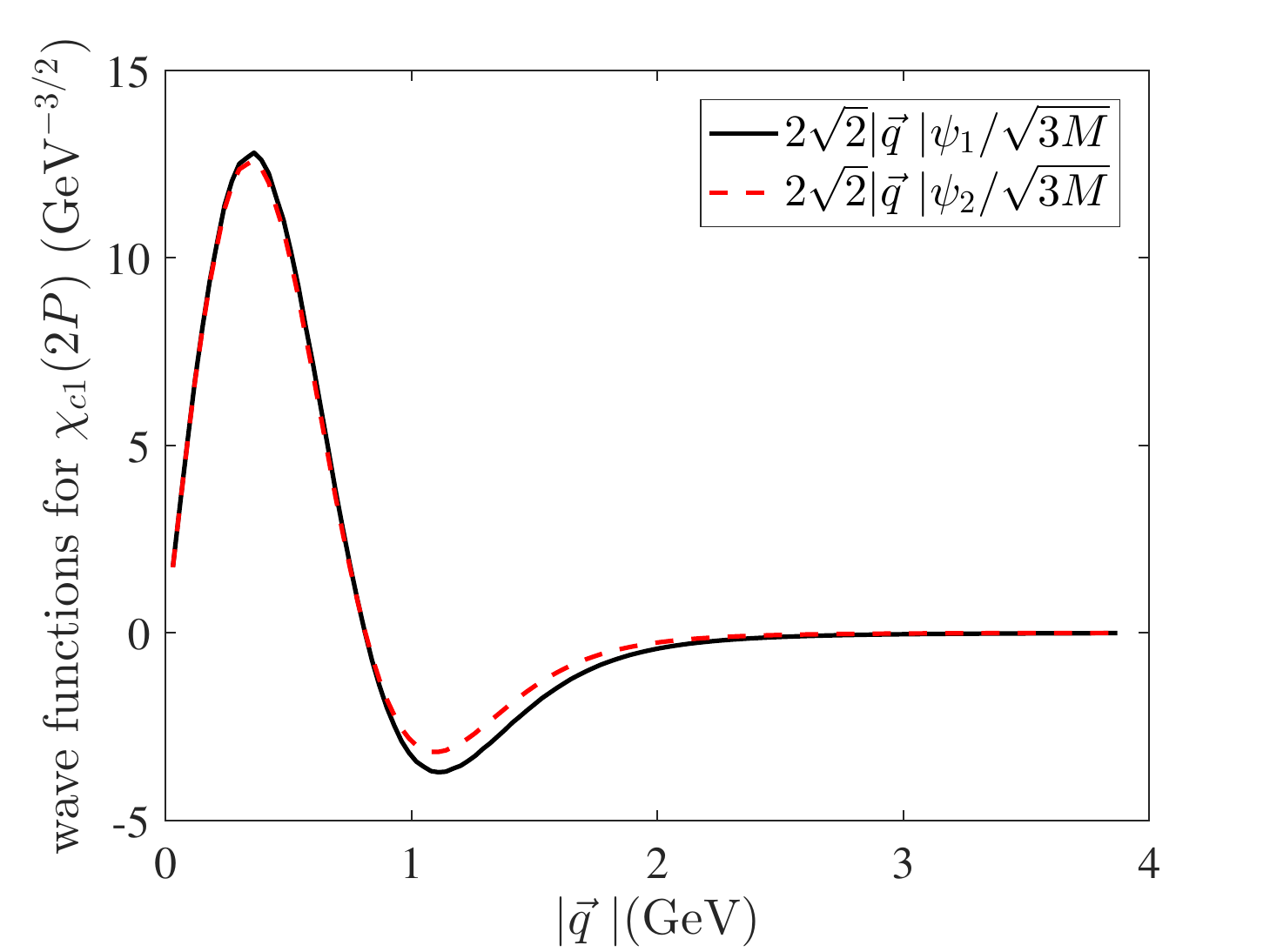}}
\subfigure[$\chi_{c2}(2P)$ 介子]{\label{fig:wfxc22p}
	\includegraphics[width=0.4\textwidth]{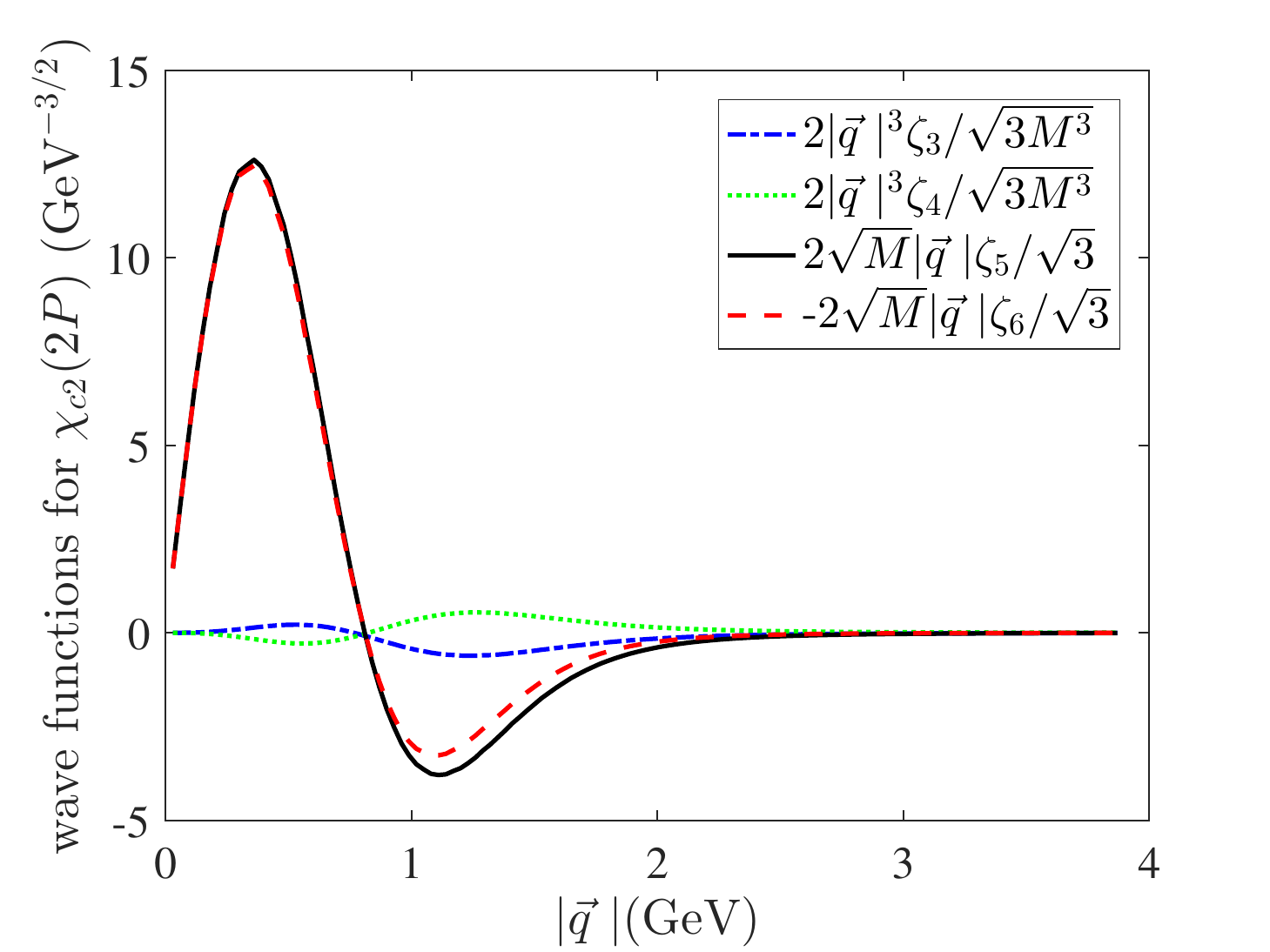}}
\caption{The normalized radial wave functions of the charmonium ($n=2$).}\label{fig:wfs2}
\end{figure}

\begin{figure}[!hbp]
\centering
\subfigure[$B_c\to\eta_c$]{\label{fig:iwf-bcetac}\label{fig:iwfetac}
			      \includegraphics[width=0.4\textwidth]{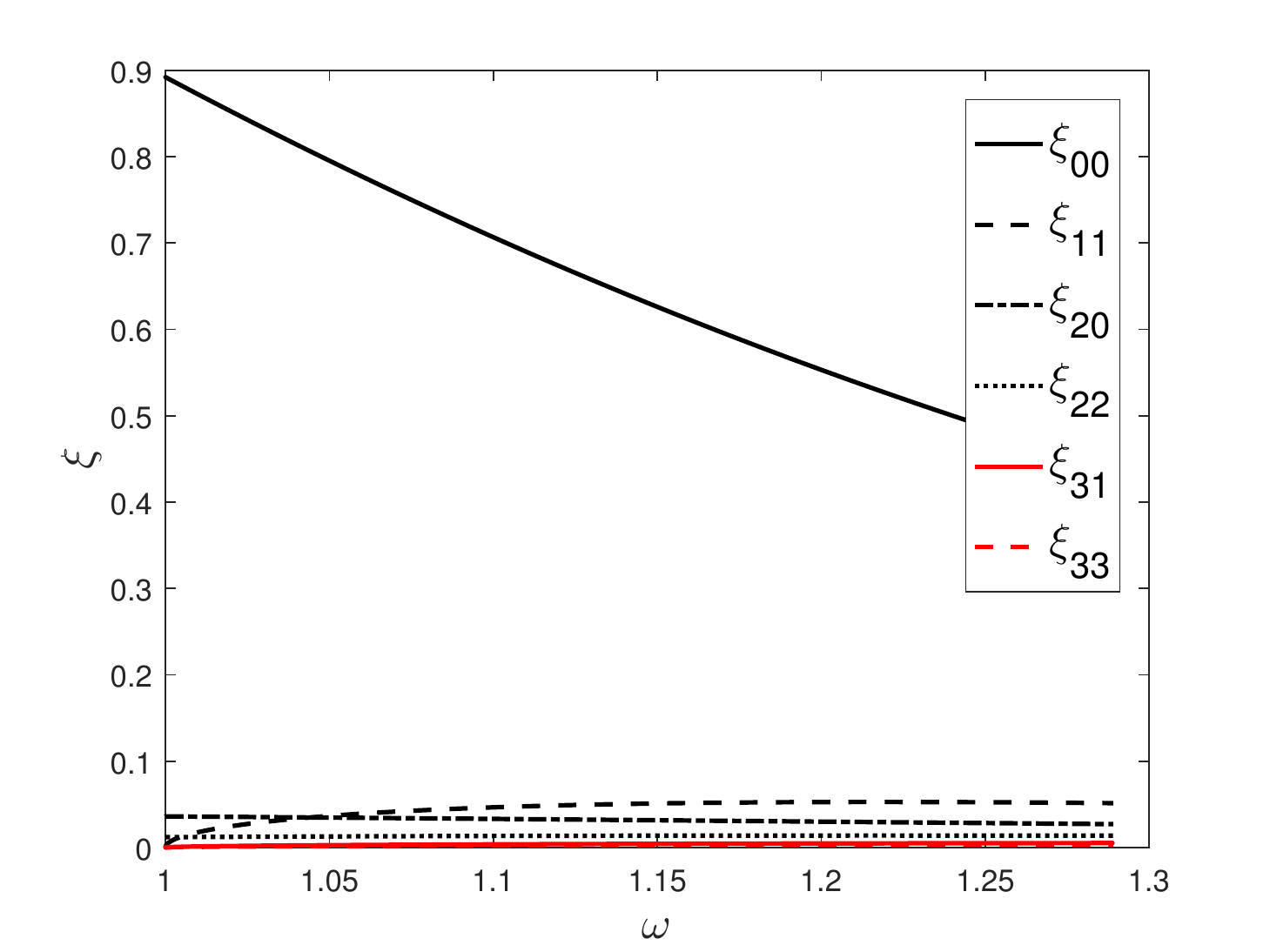}}
\subfigure[$B_c\to J/\psi$]{\label{fig:iwf-bcjpsi}\label{fig:iwfjpsi}
			      \includegraphics[width=0.4\textwidth]{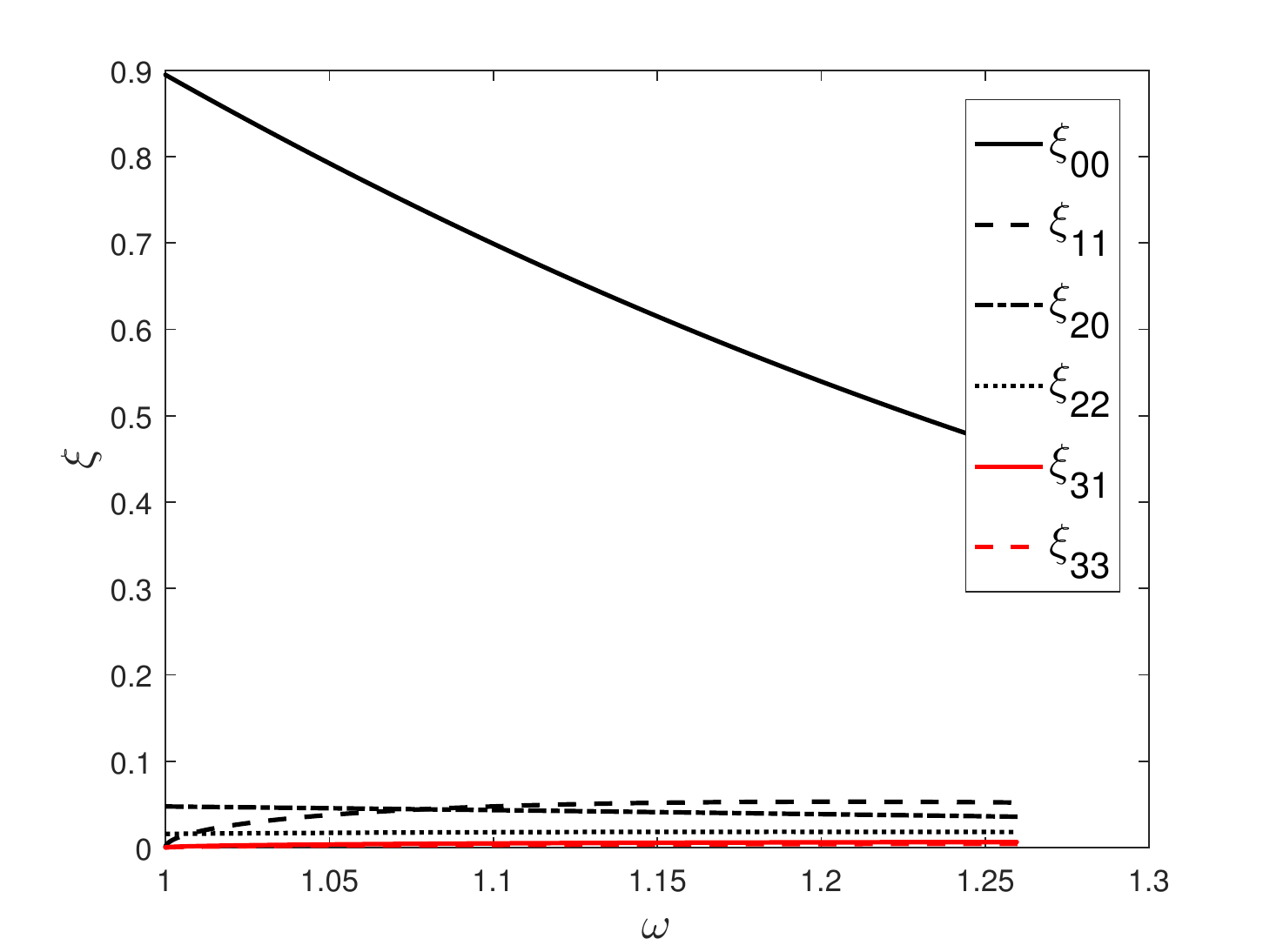}}
\subfigure[$B_c\to h_c$]{\label{fig:iwf-bchc}
			      \includegraphics[width=0.4\textwidth]{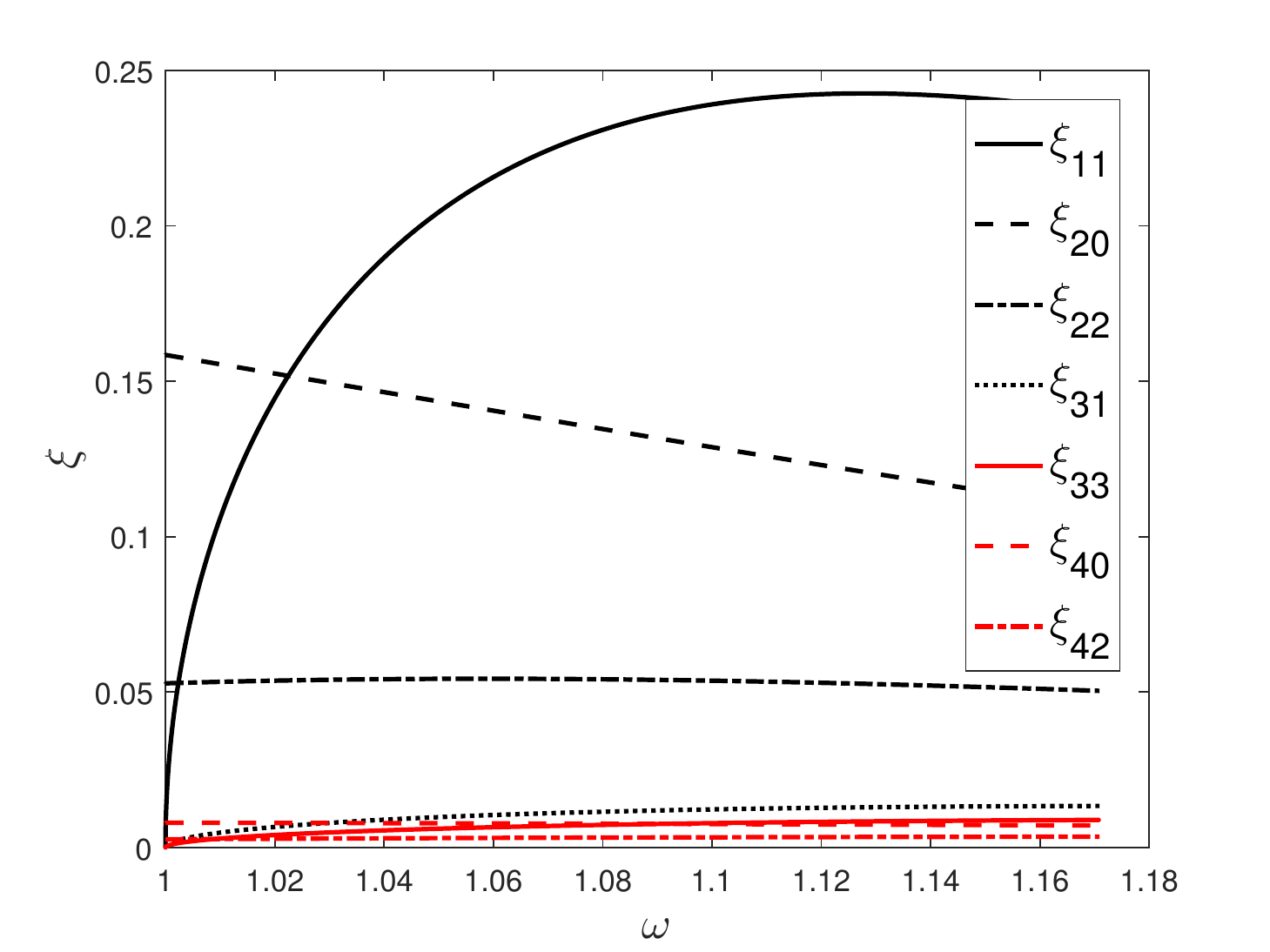}}
\subfigure[$B_c\to\chi_{c0}$]{\label{fig:iwf-bcxc0}
			      \includegraphics[width=0.4\textwidth]{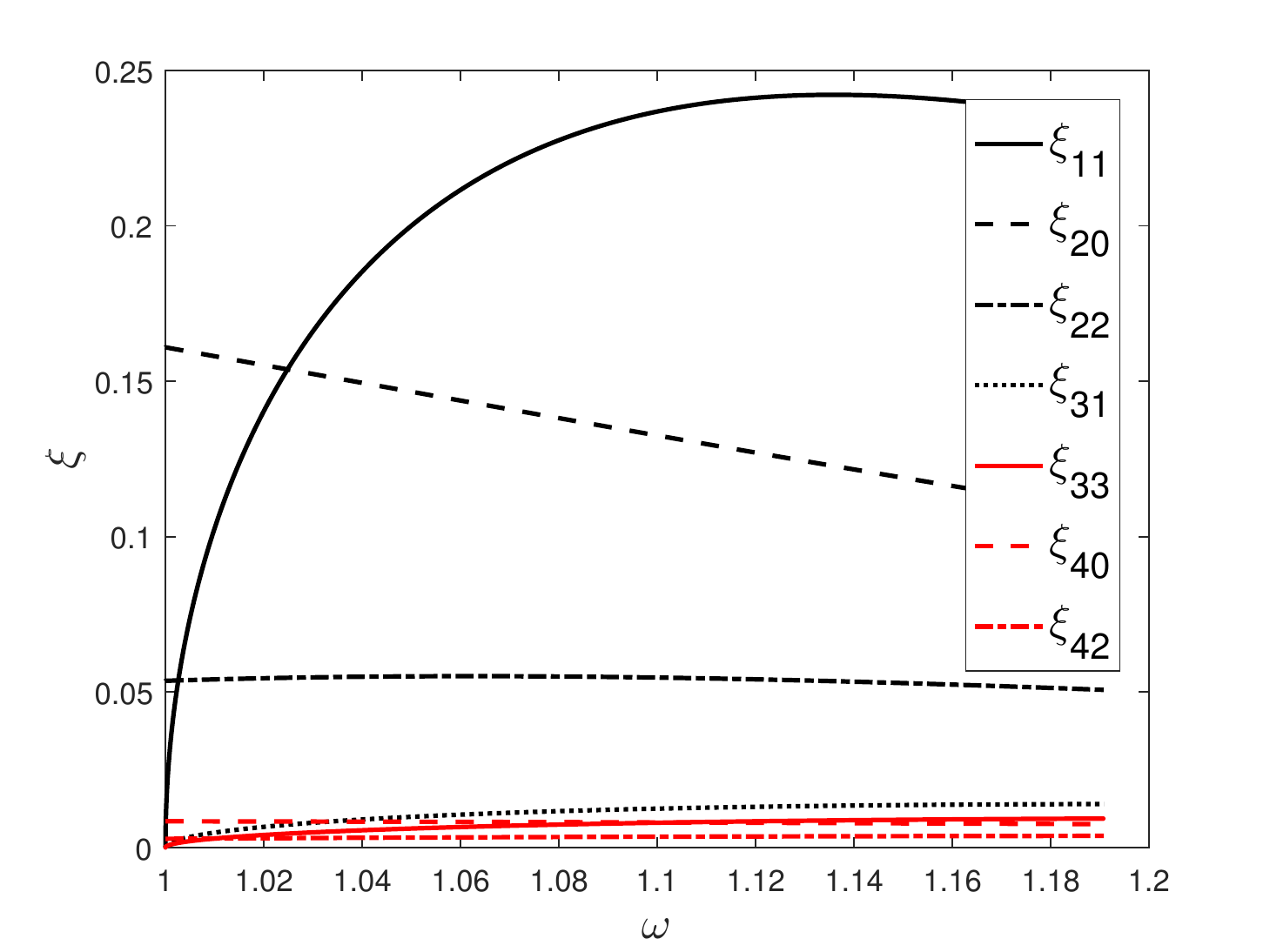}}
\subfigure[$B_c\to\chi_{c1}$]{\label{fig:iwf-bcxc1}
			      \includegraphics[width=0.4\textwidth]{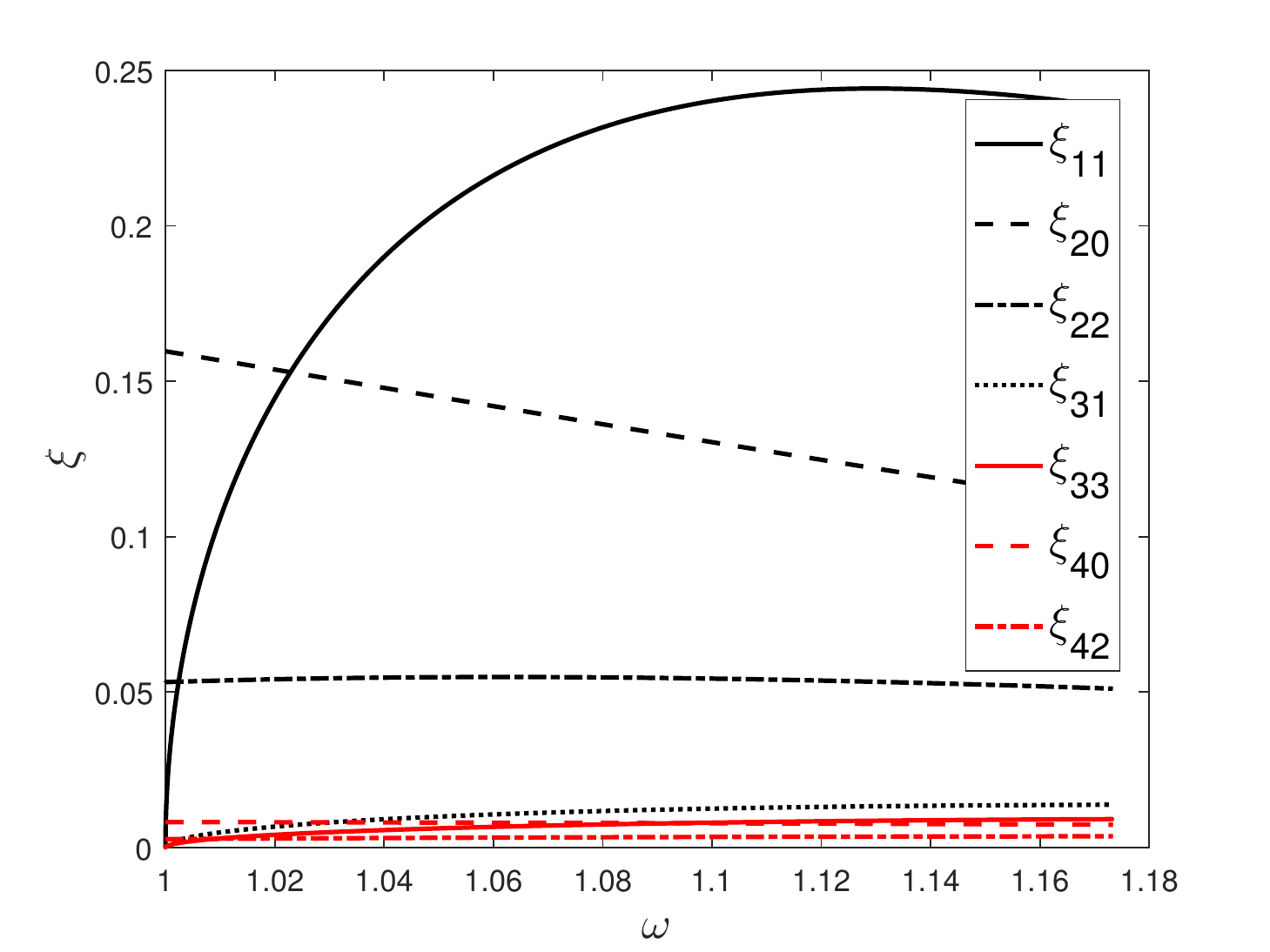}}
\subfigure[$B_c\to\chi_{c2}$]{\label{fig:iwf-bcxc2}
	\includegraphics[width=0.4\textwidth]{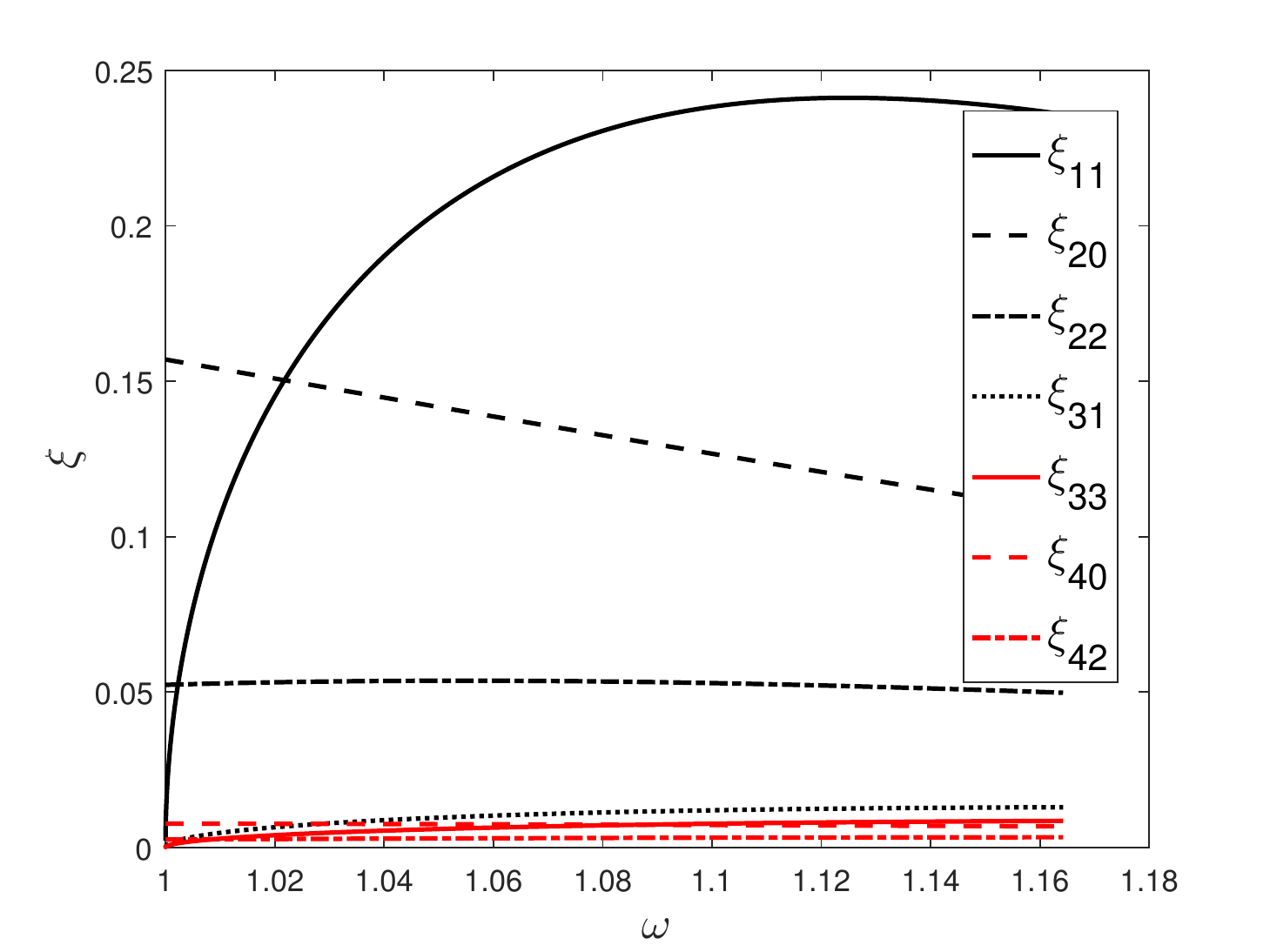}}
\caption{The IWF and high-order corrections $\xi_{qx}$ vs $\omega$ for $B_c$ to charmonium ($n=1$), where $\omega=v\cdot v_f=\frac{P\cdot P_f}{MM_f}$. The solid line is the Isgur-Wise function, the dash and dot-dash one are the first order correction functions, the dot one is the second order correction function, and so on, in every subfigure.}\label{fig:IWF}
\end{figure}

\begin{figure}[!hbp]
\centering
\subfigure[$B_c\to\eta_c(2S)$]{\label{fig:iwf-bcetac2s}\label{fig:iwfetac2s}
			      \includegraphics[width=0.4\textwidth]{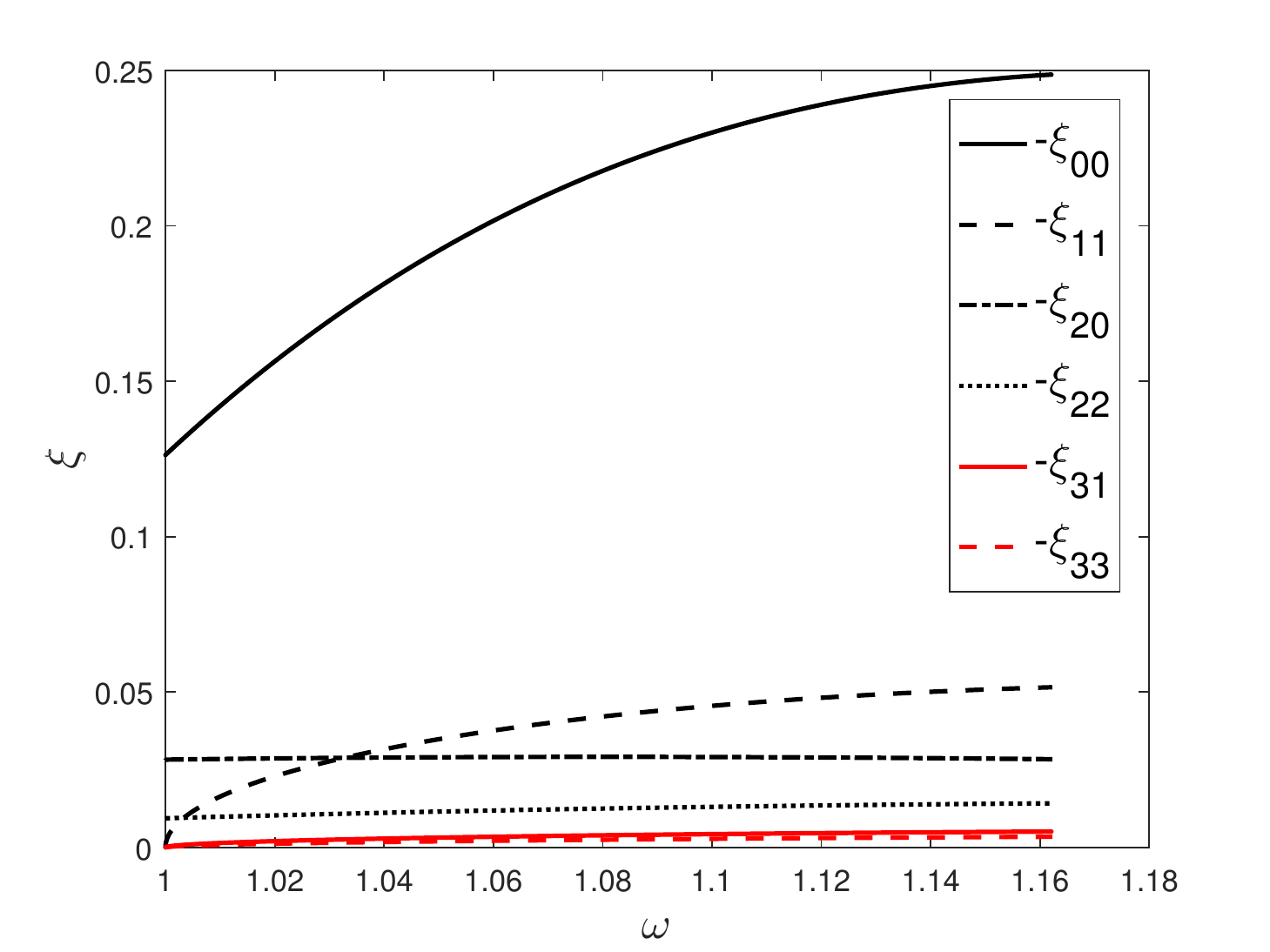}}
\subfigure[$B_c\to \psi(2S)$]{\label{fig:iwf-bcpsi2s}\label{fig:iwfpsi2s}
			      \includegraphics[width=0.4\textwidth]{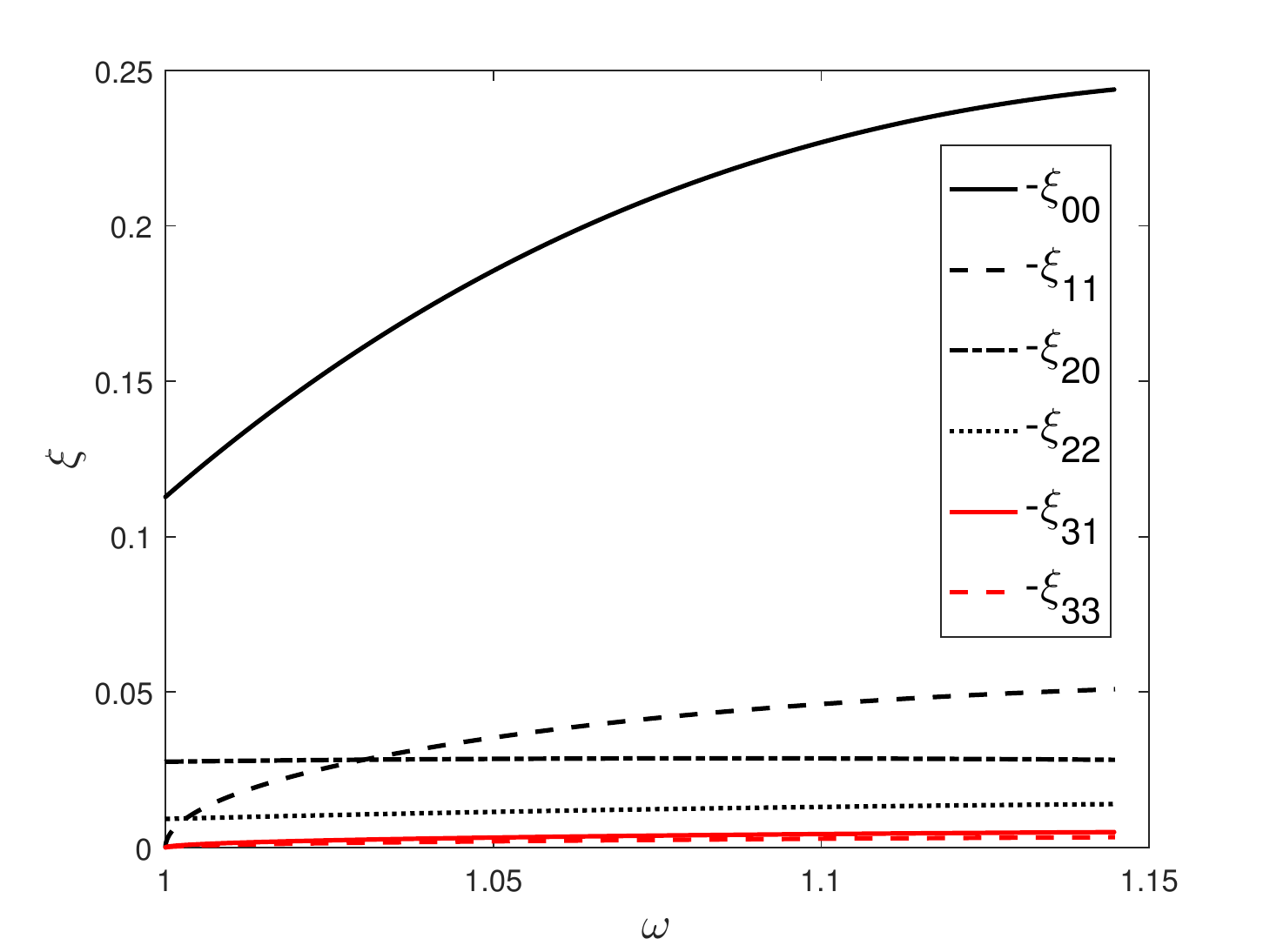}}
\subfigure[$B_c\to h_c(2P)$]{\label{fig:iwf-bchc2p}
			      \includegraphics[width=0.4\textwidth]{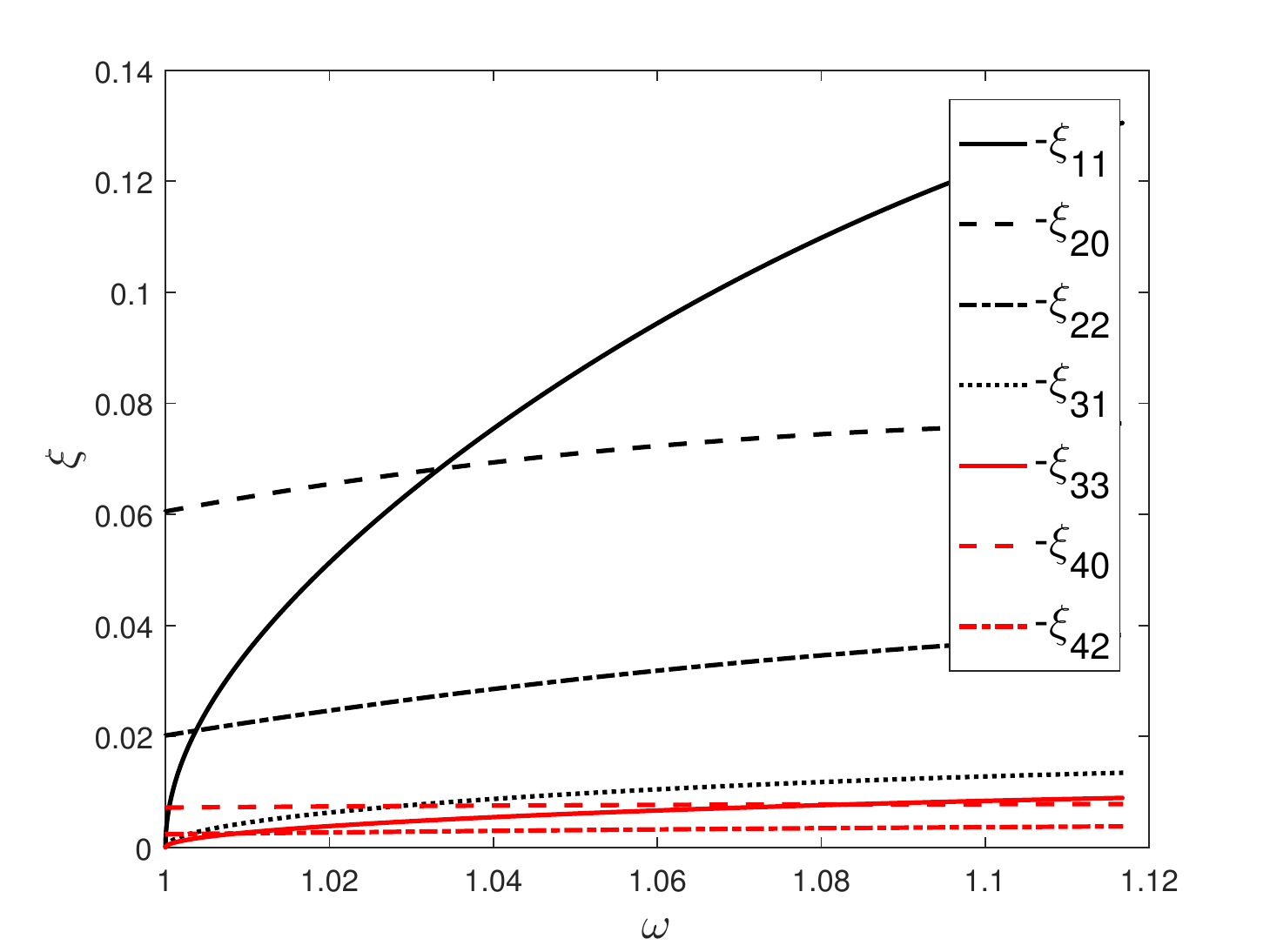}}
\subfigure[$B_c\to\chi_{c0}(2P)$]{\label{fig:iwf-bcxc02p}
			      \includegraphics[width=0.4\textwidth]{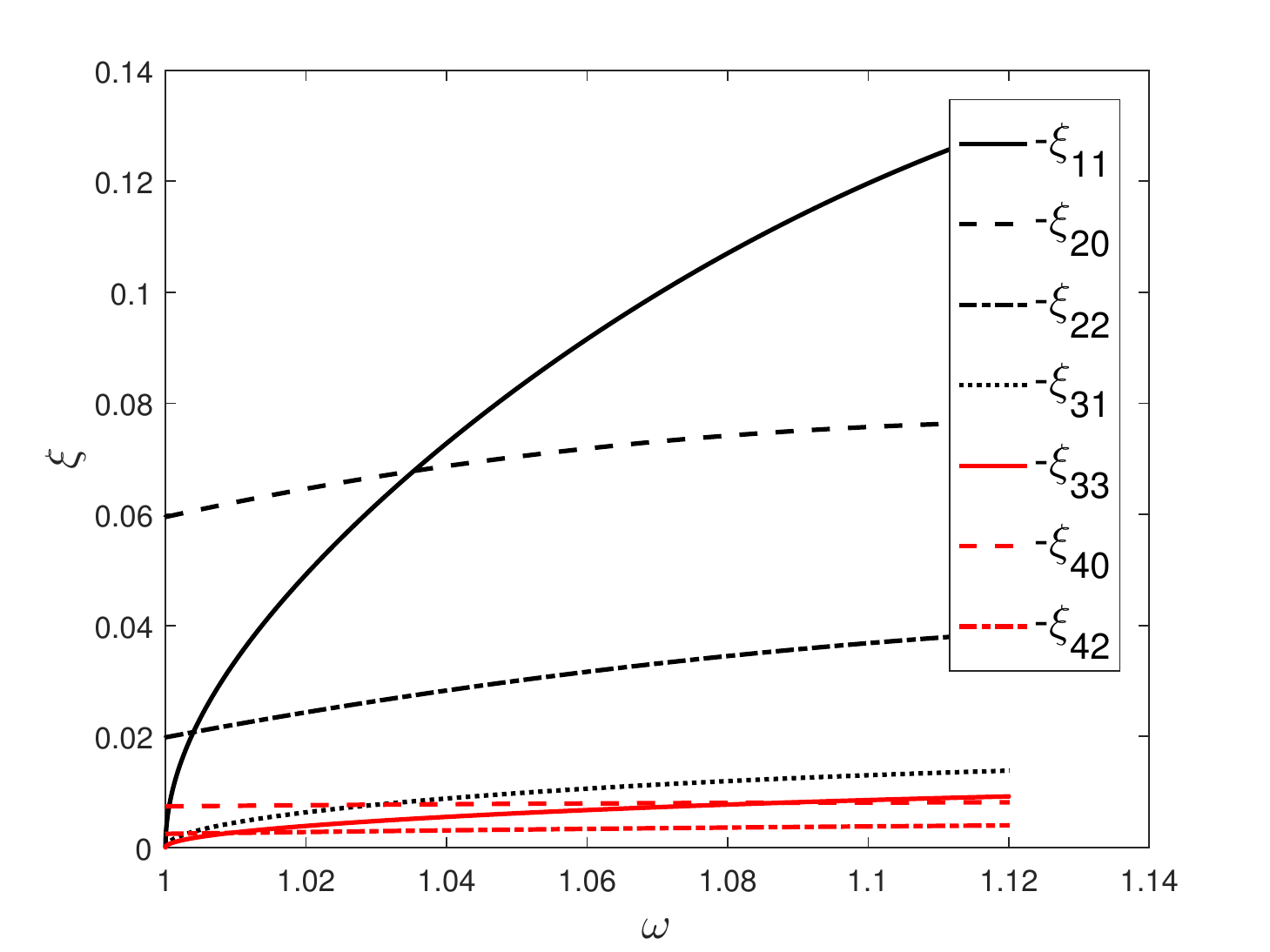}}
\subfigure[$B_c\to\chi_{c1}(2P)$]{\label{fig:iwf-bcxc12p}
			      \includegraphics[width=0.4\textwidth]{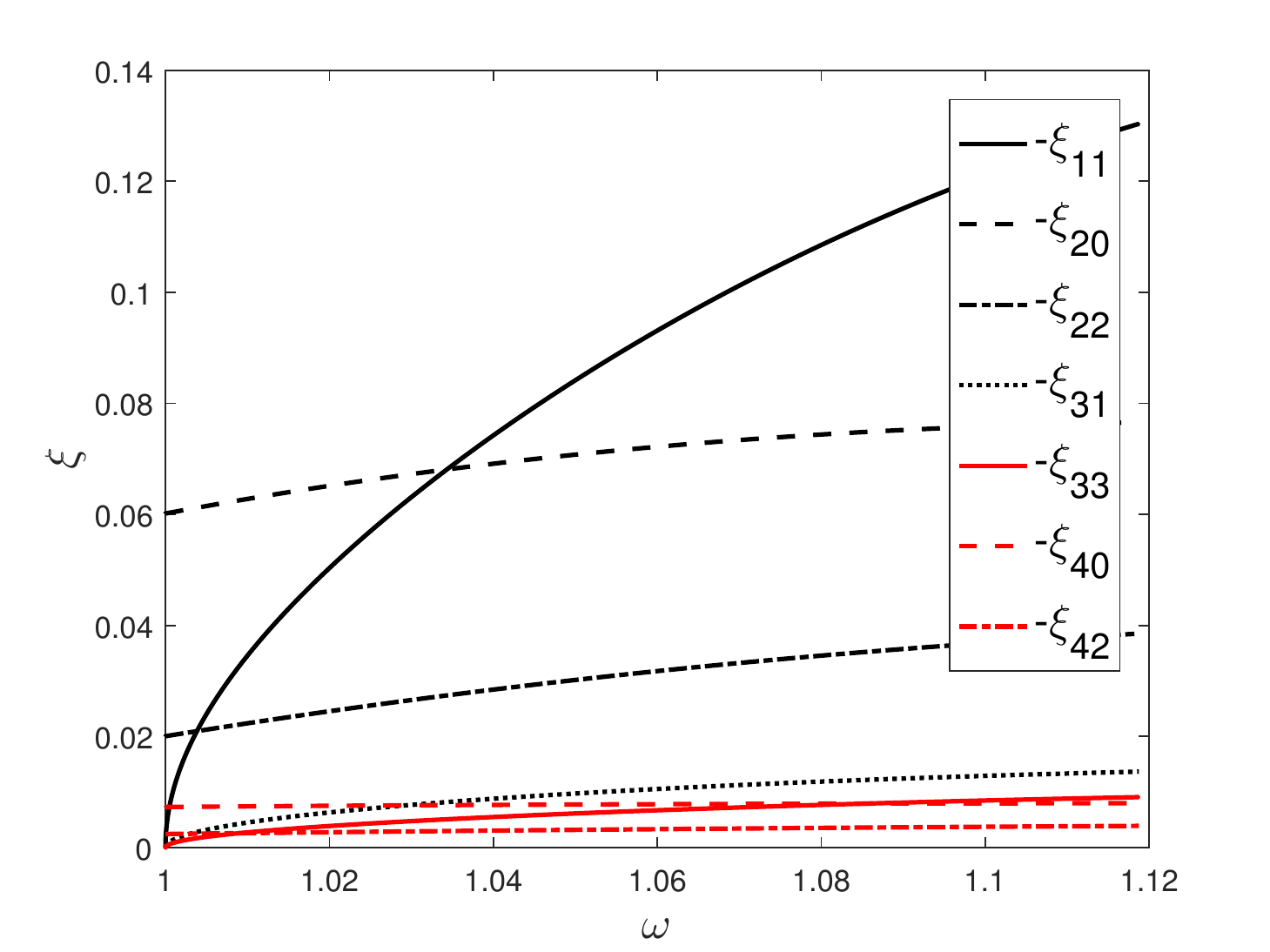}}
\subfigure[$B_c\to\chi_{c2}(2P)$]{\label{fig:iwf-bcxc22p}
	              \includegraphics[width=0.4\textwidth]{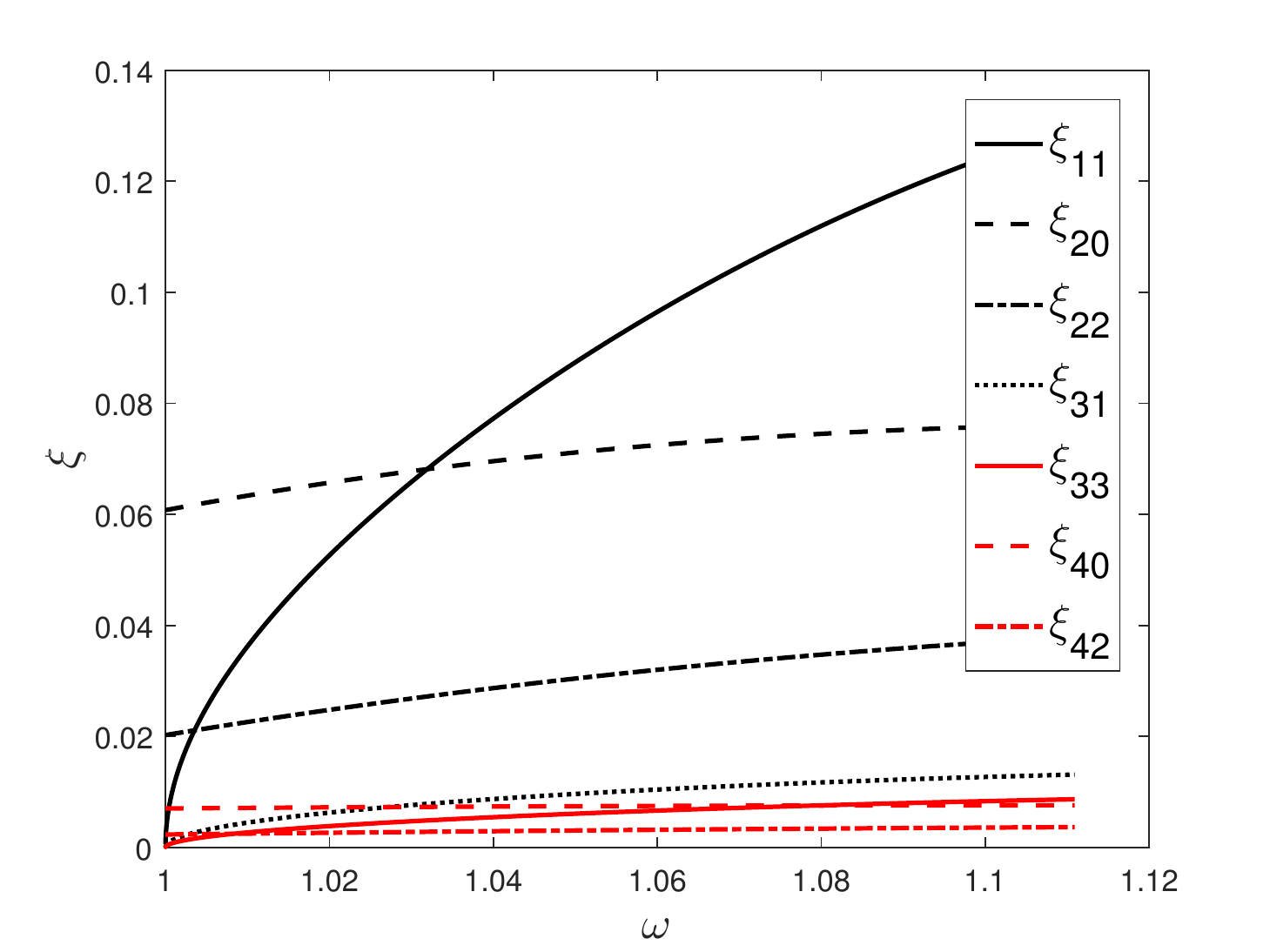}}
\caption{The IWF and high-order corrections $\xi_{qx}$ vs $\omega$ for $B_c$ to charmonium ($n=2$), where $\omega=v\cdot v_f=\frac{P\cdot P_f}{MM_f}$. The meaning of each type line is the same as that in Fig.~\ref{fig:IWF}.}\label{fig:IWF2}
\end{figure}

The parameters used in this paper: $\Gamma_{B_c}=1.298\times 10^{-12}~\mathrm{GeV}, G_F=1.166\times 10^{-5}~\mathrm{GeV^{-2}}, m_b=4.96~\mathrm{GeV}, m_c=1.62~\mathrm{GeV}$, $M_{h_c(2P)}=3.887~\mathrm{GeV}, M_{\chi_{c0}(2P)}=3.862~\mathrm{GeV}, M_{\chi_{c1}(2P)}=3.872~\mathrm{GeV}, M_{\chi_{c2}(2P)}=3.927~\mathrm{GeV}$.

After solving the corresponding full Salpeter equations, the numerical wave functions for different mesons are obtained and shown in figures \ref{fig:wfs}-\ref{fig:wfs2}. When $|\vec q\:|$ is large, the wave functions will decrease rapidly. So the weak-binding approximation Eq.~(\ref{eq:weakbind}) can be taken here, and the error from large $|\vec q\:|$ will be suppressed by the wave function. The numerical values of two dominate wave functions are almost equivalent for each meson, so the approximation that the four overlapping integrals in Eq.~(\ref{eq:overlap}) are replaced by their average is reasonable. For $1^{--}$ or $2^{++}$ state, there are two other minor wave functions $g_3,g_4$, and $g_3\approx -g_4$. Taking the approximation $g_3=-g_4$ and weak-binding approximation Eq.~(\ref{eq:weakbind}), these two minor wave functions $g_3,g_4$ only appear in the $\mathcal O(q^4)$ or higher order in the $1^{--}$ state BS wave function. Within the precision $\mathcal O(q^3)$ of this study for process $0^-\to 1^{--}$, these two minor wave functions $g_3,g_4$ disappear. It is same for $2^{++}$ state wave function. These are consistent with Eq.~(\ref{eq:nrwf1}) and (\ref{eq:nrwf2}), in which there is only one radial wave function.

The behaviors of the Isgur-Wise function and high-order corrections, i.e., the overlapping integrals of the wave functions of the initial and final bound states, are computed numerically and plotted in Fig.~\ref{fig:IWF}-\ref{fig:IWF2}, where $\omega=v\cdot v_f=\frac{P\cdot P_f}{MM_f}$. These IWFs can be classified into four categories according to the configurations $nL$ of initial and final states. They belong to the modes $1S\to 1S$, $1S\to 1P$, $1S\to 2S$ and $1S\to 2P$ respectively. With the same configurations of initial and final states, for example, in the processes $B_c\to\eta_c$ and $B_c\to J/\psi$, the behaviors of IWFs are virtually identical except the ranges of $\omega=v\cdot v_f$ slightly differ. Because these decay processes are just related by a rotation of the heavy-quark spin or the meson spin, and this rotation is a symmetry transformation in the infinite-mass limit. Note that the infinite-mass limit is not used in this paper, but this spin-symmetry reflected in the results automatically, as Fig.~\ref{fig:IWF}-\ref{fig:IWF2} shows. This indicates that spin-symmetry still mantains though the initial and final states are both the double-heavy mesons. When the configuration of initial or final state changes, for example, the final $\eta_c$ turns into $\eta_c(2S)$, the behaviors of IWFs become significantly different from before. Next we will discuss these four modes one by one.

The mode $1S\to 1S$ has been extensively studied in HQET. $B_c$ and $\eta_c$ are related by the replacement $b\to c$, while $\eta_c$ and $J/\psi$ are related by the transformation $c^\Uparrow\to c^\Downarrow$ here. These two rotations (flavour and spin rotations) are symmetry transformations in the infinite-mass limit. So the radial wave functions of these mesons will be identical in this limit, and the corresponding IWF whose general form is the overlapping integral of the initial and final wave functions will be the same as the normalization formula at zero recoil. It is very natural that $\xi(1)=1$ in HQET. In this paper we solve the full Salpeter equations without the infinite-mass limit, and the normalized radial wave function is approximated to $2\sqrt{M}f$ for $1S$ state. The normalized wave functions have little difference for $\eta_c$ and $J/\psi$ (two dominate wave functions), which is consistent with Eq.~(\ref{eq:nrwf1}). But the discrepancy between $B_c$ and the former two is in the order of 30\% (peak value), as Fig.~\ref{fig:wfbc}-\ref{fig:wfjpsi} shows. This indicates that in the double-heavy system the spin-symmetry keeps, while the flavour-symmetry breaks. The masses of quark and antiquark are in the same order of magnitude, and therefore the change of flavour will lead to a great impact. Although the behaviors of IWF $\xi_{00}$ are the same as $\xi$ in HQET, they are not strict unity at zero recoil in this paper, as Fig.~\ref{fig:iwfetac}-\ref{fig:iwfjpsi} shows. The relativistic correction reflected in IWF $\xi_{00}$ is around 10\% at zero recoil. In the mode $1S\to 1S$, it is convenient to fit the IWF as
\begin{equation}
\xi_{00}(\omega)=\xi_{00}(1)\left[1-\rho^2(\omega-1)+c(\omega-1)^2\right],\label{eq:fit}
\end{equation}
where $\rho^2$ is the slope parameter and $c$ is the curvature parameter which characterizes the shape of the IWF.
The slope and curvature by fitting are 2.25 and 1.74 respectively in $B_c\to\eta_c$, and they are 2.38 and 1.98 respectively in $B_c\to J/\psi$. The result is agree with the rule that the slope is bigger as the (reduced) mass is heavier \cite{Das:2016hmv,Hassanabadi:2014isa}. The other functions $\xi_{qx}$ are the relativistic corrections to IWF $\xi_{00}$. The more $\vec q\:'$ the correction function contains, the less contribution it makes. We may call the correction function with one relative momentum $\vec q\:'$ as the first order correction, the correction function with two $\vec q\:'$ as the second order correction, and so on. The values of $\xi_{11}$ is about 1/20 of $\xi_{00}$. Because the decay width is proportional to modular square of amplitude, the first order correction may reach 1/10 of the leading order result. It is important for accurate calculation. Our previous study shows that the higher order relativistic corrections also have considerable contributions, and the total relativistic correction can reach around 20\% at the level of decay width \cite{Geng:2018qrl}.

In the mode $1S\to 1P$, the configuration of initial state is $1S$, while the configuration of final state is $1P$. Their orbital angular momenta are different, so the symmetry transformations exist only between the final states, i.e., spin rotations. $\chi_{c0}$, $\chi_{c1}$ and $\chi_{c2}$ are spin triplet states that are related by the rotation of total spin component (the component of total spin in the direction of orbital angular momentum), while $h_c$ and the former three are related by the transformation $c^\Uparrow\to c^\Downarrow$. These two spin rotations are symmetry transformations in the infinite-mass limit, so the normalized radial wave functions of these mesons will be identical. In this paper we study these mesons without the infinite-mass limit, and their normalized radial wave functions are approximated to $\frac{2|\vec q\:|h}{\sqrt{3M}},\frac{2|\vec q\:|\phi}{\sqrt{M}},\frac{2\sqrt{2}|\vec q\:|\psi}{\sqrt{3M}}$ and $\frac{2\sqrt{M}|\vec q\:|\zeta}{\sqrt 3}$ respectively. Their numerical results are almost the same, as Figs.~\ref{fig:wfhc}-\ref{fig:wfxc1} shows, which is consistent with Eq.~(\ref{eq:nrwf2}). This indicates that the spin-symmetry keeps in the P-wave charmonium though the quark and anti-quark have the same masses. Because P-wave function contains a $\vec q\:$, $\xi_{00}$ disappears, and IWF is $\xi_{11}$ which behavior is obvious different from $\xi_{00}$. Due to the presence of $\cos\theta$, see Eq.~(\ref{eq:iwother}), the IWF $\xi_{11}$ is zero at zero recoil. And it is enhanced kinematically, as Figs.~\ref{fig:iwf-bchc}-\ref{fig:iwf-bcxc1} shows. This behavior is agree with Ref.~\cite{Chang:2001pm}. There is a kinematically suppressed factor $1/|\vec v_f|$ in the form factors Eq.~(\ref{eq:heavylimit}), and therefore the behaviors of the leading order form factors are not purely dependent on $\xi_{11}$. The $\xi_{20}$ and $\xi_{22}$ are comparable to the leading order $\xi_{11}$, especially at zero recoil. $\xi_{22}$ is smaller than $\xi_{20}$ due to the factor $\cos^2\theta$. They decrease slowly when the momentum recoil increases and therefore the relativistic corrections may be comparable to the nonrelativistic results in this mode. Although the other correction functions seem to be very small, they are still important for accurate calculation, just as the mode $1S\to 1S$. For the final states as $h_c$, $\chi_{c0}$, $\chi_{c1}$ and $\chi_{c2}$, the total relativistic corrections are 50\%, 64\%, 34\% and 14\% at the level of decay width respectively\cite{Geng:2018qrl}. The total relativistic correction of $B_c\to\chi_{c2}$ is unusually small, because the different orders corrections cancel each other out. This can be seen in the following analysis of form factors.

In the mode $1S\to 2S$, the configuration of initial state is different from the final state. Similarly, the only symmetry transformation is the spin rotation $c^\Uparrow\to c^\Downarrow$ which relates $\eta_c(2S)$ with $\psi(2S)$. Their normalized wave functions are almost the same, as Fig.~\ref{fig:wfetac2s}-\ref{fig:wfpsi2s} shows, which is consistent with Eq.~(\ref{eq:nrwf1}). The numerical results of IWFs in this mode is negative. Though the negative and positive of IWFs dose not affect the result of width, but it indicates the negative parts of $2S$-wave functions play a primary role. The overlapping integral of wave functions $\int\ud\vec q=\int\vec q\:^2\sin\theta\ud|\vec q\:|\ud\theta\ud\phi$ contains a factor $\vec q\:^2$. It is suppressed when $|\vec q\:|<1$ while is enhanced when $|\vec q\:|>1$. The negative parts of $2S$-wave functions are mainly in the range of $|\vec q\:|>1$, so the negative parts play a primary role in the overlapping integral. The IWF $\xi_{00}$ is increasing together with the momentum recoil, as Figs.~\ref{fig:iwf-bcetac2s}-\ref{fig:iwf-bcpsi2s} shows, and the leading order form factors have the same behaviors due to Eq.~(\ref{eq:limit0-}) and (\ref{eq:limit1--}). The behaviors of the other correction functions in this mode are similar to $1S\to 1S$, but they make more contributions here. For example, $\xi_{11}$ and $\xi_{20}$ are about one fifth and one eighth of $\xi_{00}$ at the maximum recoil, respectively. So the relativistic corrections become greater, and our previous study shows they are about 19\%--28\% larger than those in the mode $1S\to 1S$ \cite{Geng:2018qrl}.

Comparing to the mode $1S\to 1P$, the analysis about the symmetry and normalized wave functions is the same in $1S\to 2P$, as Figs.~\ref{fig:iwf-bchc2p}-\ref{fig:iwf-bcxc12p} shows. The IWF $\xi_{11}$ is also zero at zero recoil, but increases more slowly. The $\xi_{20}$ and $\xi_{22}$ are comparable to IWF $\xi_{11}$, and are no more decreasing but increasing as the momentum recoil is increasing. So the relativistic corrections may be more significant in this mode. They are about 10\%--16\% larger than those 
in the mode $1S\to 1P$ \cite{Geng:2018qrl}.

\begin{figure}[!hbp]
\centering
\subfigure[$B_c\to\eta_c:S_{+}$]{\label{fig:s+-bcetac}
			      \includegraphics[width=0.4\textwidth]{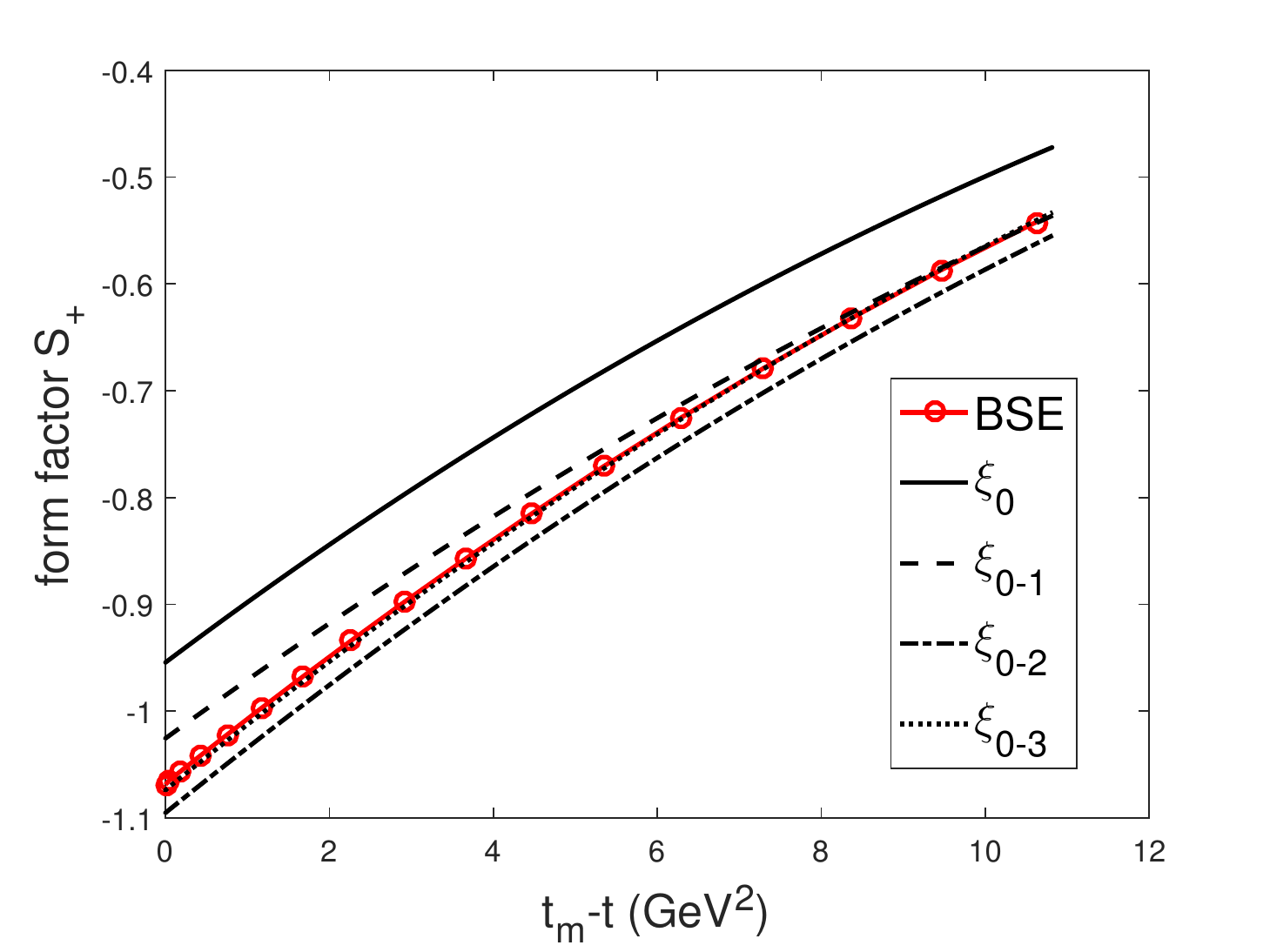}}
\subfigure[$B_c\to\eta_c:S_{-}$]{\label{fig:s--bcetac}
			      \includegraphics[width=0.4\textwidth]{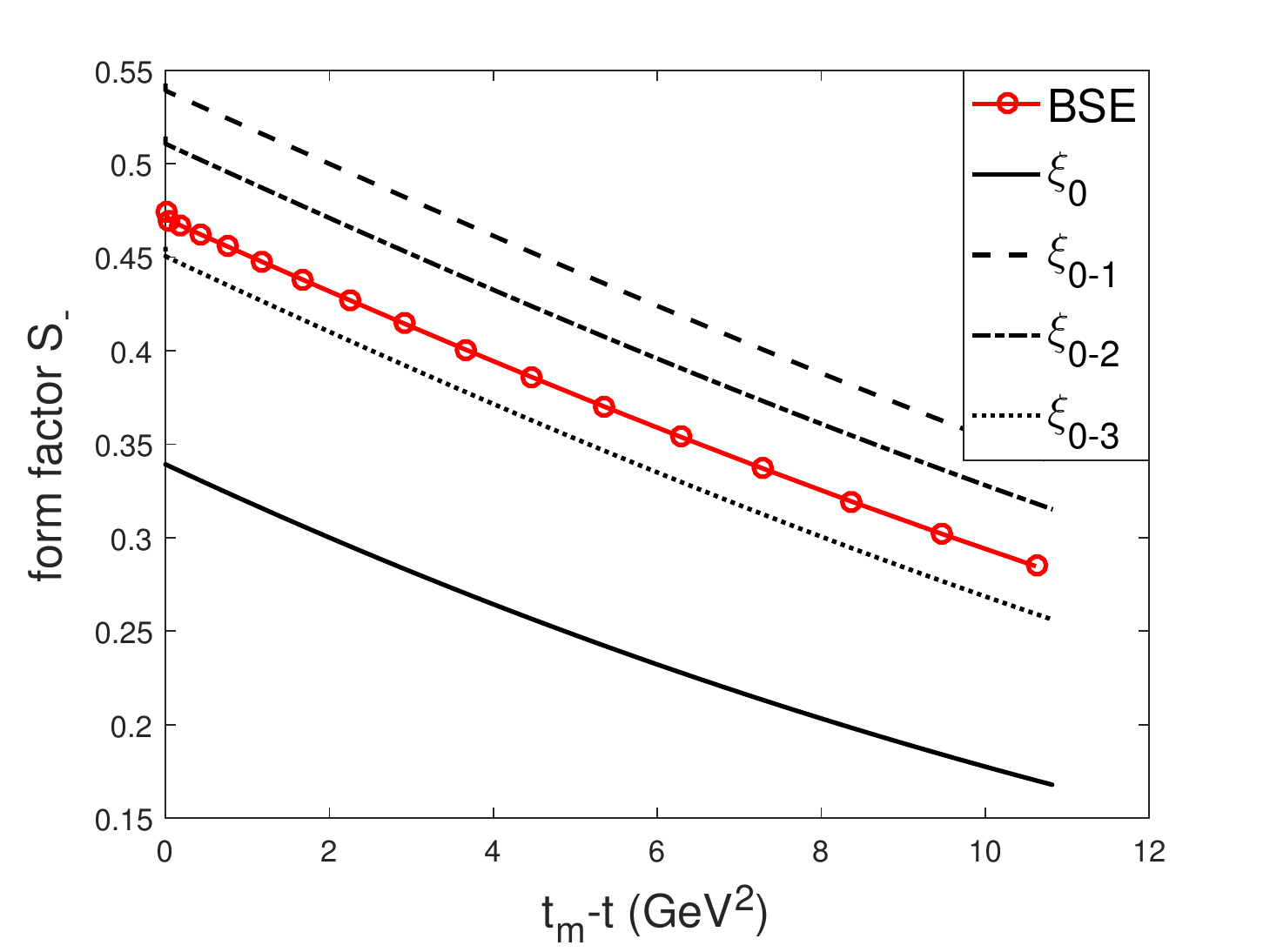}}
\subfigure[$B_c\to J/\psi:t_{1}$]{\label{fig:t1-bcjpsi}
			      \includegraphics[width=0.4\textwidth]{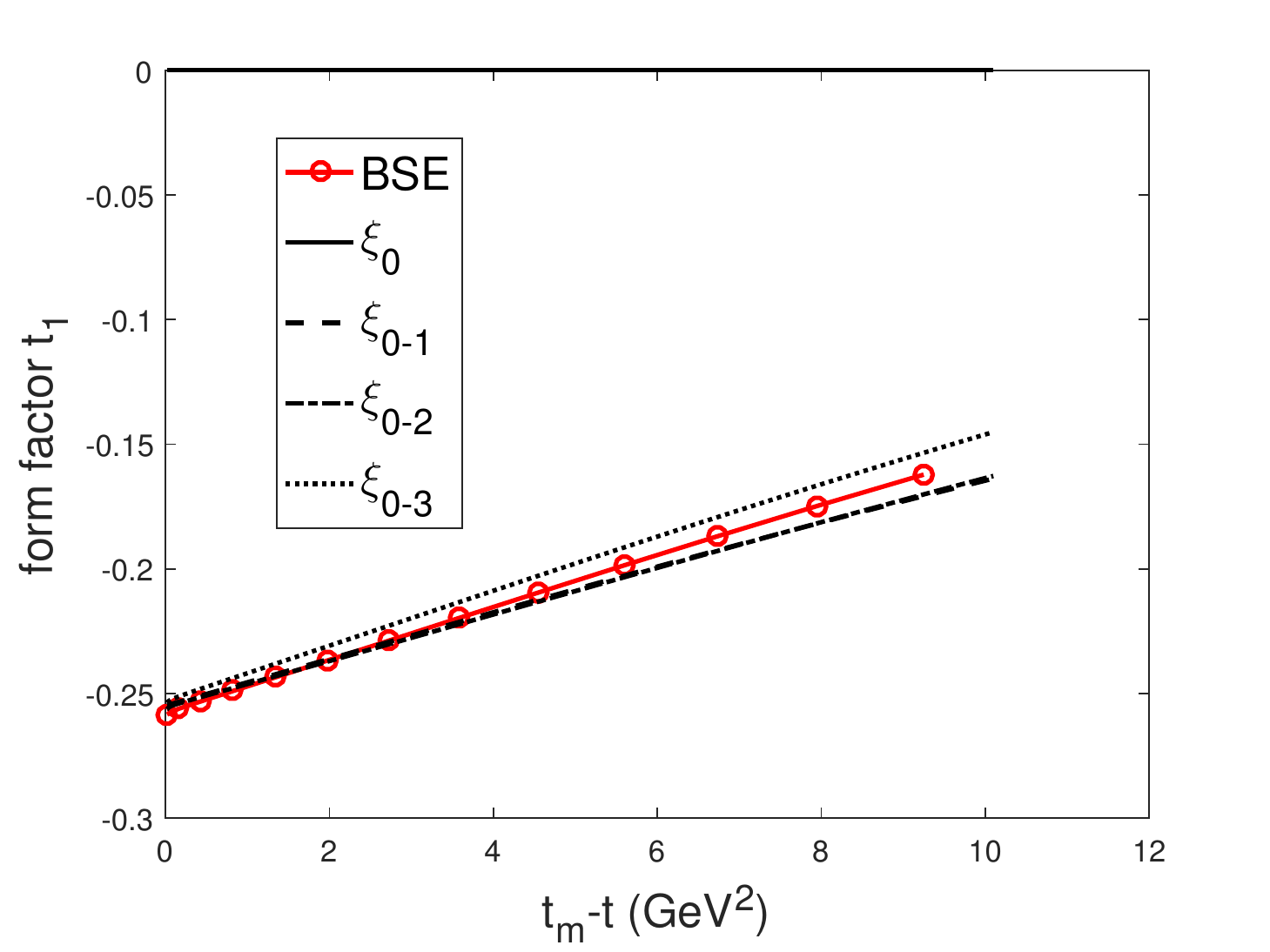}}
\subfigure[$B_c\to J/\psi:t_{2}$]{\label{fig:t2-bcjpsi}
			      \includegraphics[width=0.4\textwidth]{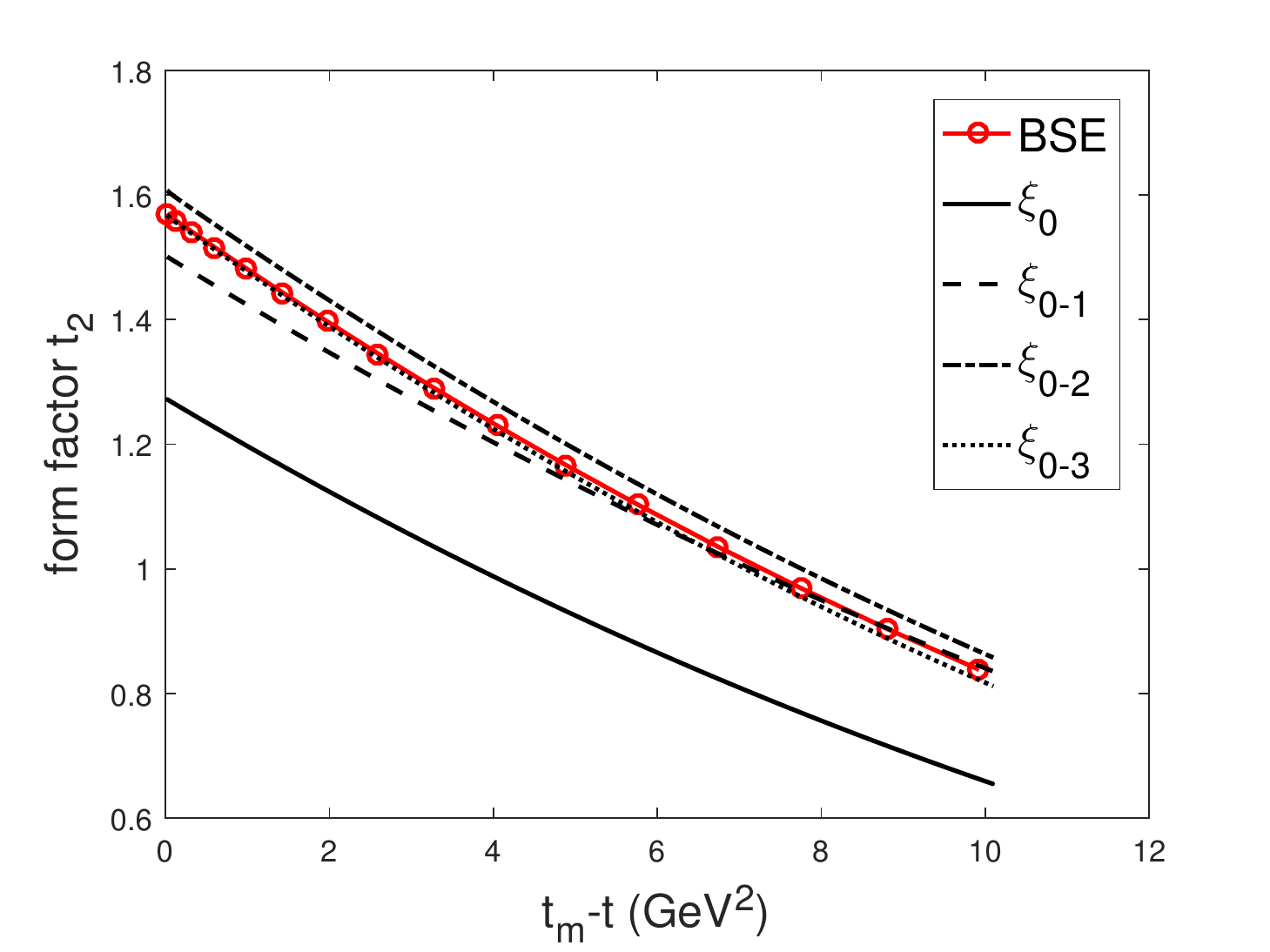}}
\subfigure[$B_c\to J/\psi:t_{3}$]{\label{fig:t3-bcjpsi}
			      \includegraphics[width=0.4\textwidth]{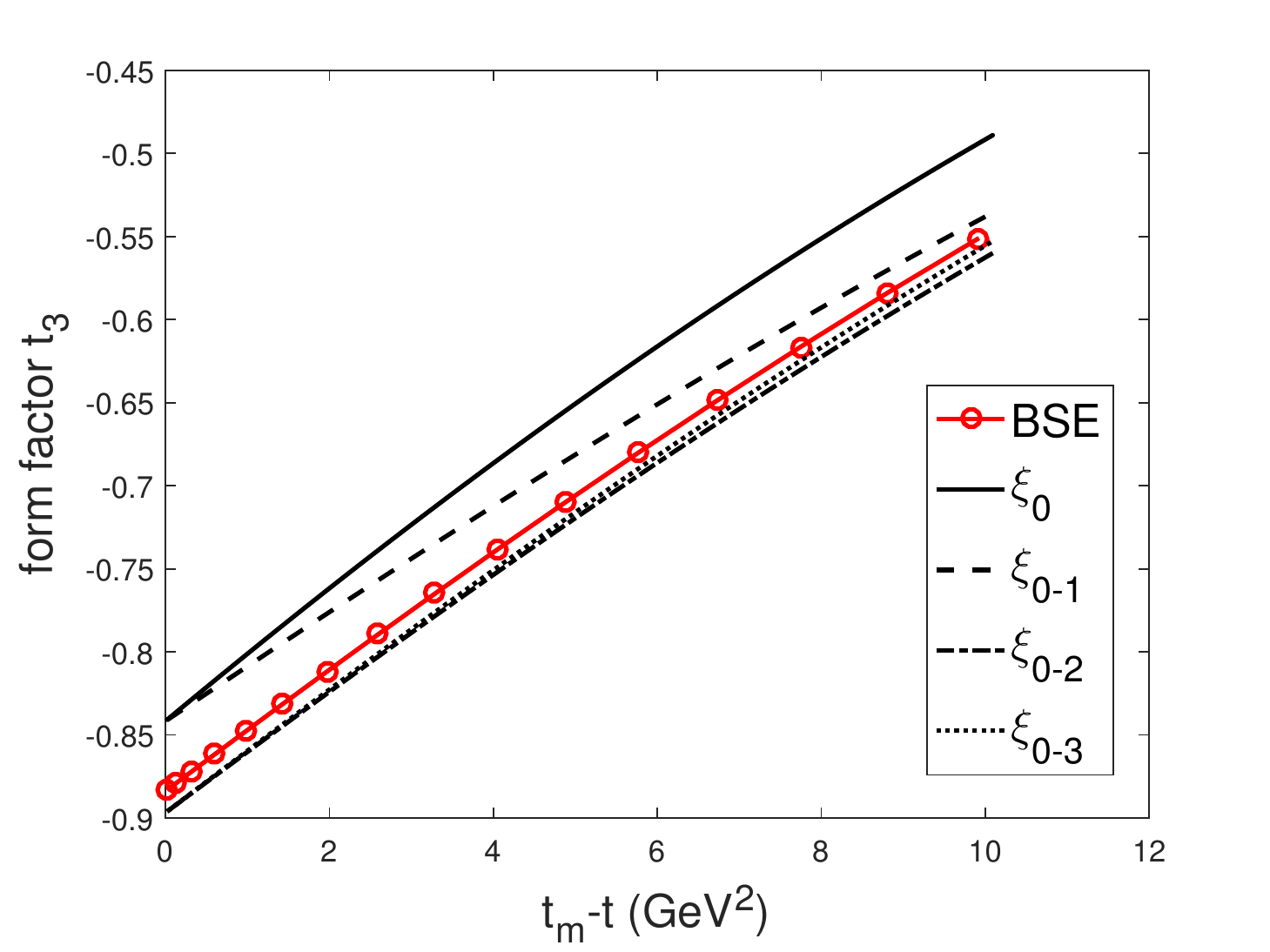}}
\subfigure[$B_c\to J/\psi:t_{4}$]{\label{fig:t4-bcjpsi}
			      \includegraphics[width=0.4\textwidth]{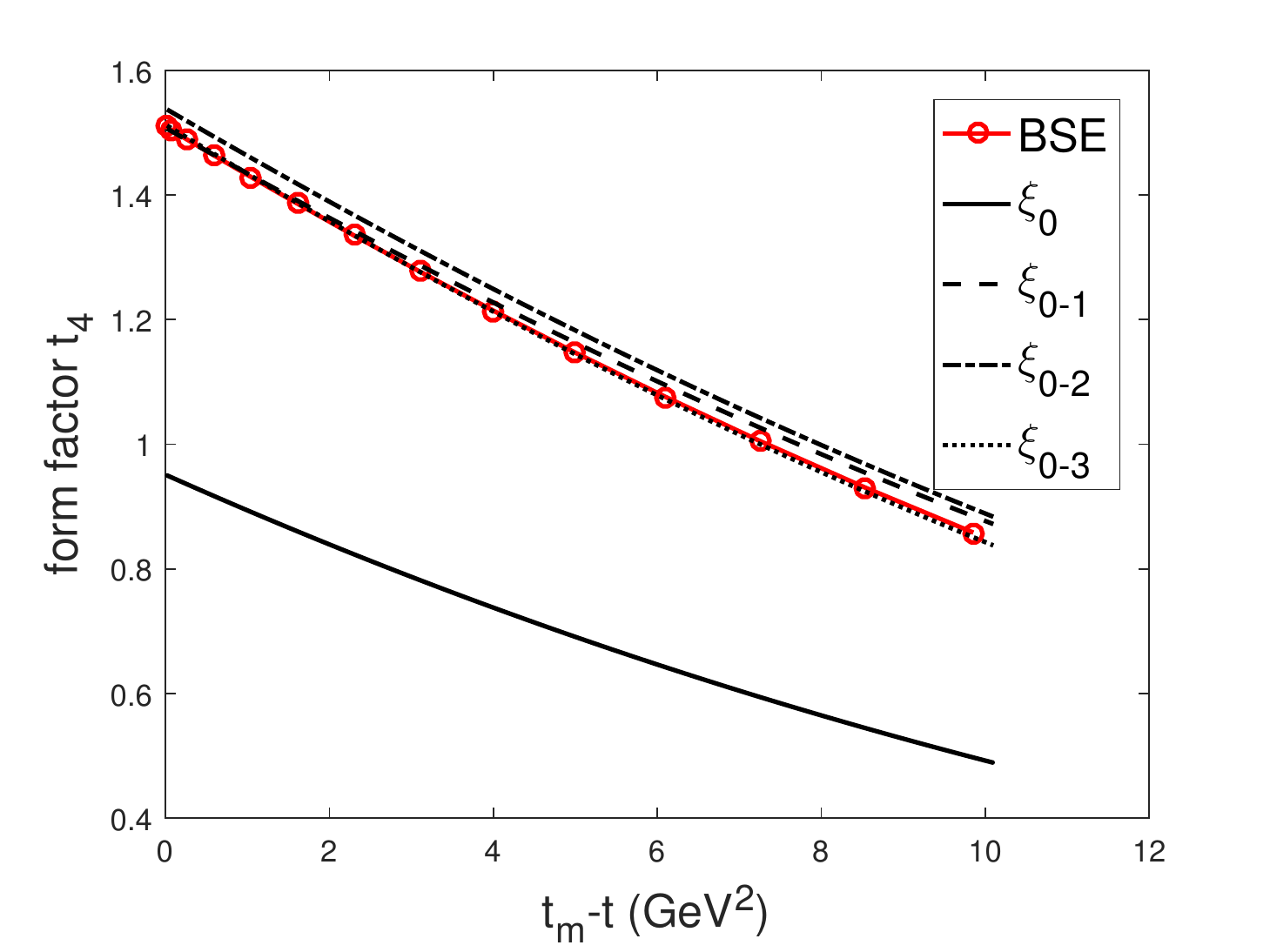}}
\caption{The form factors of $B_c\to\eta_c,J/\psi$ calculated by IWFs and instantaneous BS method, where $t\equiv(P-P_f)^2$ is the momentum transfer, and $t_m-t=2MM_f(v\cdot v'-1)$. The circle-solid line denotes the form factor calculated by instantaneous Bethe-Salpeter method directly; the solid line denotes the leading order of form factor calculated only by IWF; the dash line denotes the result with IWF and first order correction; the dot-dash line denotes the result with IWF, the first and second order corrections; the dot line denotes the result with IWF, the first, second and third order corrections.}\label{fig:ffs1}
\end{figure}

\begin{figure}[!hbp]
\centering
\subfigure[$B_c\to h_c:t_{1}$]{\label{fig:t1-bchc}
			      \includegraphics[width=0.4\textwidth]{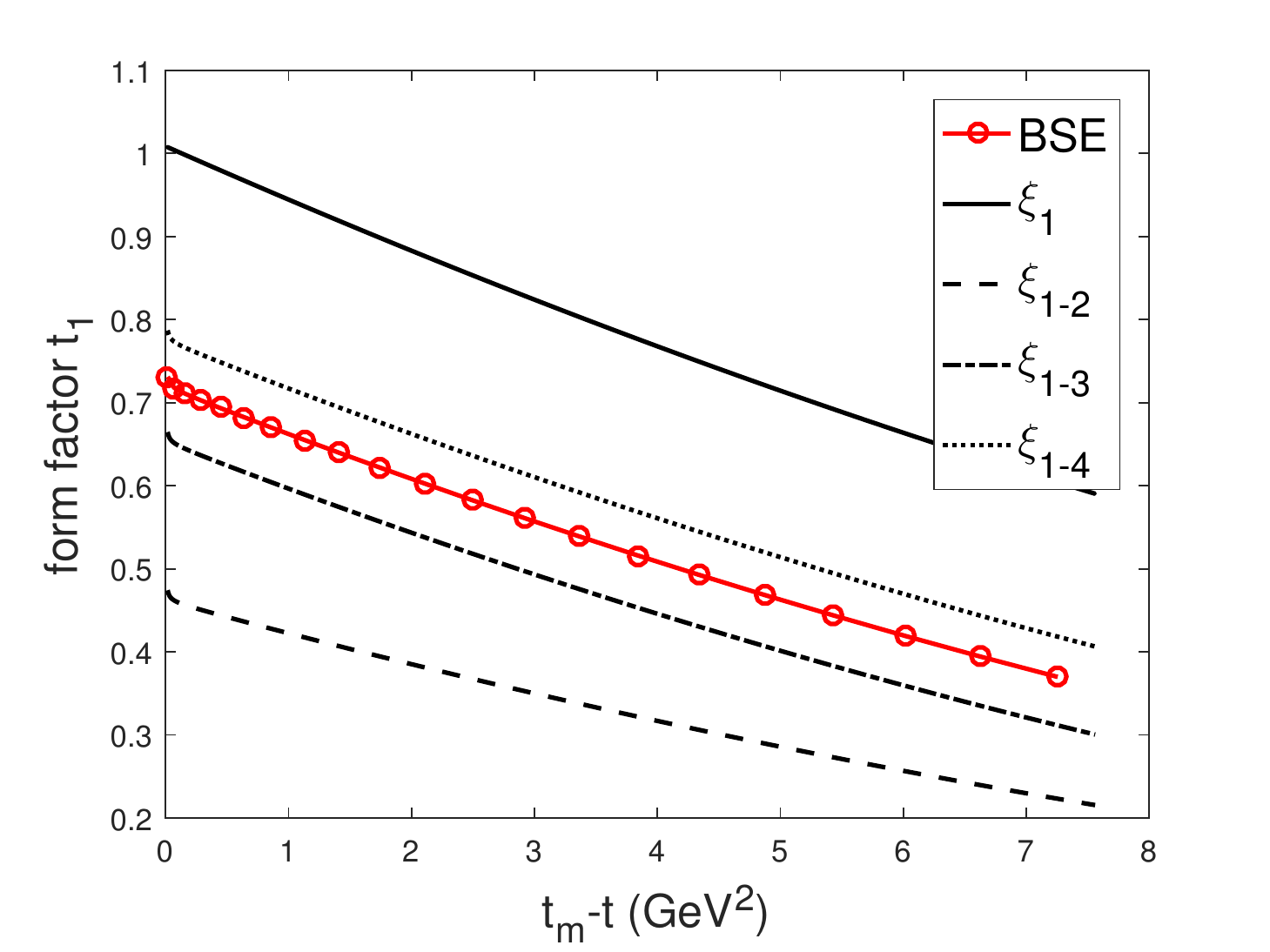}}
\subfigure[$B_c\to h_c:t_{2}$]{\label{fig:t2-bchc}
			      \includegraphics[width=0.4\textwidth]{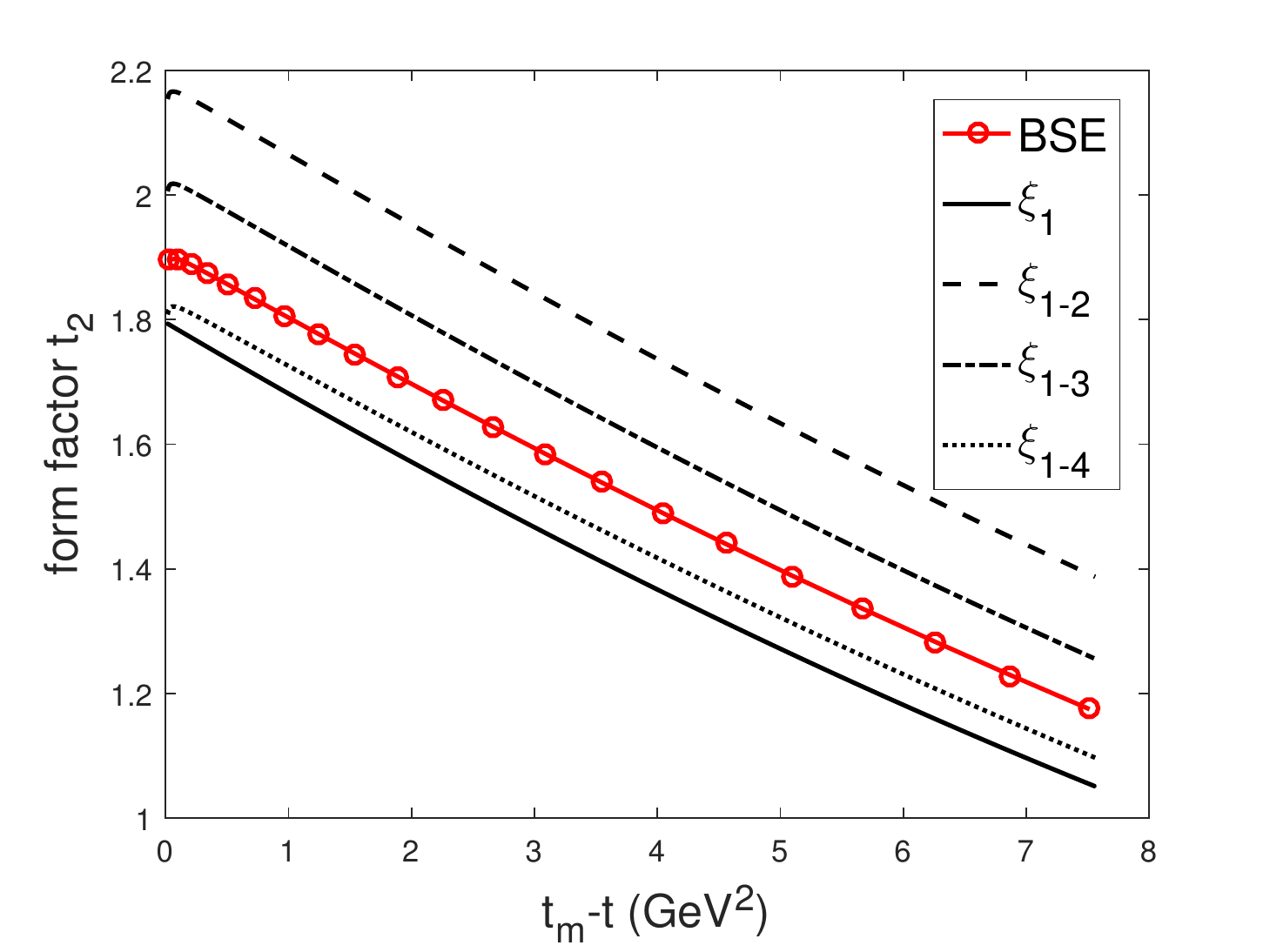}}
\subfigure[$B_c\to h_c:t_{3}$]{\label{fig:t3-bchc}
			      \includegraphics[width=0.4\textwidth]{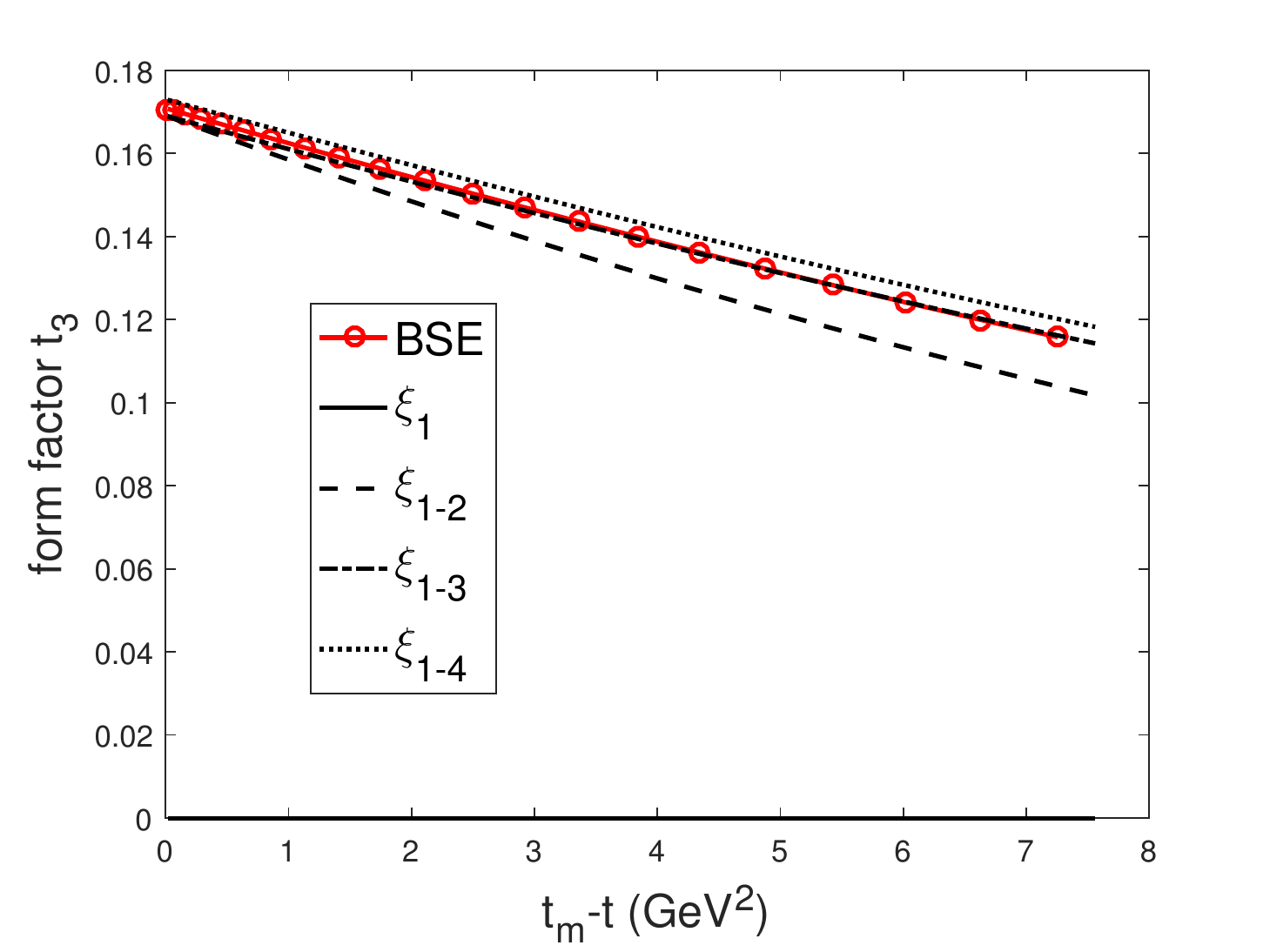}}
\subfigure[$B_c\to h_c:t_{4}$]{\label{fig:t4-bchc}
			      \includegraphics[width=0.4\textwidth]{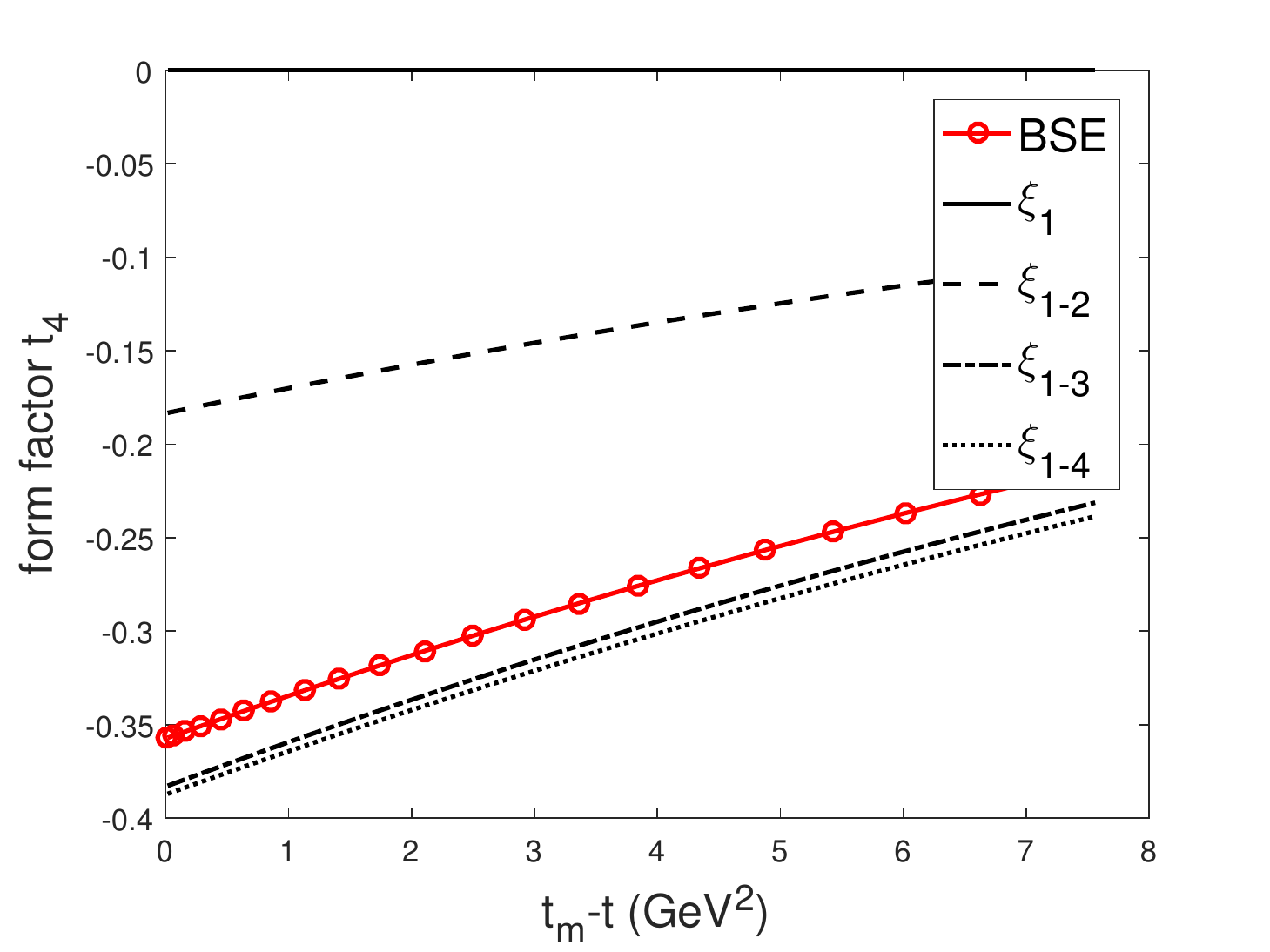}}
\subfigure[$B_c\to \chi_{c0}:S_{+}$]{\label{fig:s+-bcxc0}
			      \includegraphics[width=0.4\textwidth]{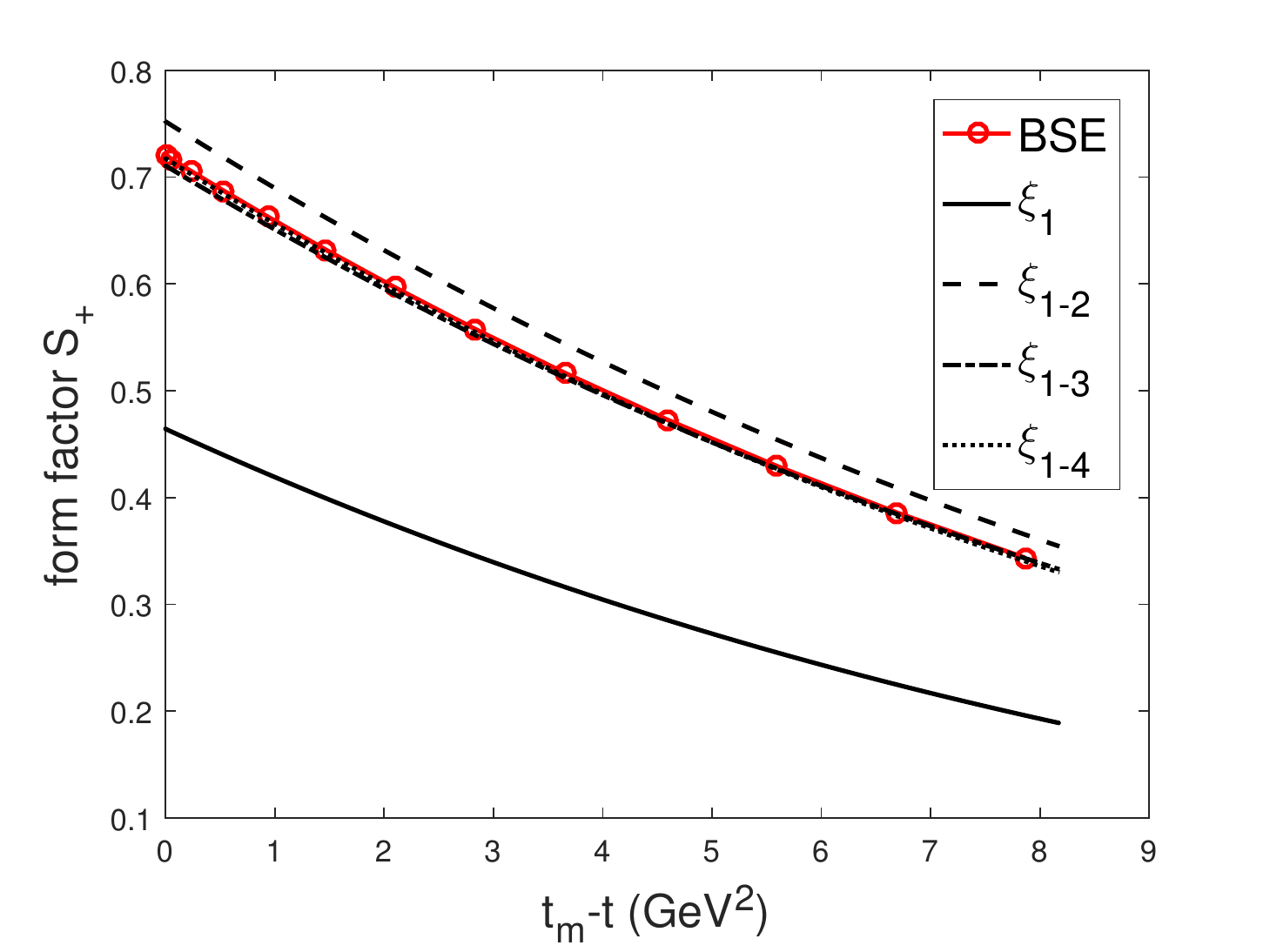}}
\subfigure[$B_c\to \chi_{c0}:S_{-}$]{\label{fig:s--bcxc0}
			      \includegraphics[width=0.4\textwidth]{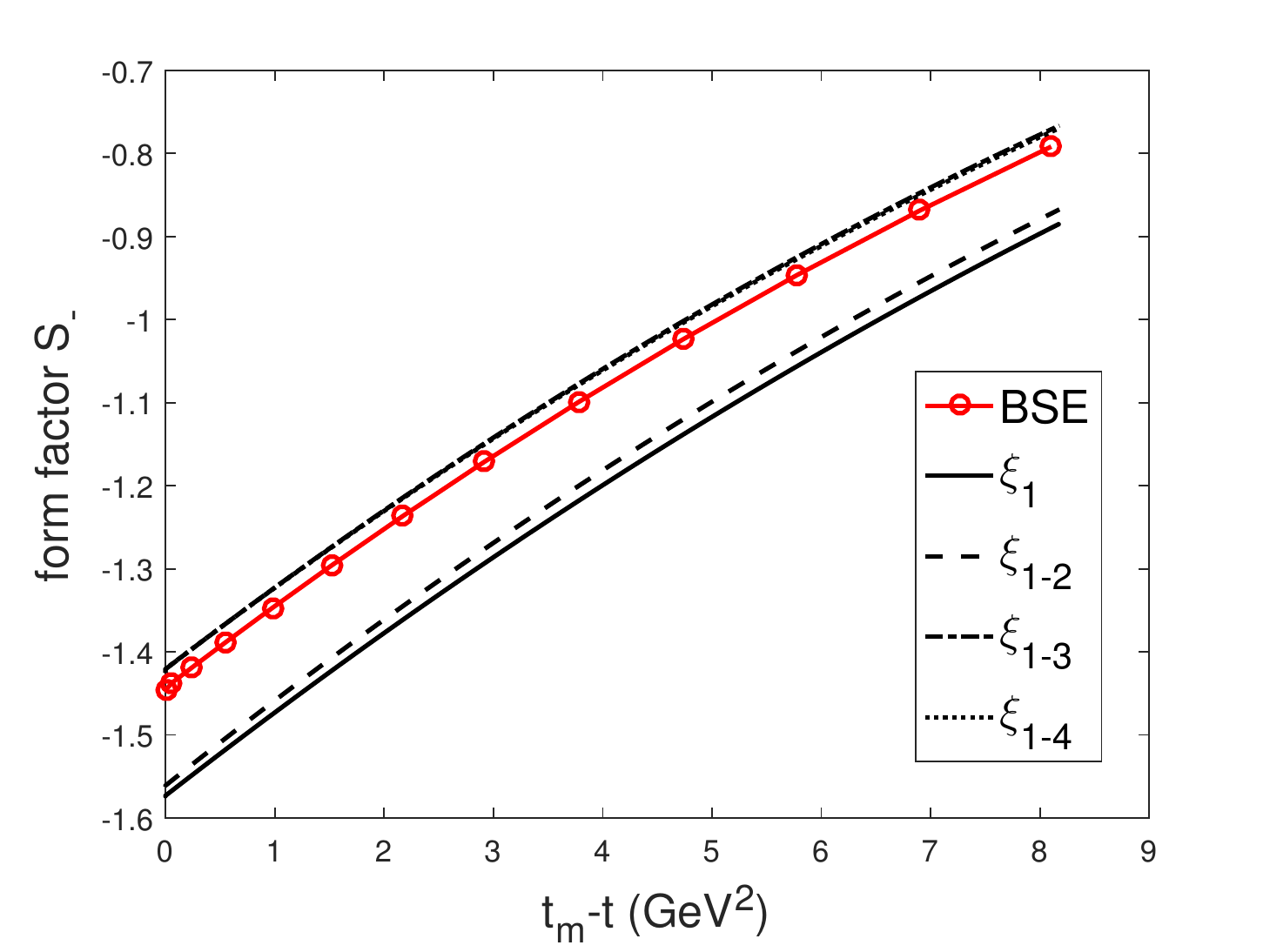}}
\caption{The form factors of $B_c\to h_c,\chi_{c0}$ calculated by IWFs and instantaneous Bethe-Salpeter method, where $t\equiv(P-P_f)^2$ is the momentum transfer, and $t_m-t=2MM_f(v\cdot v'-1)$. The meaning of each type line is the same as that in Fig.~\ref{fig:ffs1}.}\label{fig:ffs2}
\end{figure}

\begin{figure}[!hbp]
\centering
\subfigure[$B_c\to \chi_{c1}:t_{1}$]{\label{fig:t1-bcxc1}
			      \includegraphics[width=0.4\textwidth]{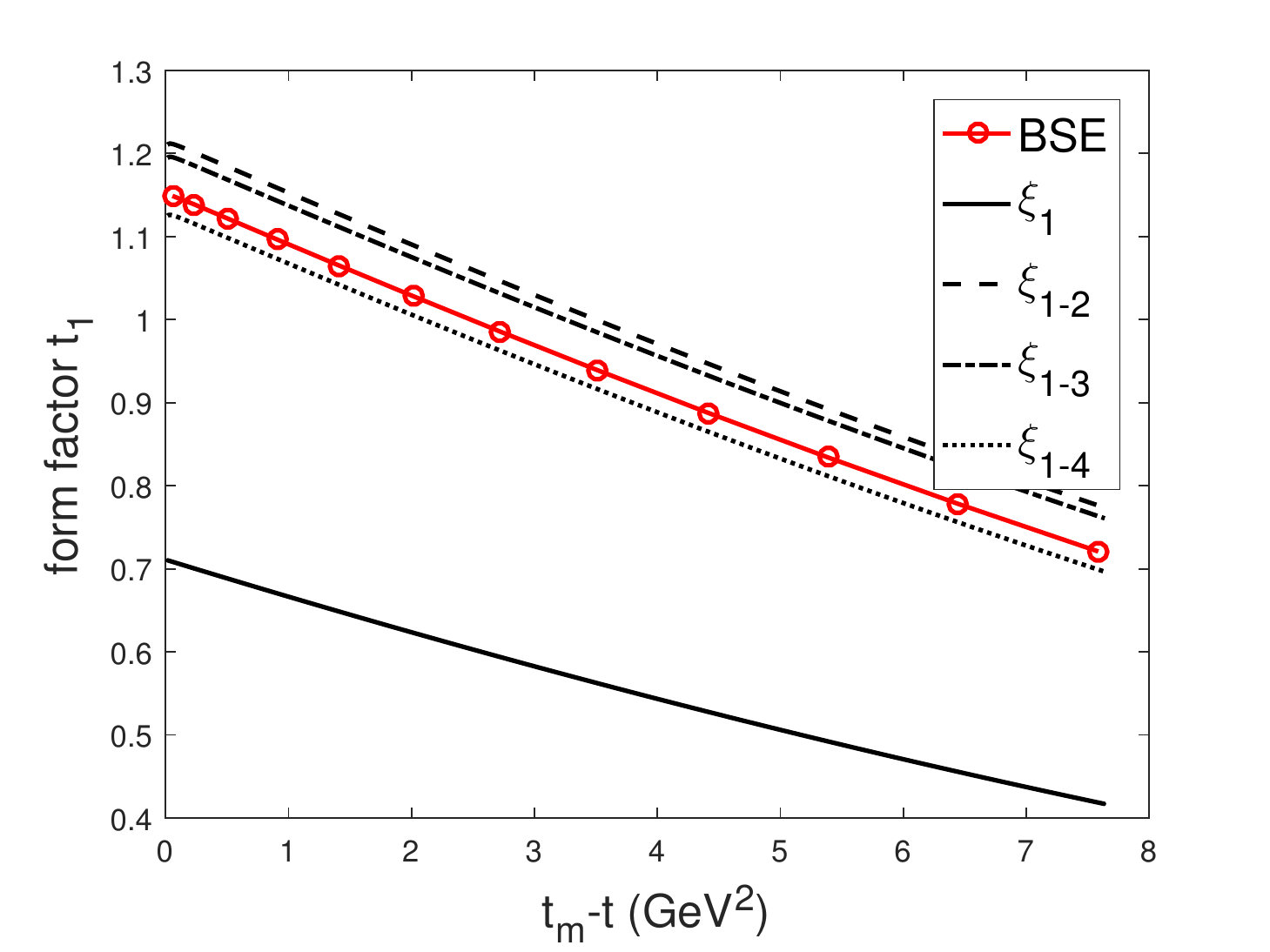}}
\subfigure[$B_c\to \chi_{c1}:t_{2}$]{\label{fig:t2-bcxc1}
			      \includegraphics[width=0.4\textwidth]{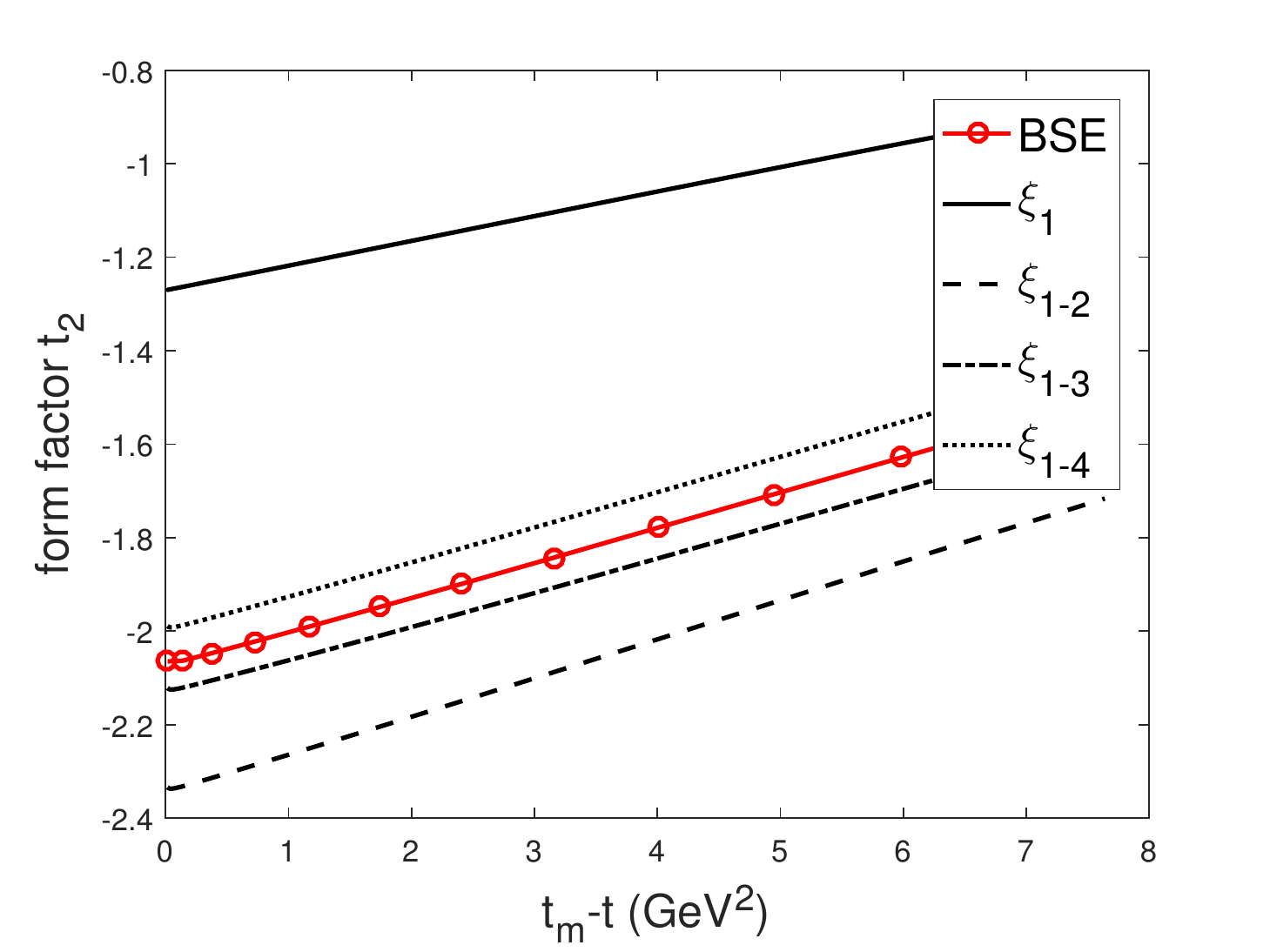}}
\subfigure[$B_c\to \chi_{c1}:t_{3}$]{\label{fig:t3-bcxc1}
			      \includegraphics[width=0.4\textwidth]{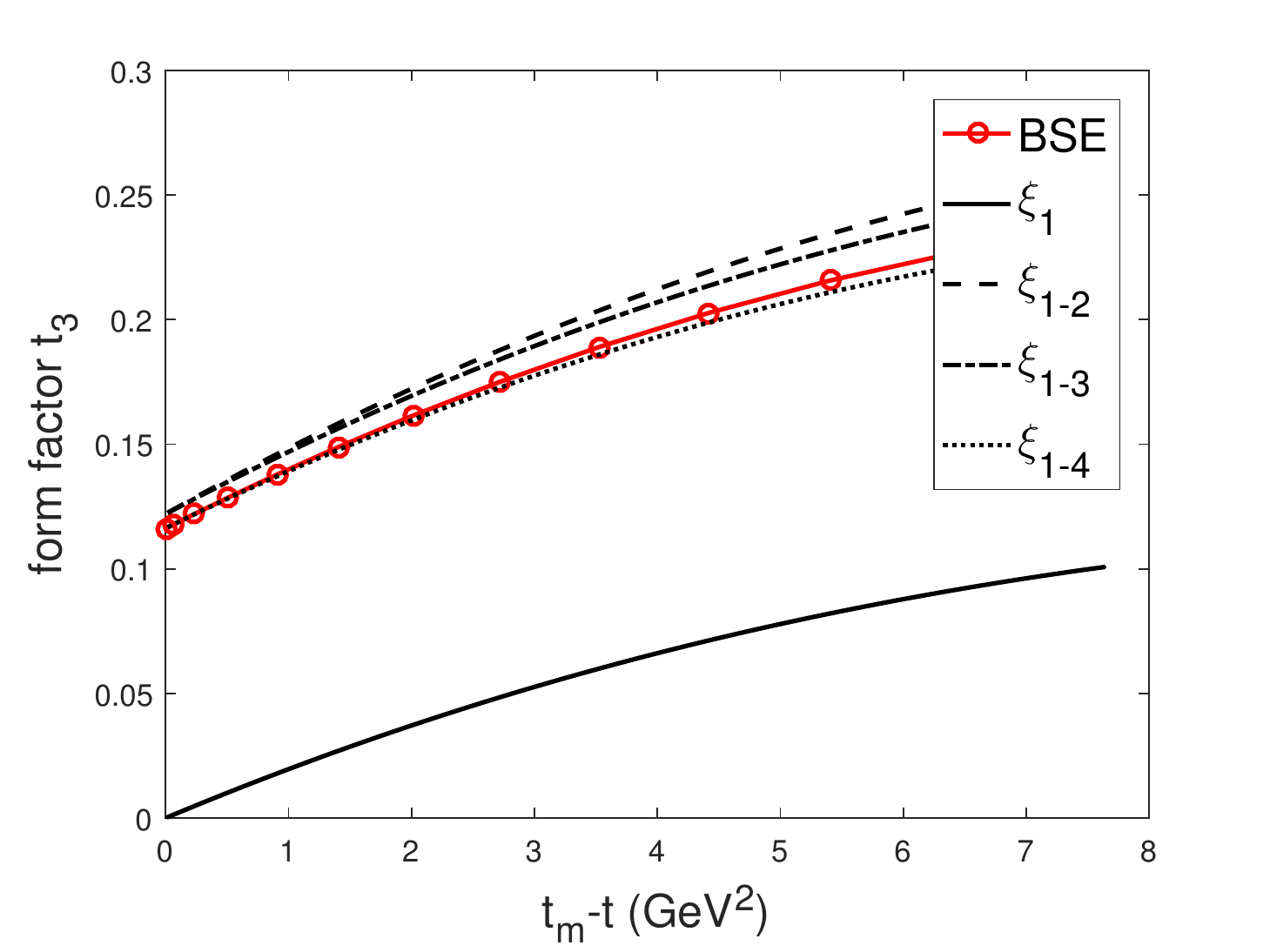}}
\subfigure[$B_c\to \chi_{c1}:t_{4}$]{\label{fig:t4-bcxc1}
			      \includegraphics[width=0.4\textwidth]{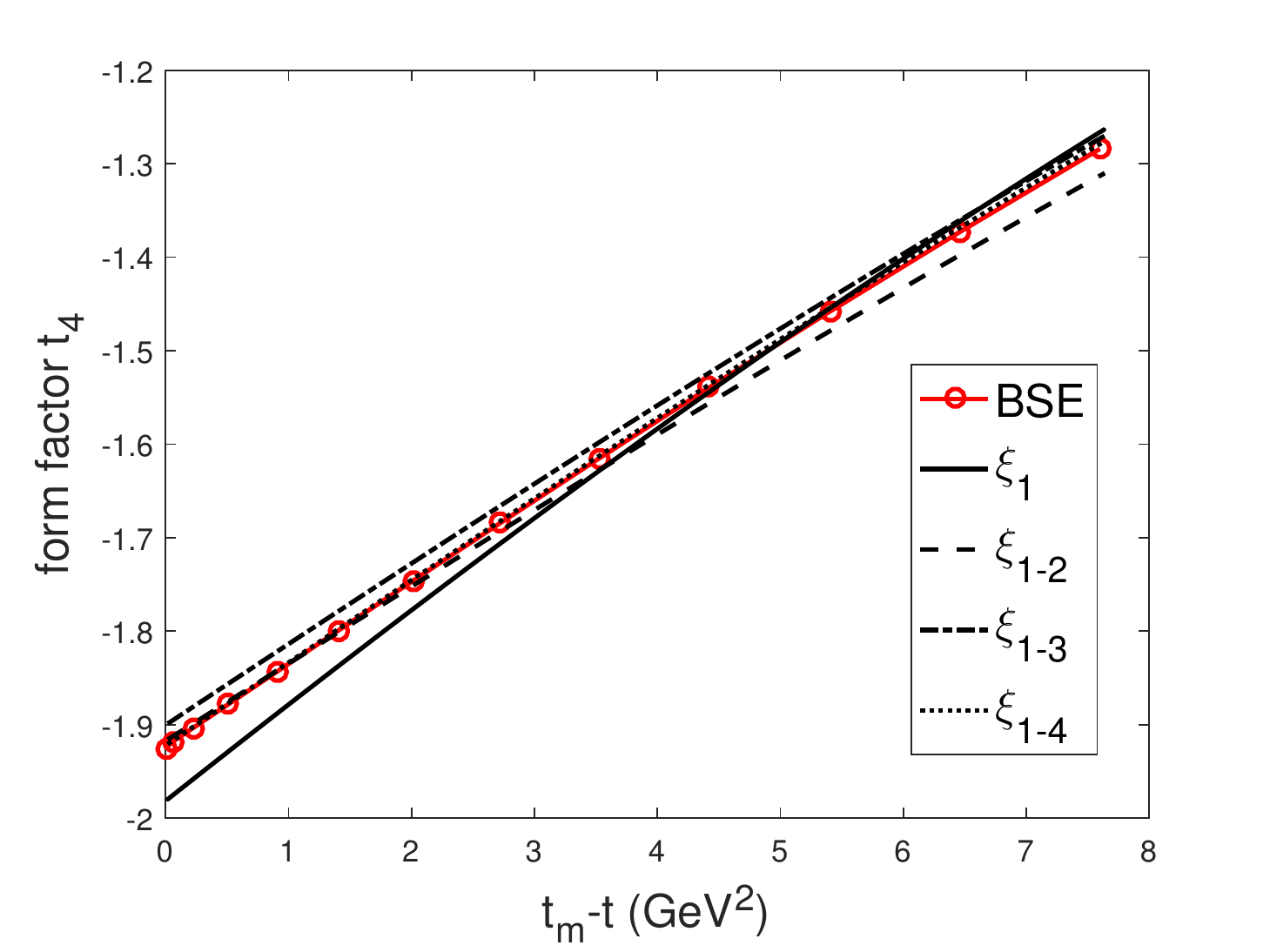}}
\caption{The form factors of $B_c\to \chi_{c1}$ calculated by IWFs and instantaneous Bethe-Salpeter method, where $t\equiv(P-P_f)^2$ is the momentum transfer, and $t_m-t=2MM_f(v\cdot v'-1)$. The meaning of each type line is the same as that in Fig.~\ref{fig:ffs1}.}\label{fig:ffs3}
\end{figure}

\begin{figure}[!hbp]
	\centering
	\subfigure[$B_c\to \chi_{c2}:t_{1}$]{\label{fig:t1-bcxc2}
		\includegraphics[width=0.4\textwidth]{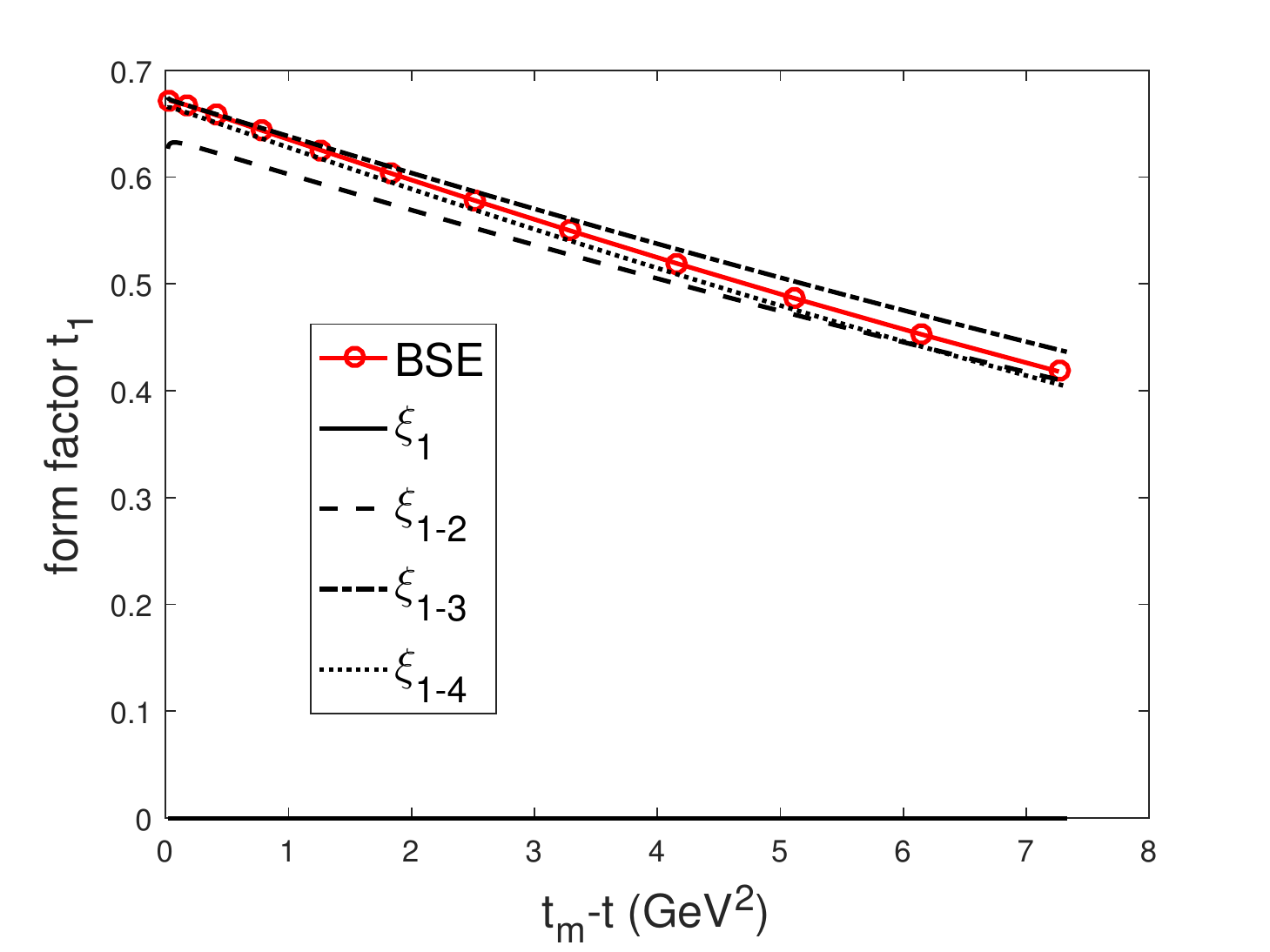}}
	\subfigure[$B_c\to \chi_{c2}:t_{2}$]{\label{fig:t2-bcxc2}
		\includegraphics[width=0.4\textwidth]{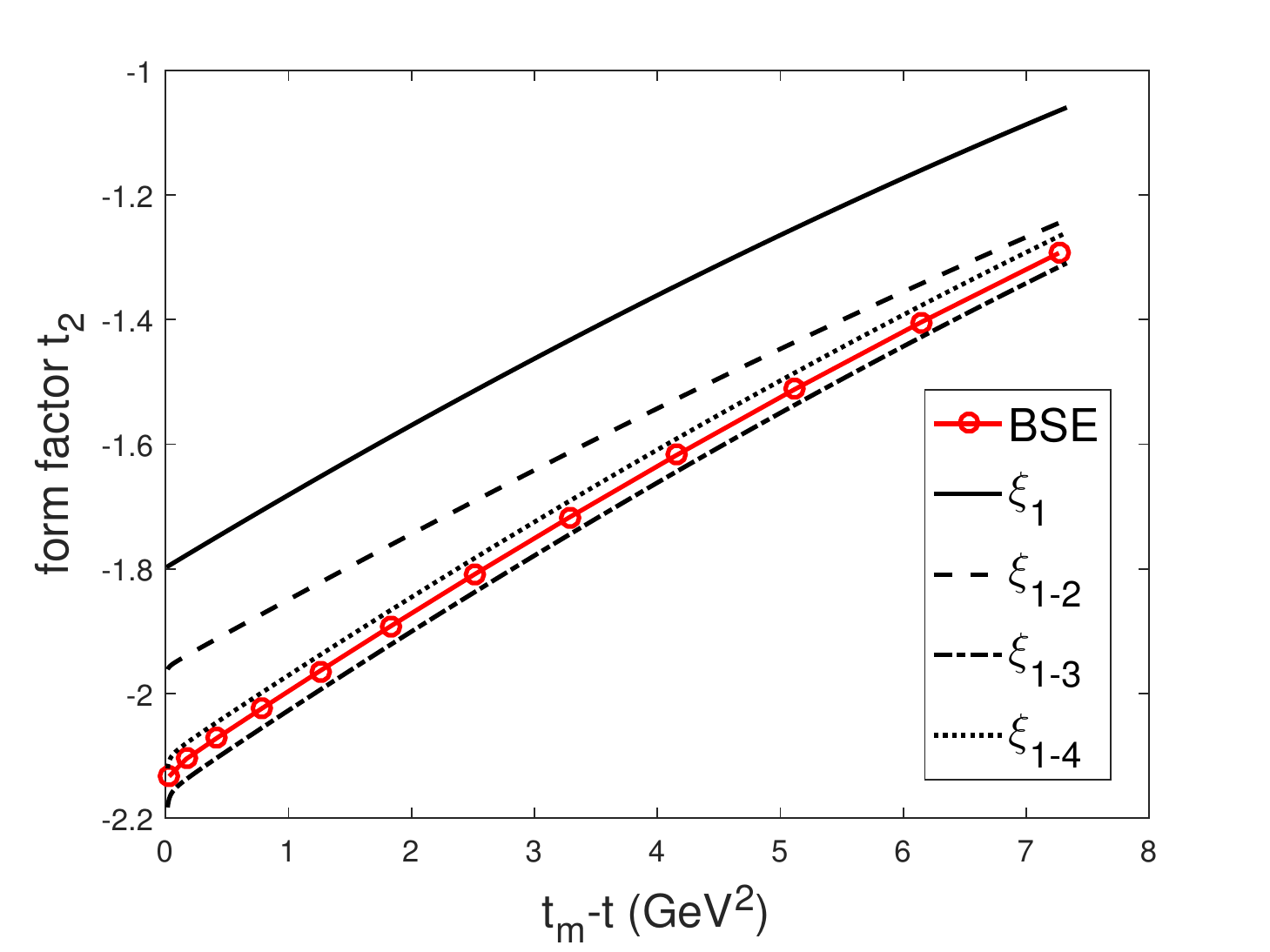}}
	\subfigure[$B_c\to \chi_{c2}:t_{3}$]{\label{fig:t3-bcxc2}
		\includegraphics[width=0.4\textwidth]{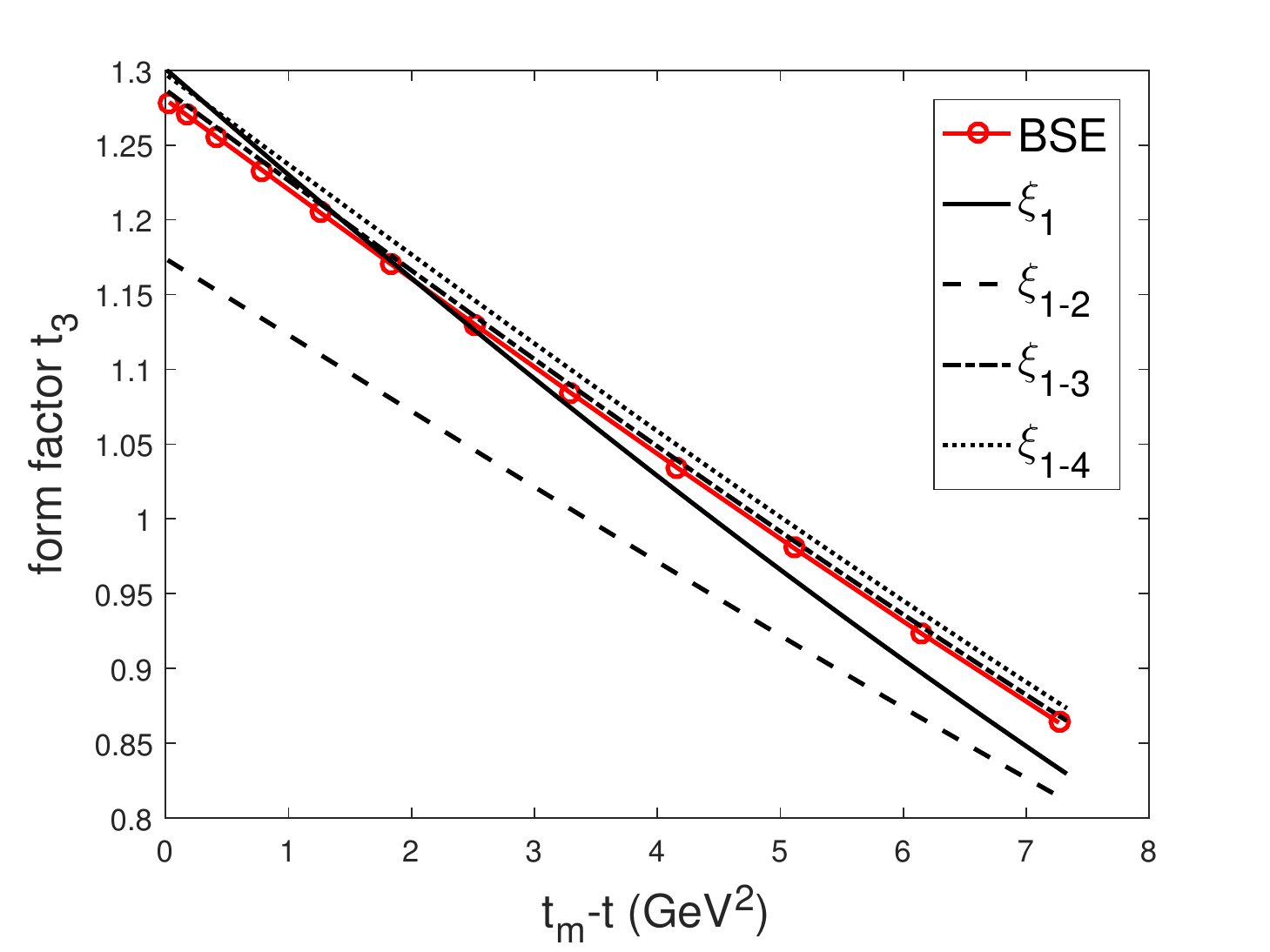}}
	\subfigure[$B_c\to \chi_{c2}:t_{4}$]{\label{fig:t4-bcxc2}
		\includegraphics[width=0.4\textwidth]{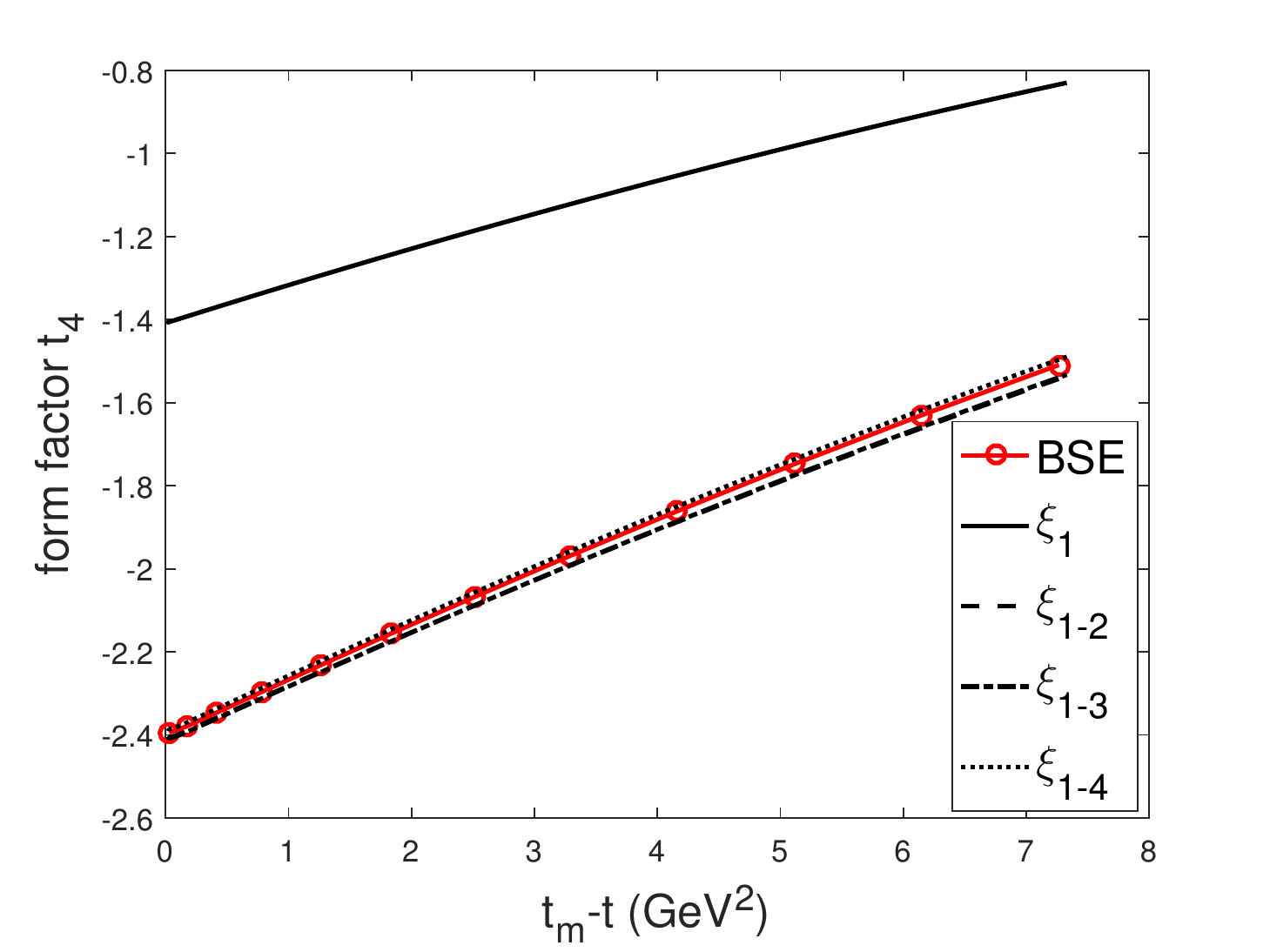}}
	\caption{The form factors of $B_c\to \chi_{c2}$ calculated by IWFs and instantaneous Bethe-Salpeter method, where $t\equiv(P-P_f)^2$ is the momentum transfer, and $t_m-t=2MM_f(v\cdot v'-1)$. The meaning of each type line is the same as that in Fig.~\ref{fig:ffs1}.}\label{fig:ffs3}
\end{figure}

\begin{figure}[!hbp]
\centering
\subfigure[$B_c\to\eta_c(2S):S_{+}$]{\label{fig:s+-bcetac2s}
			      \includegraphics[width=0.4\textwidth]{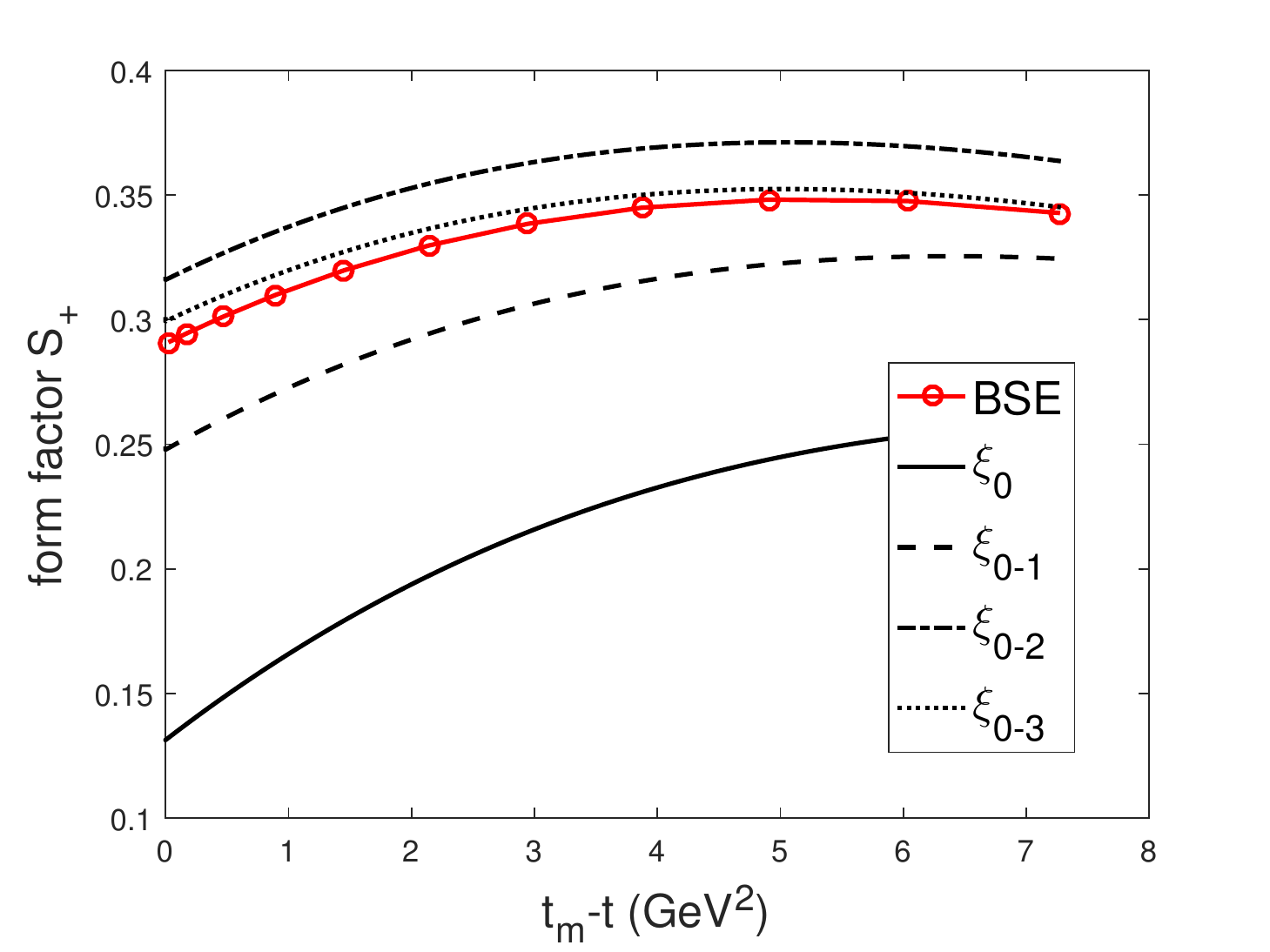}}
\subfigure[$B_c\to\eta_c(2S):S_{-}$]{\label{fig:s--bcetac2s}
			      \includegraphics[width=0.4\textwidth]{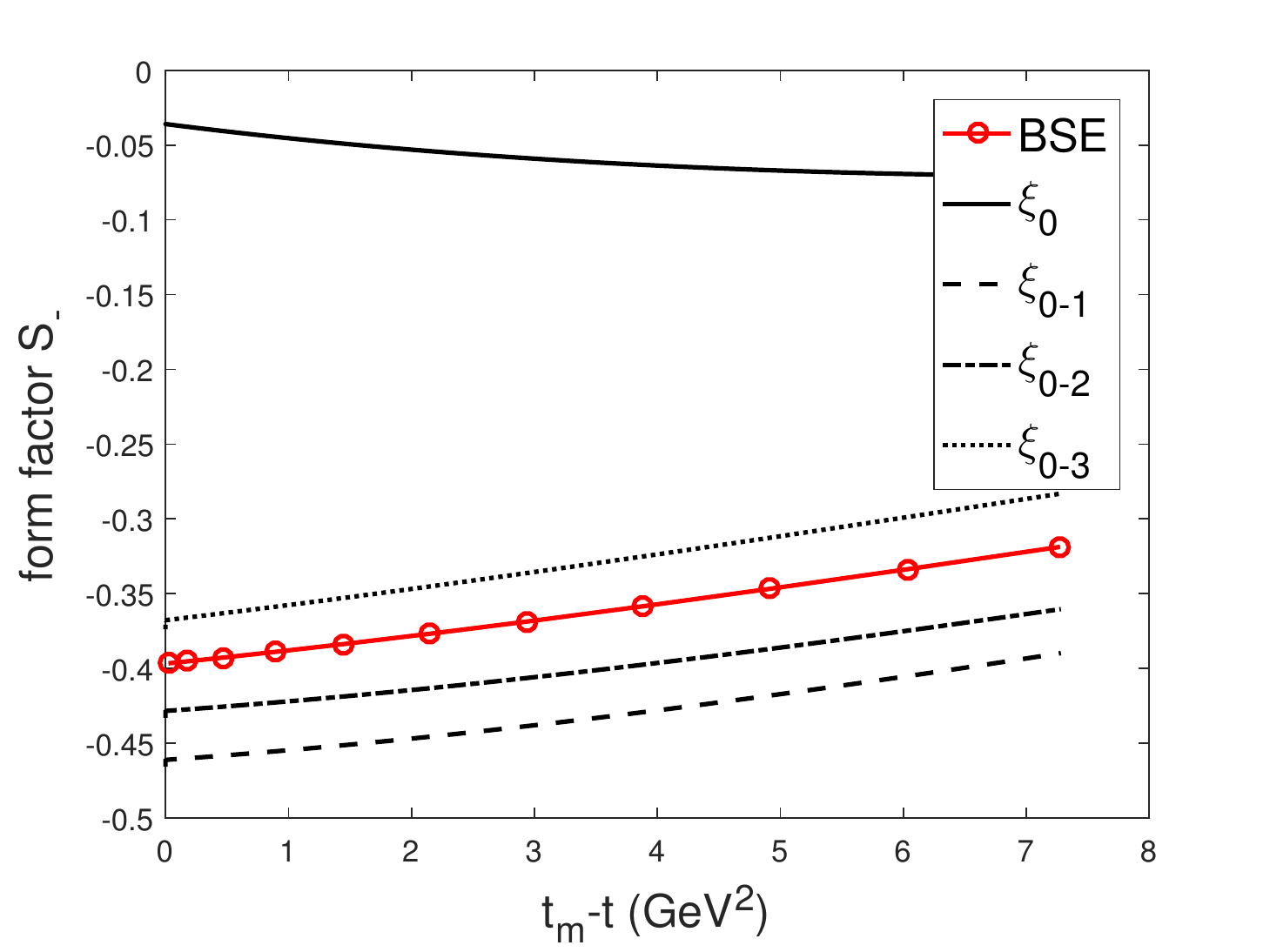}}
\subfigure[$B_c\to \psi(2S):t_{1}$]{\label{fig:t1-bcpsi2s}
			      \includegraphics[width=0.4\textwidth]{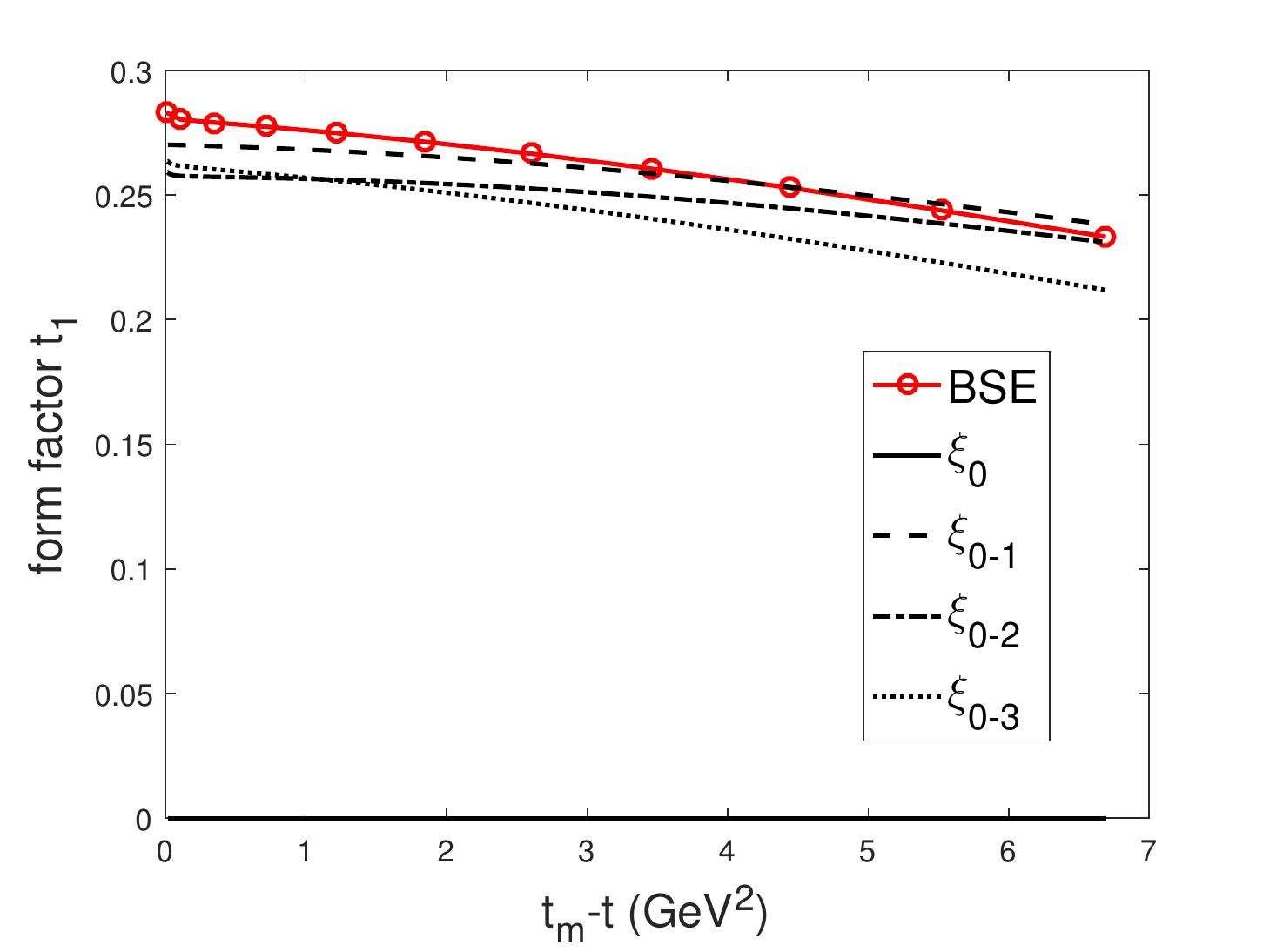}}
\subfigure[$B_c\to \psi(2S):t_{2}$]{\label{fig:t2-bcpsi2s}
			      \includegraphics[width=0.4\textwidth]{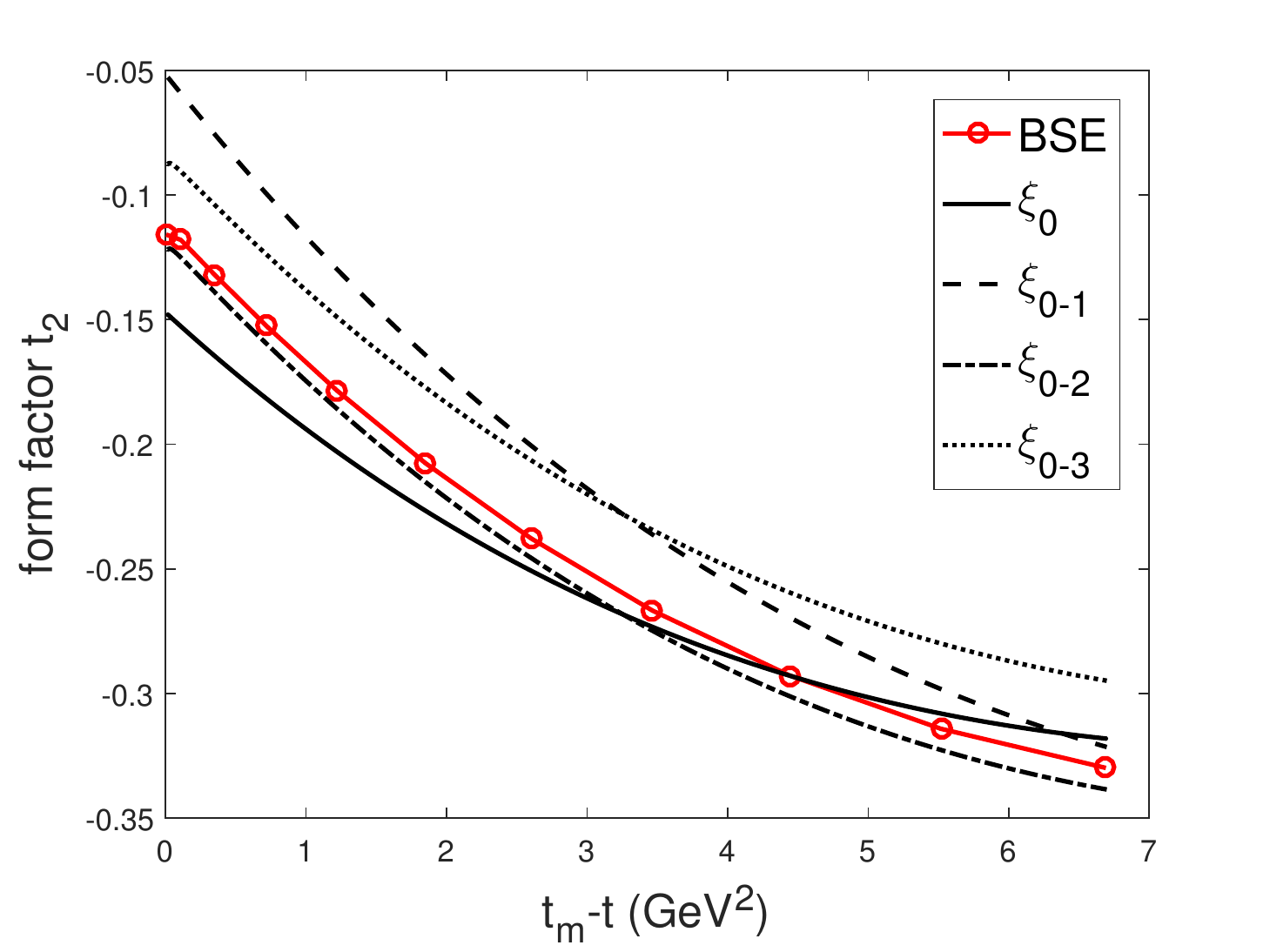}}
\subfigure[$B_c\to \psi(2S):t_{3}$]{\label{fig:t3-bcpsi2s}
			      \includegraphics[width=0.4\textwidth]{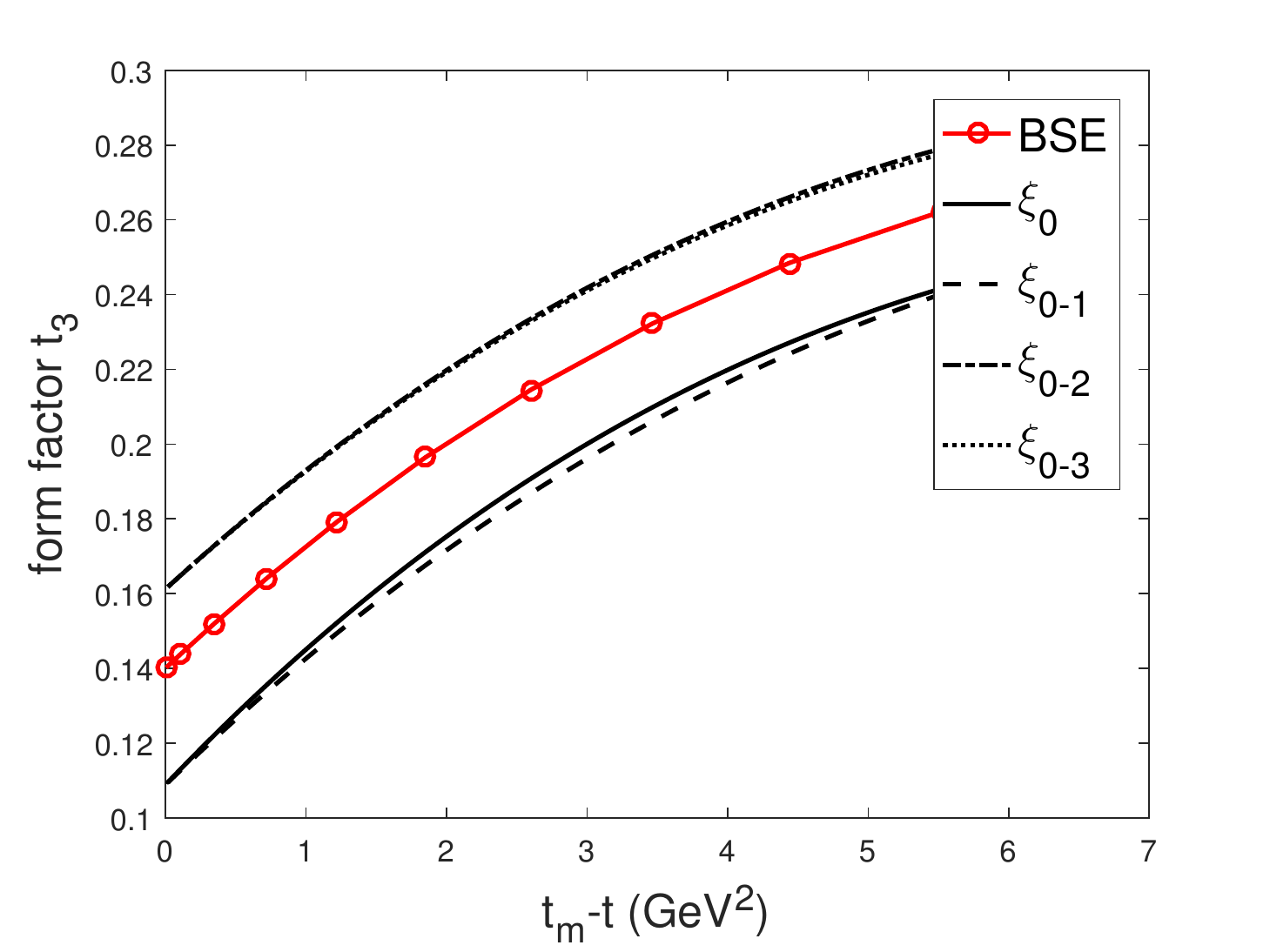}}
\subfigure[$B_c\to \psi(2S):t_{4}$]{\label{fig:t4-bcpsi2s}
			      \includegraphics[width=0.4\textwidth]{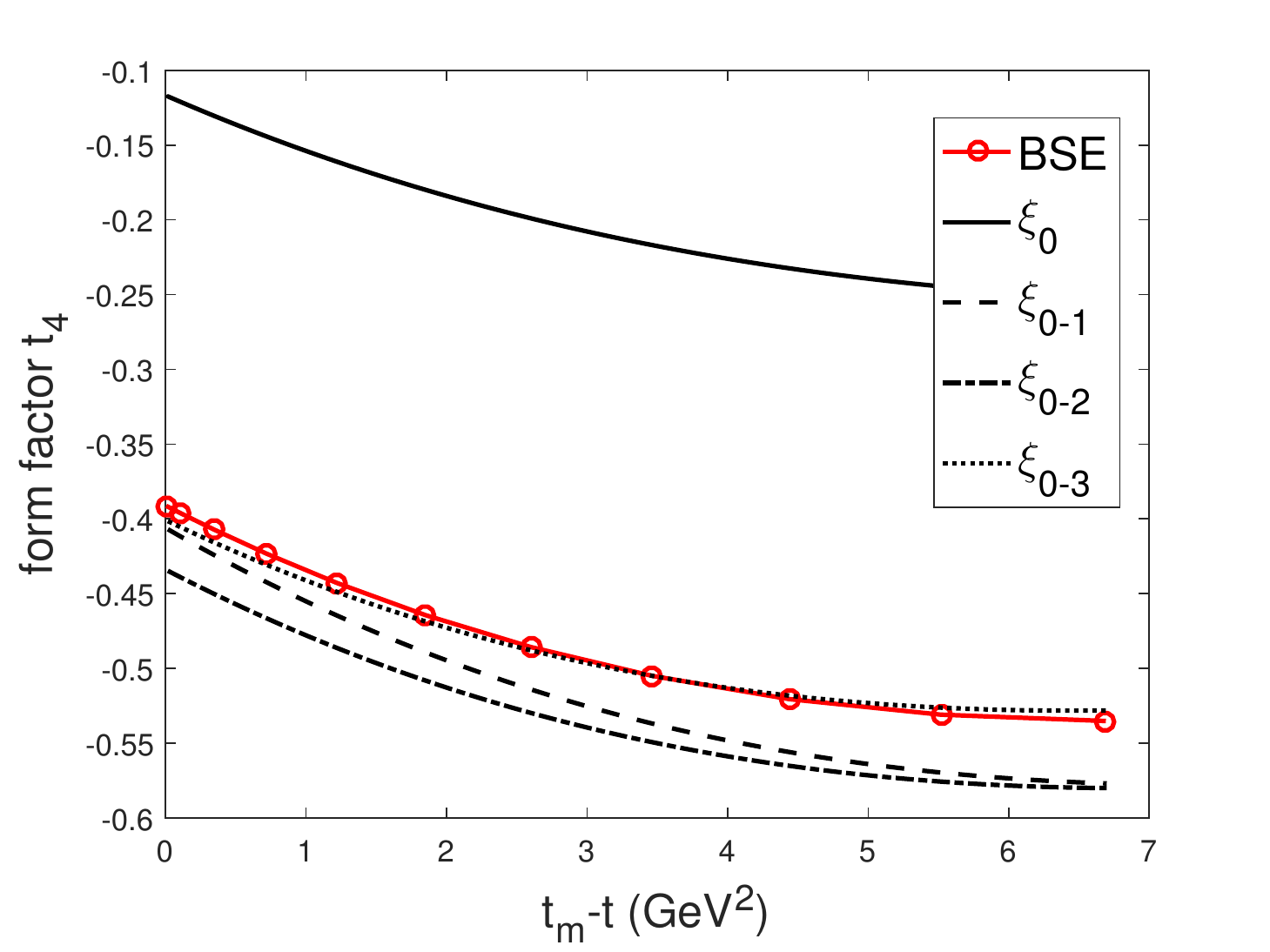}}
\caption{The form factors of $B_c\to\eta_c(2S),\psi(2S)$ calculated by IWFs and instantaneous Bethe-Salpeter method, where $t\equiv(P-P_f)^2$ is the momentum transfer, and $t_m-t=2MM_f(v\cdot v'-1)$. The meaning of each type line is the same as that in Fig.~\ref{fig:ffs1}.}\label{fig:ffs4}
\end{figure}

\begin{figure}[!hbp]
\centering
\subfigure[$B_c\to h_c(2P):t_{1}$]{\label{fig:t1-bchc2p}
			      \includegraphics[width=0.4\textwidth]{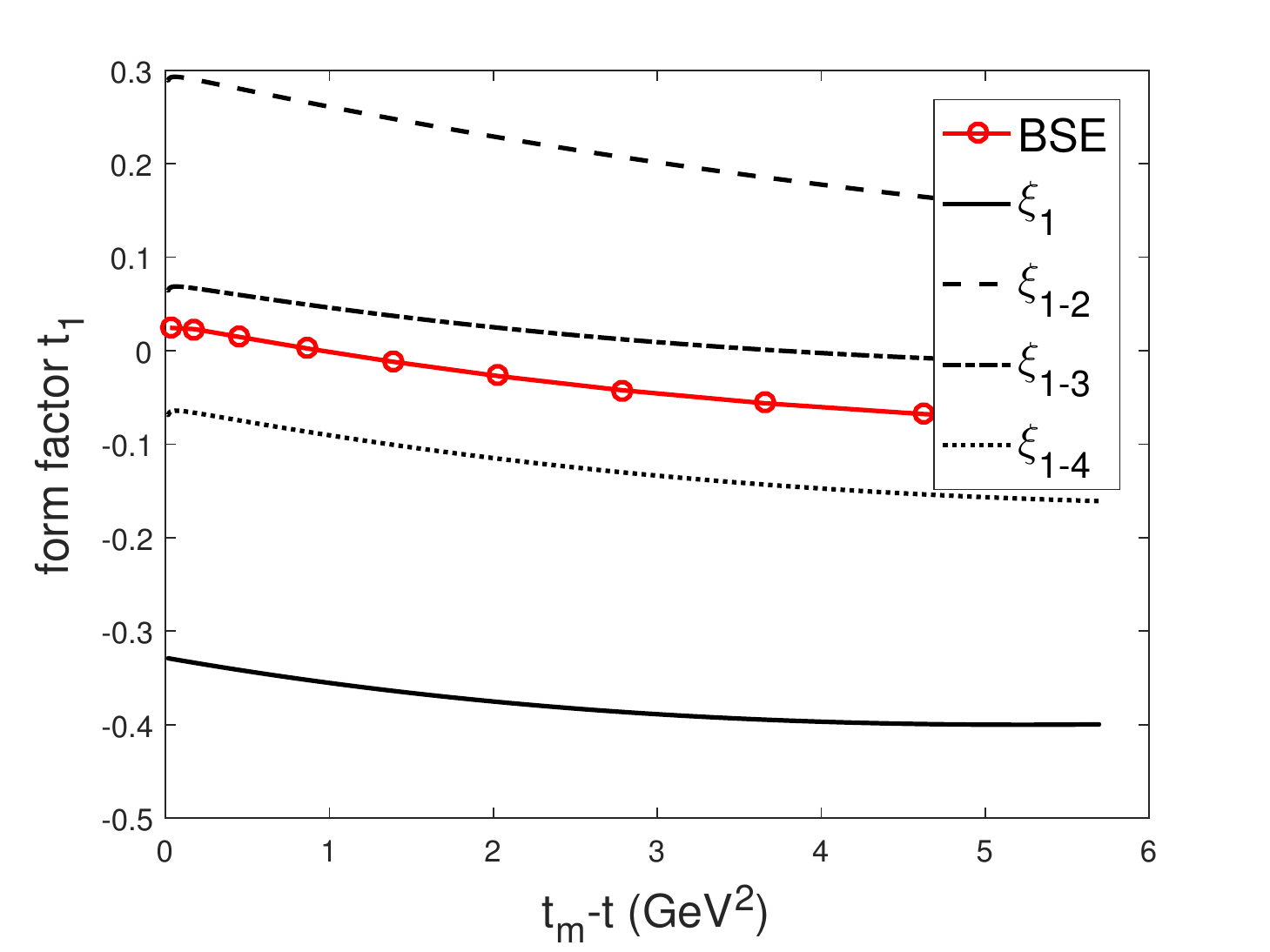}}
\subfigure[$B_c\to h_c(2P):t_{2}$]{\label{fig:t2-bchc2p}
			      \includegraphics[width=0.4\textwidth]{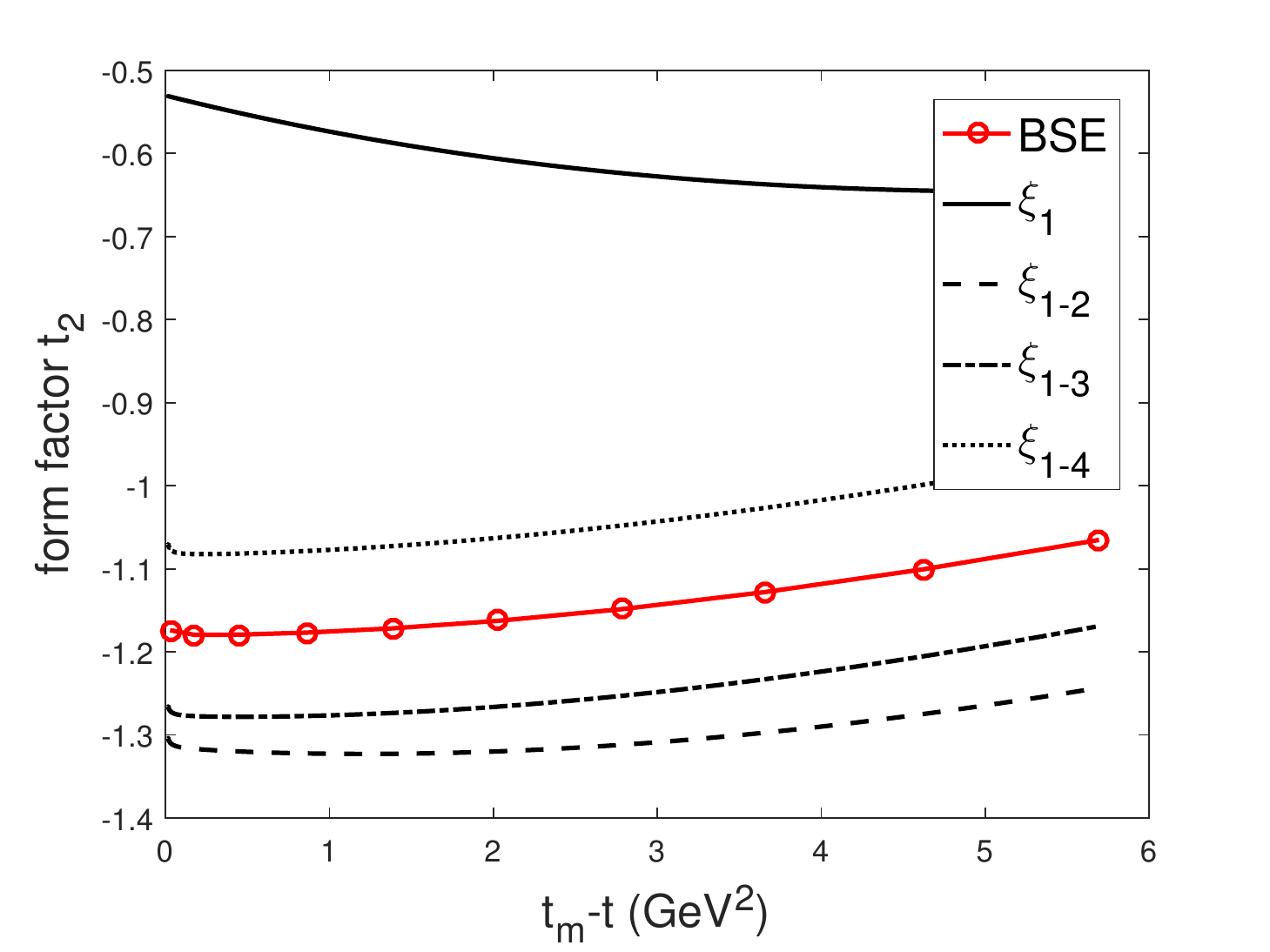}}
\subfigure[$B_c\to h_c(2P):t_{3}$]{\label{fig:t3-bchc2p}
			      \includegraphics[width=0.4\textwidth]{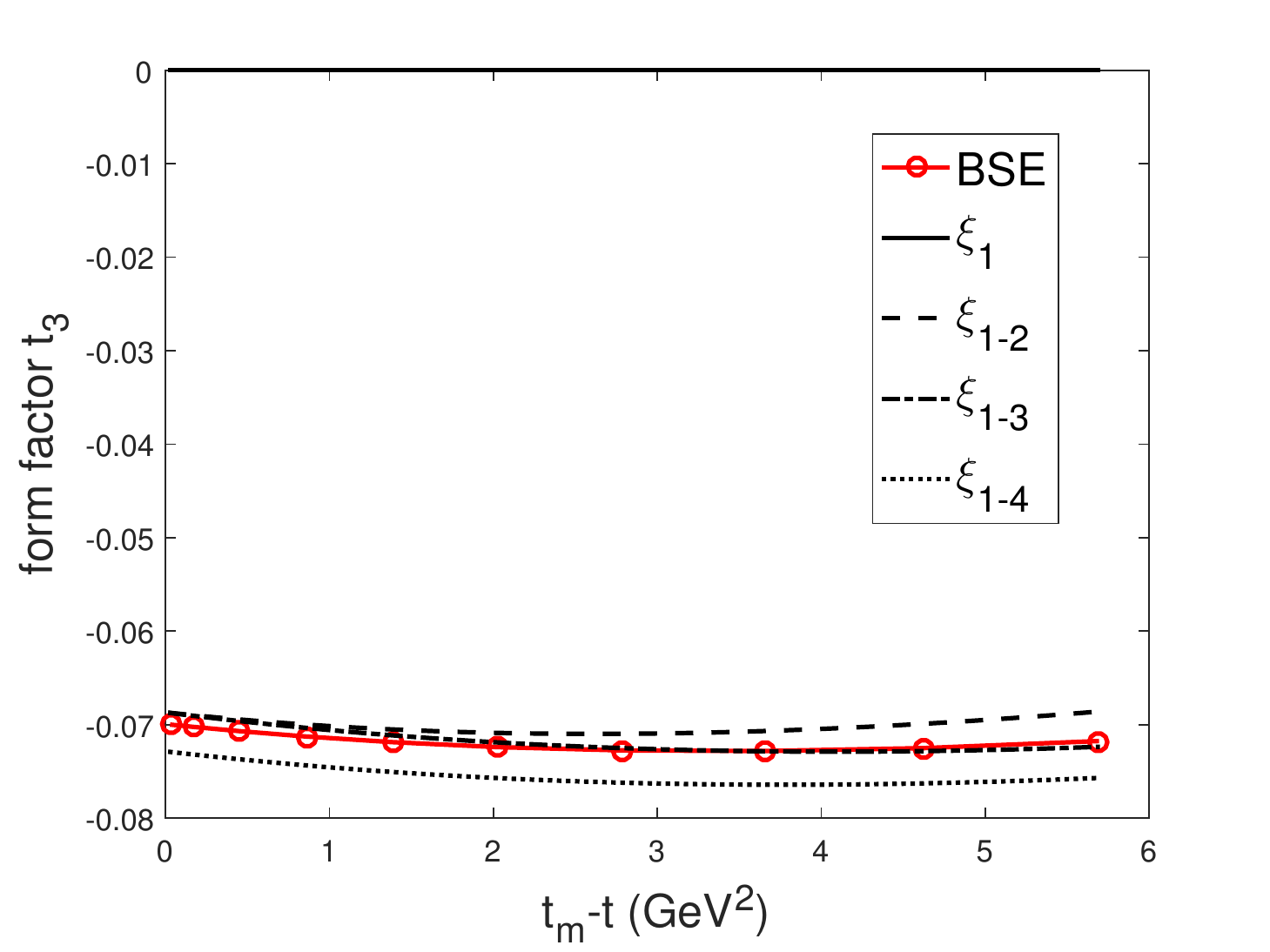}}
\subfigure[$B_c\to h_c(2P):t_{4}$]{\label{fig:t4-bchc2p}
			      \includegraphics[width=0.4\textwidth]{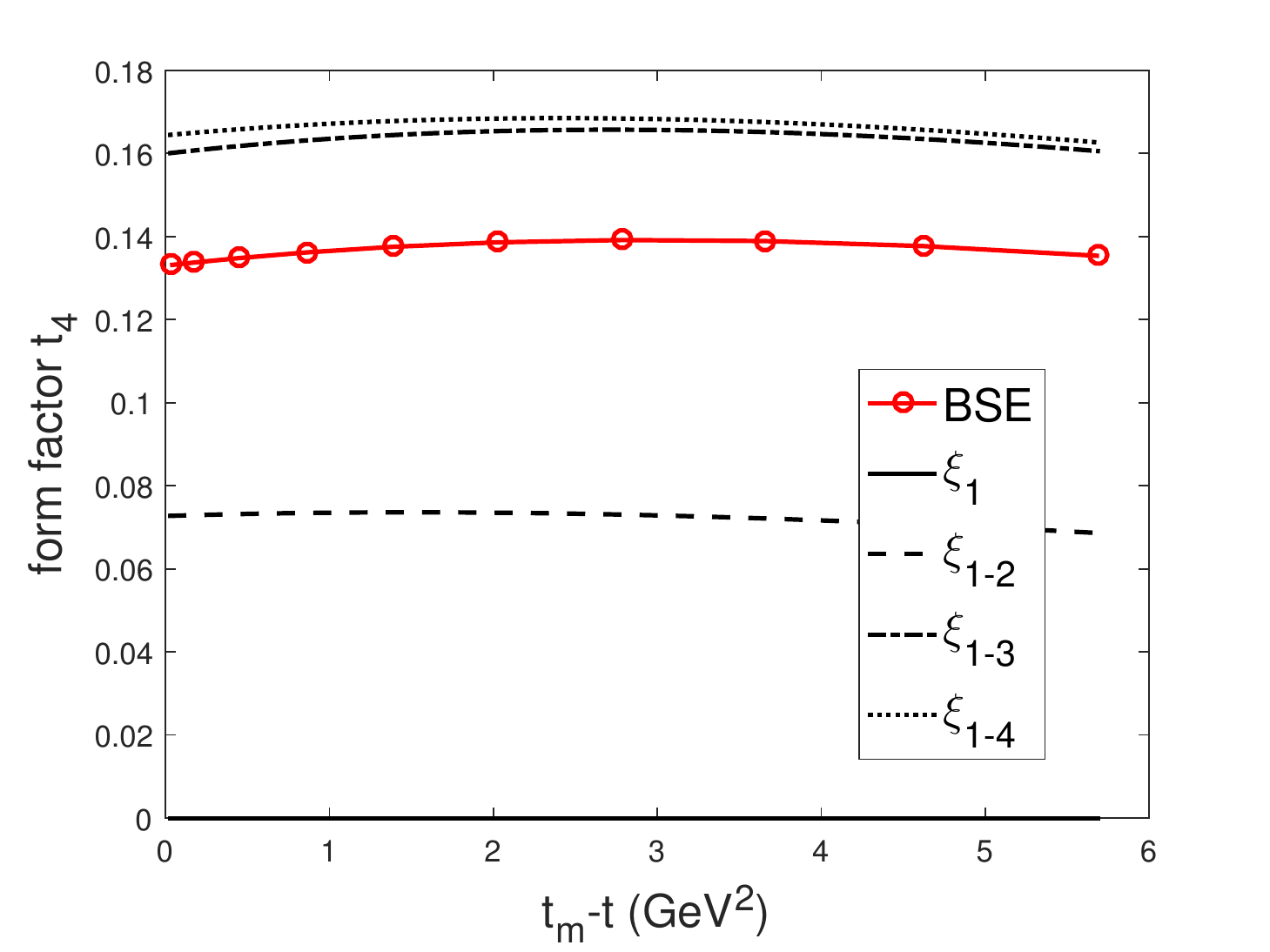}}
\subfigure[$B_c\to \chi_{c0}(2P):S_{+}$]{\label{fig:s+-bcxc02p}
			      \includegraphics[width=0.4\textwidth]{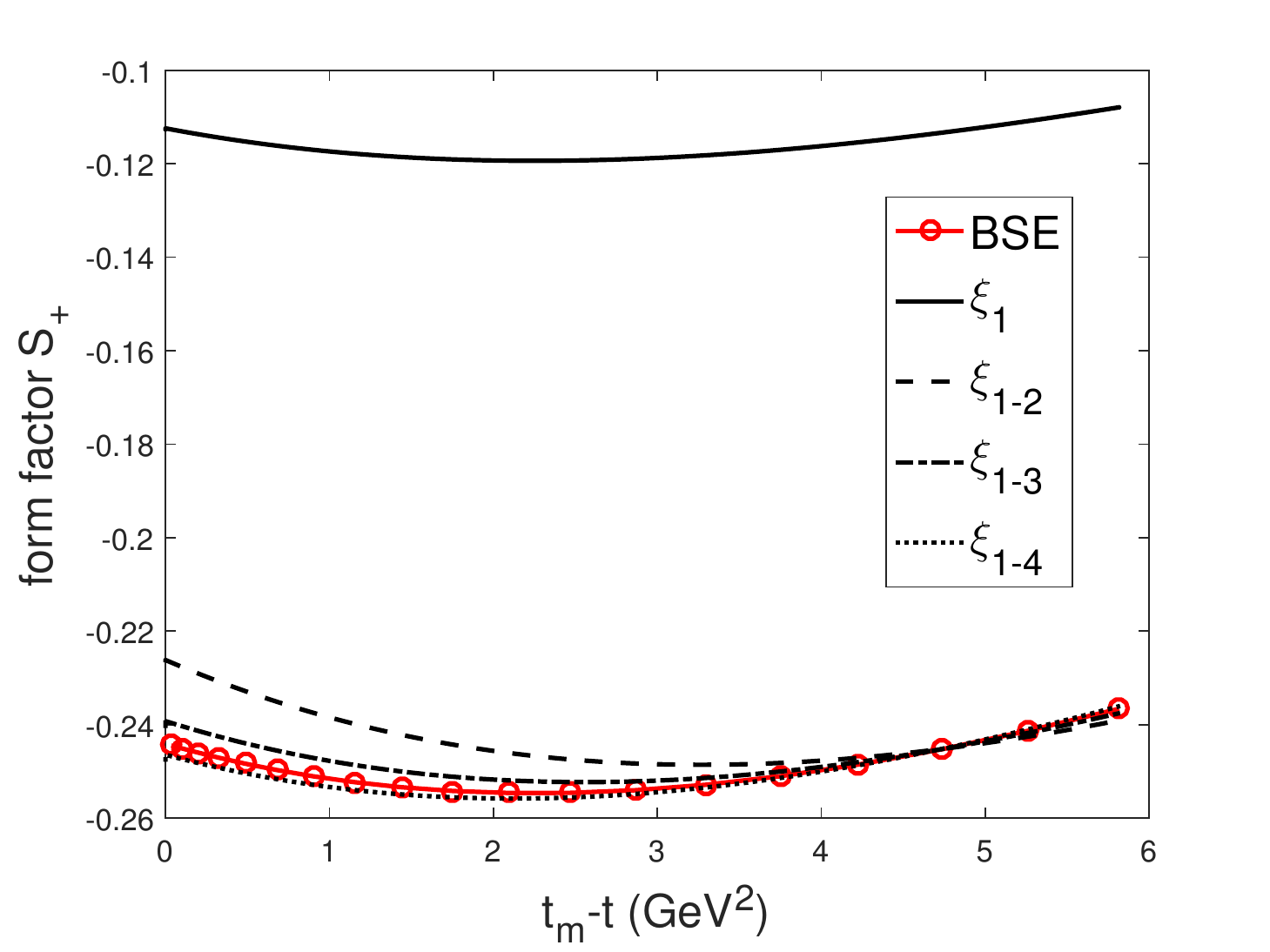}}
\subfigure[$B_c\to \chi_{c0}(2P):S_{-}$]{\label{fig:s--bcxc02p}
			      \includegraphics[width=0.4\textwidth]{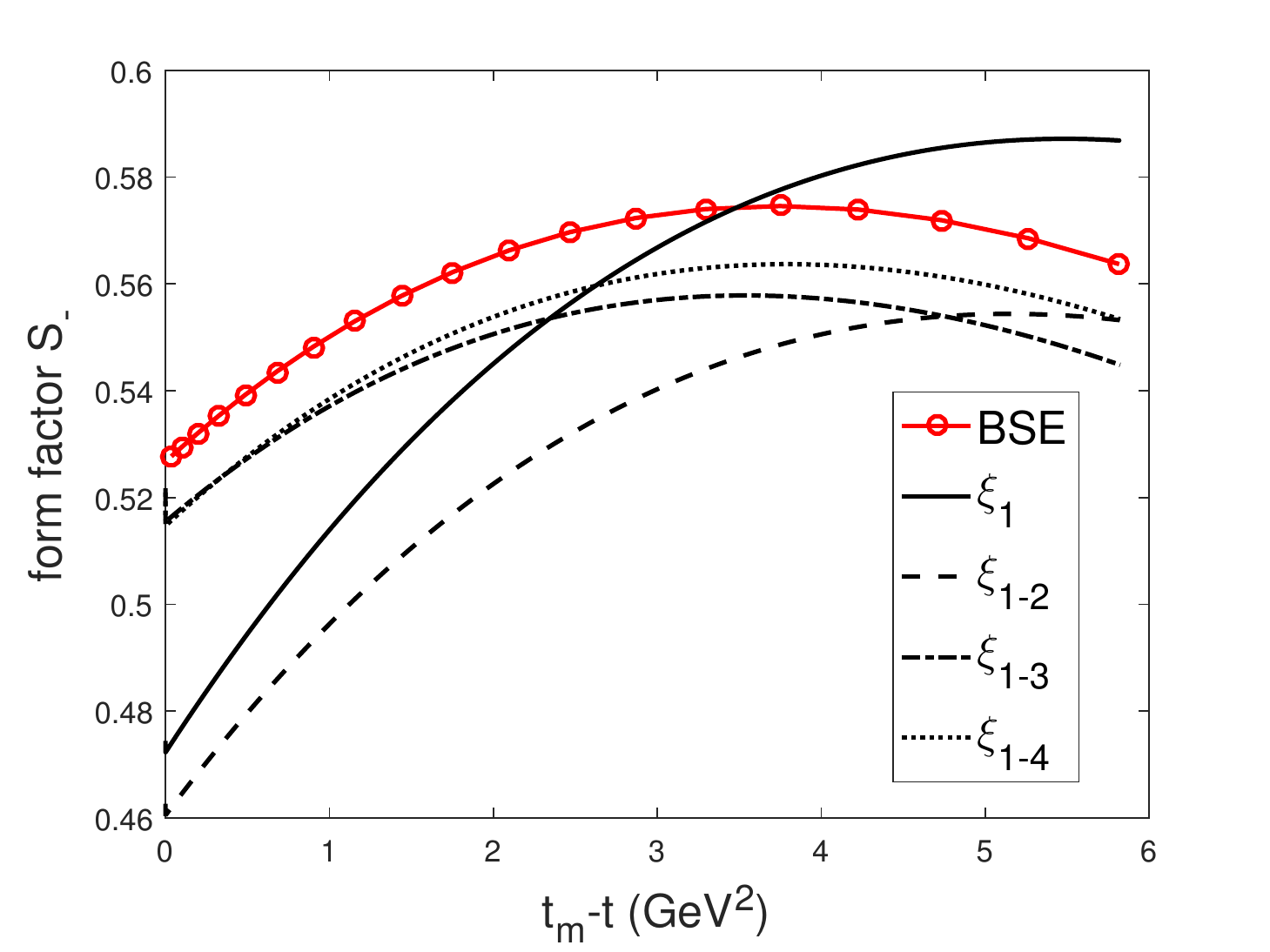}}
\caption{The form factors of $B_c\to h_c(2P),\chi_{c0}(2P)$ calculated by IWFs and instantaneous Bethe-Salpeter method, where $t\equiv(P-P_f)^2$ is the momentum transfer, and $t_m-t=2MM_f(v\cdot v'-1)$. The meaning of each type line is the same as that in Fig.~\ref{fig:ffs1}. }\label{fig:ffs5}
\end{figure}

\begin{figure}[!hbp]
\centering
\subfigure[$B_c\to \chi_{c1}(2P):t_{1}$]{\label{fig:t1-bcxc12p}
			      \includegraphics[width=0.4\textwidth]{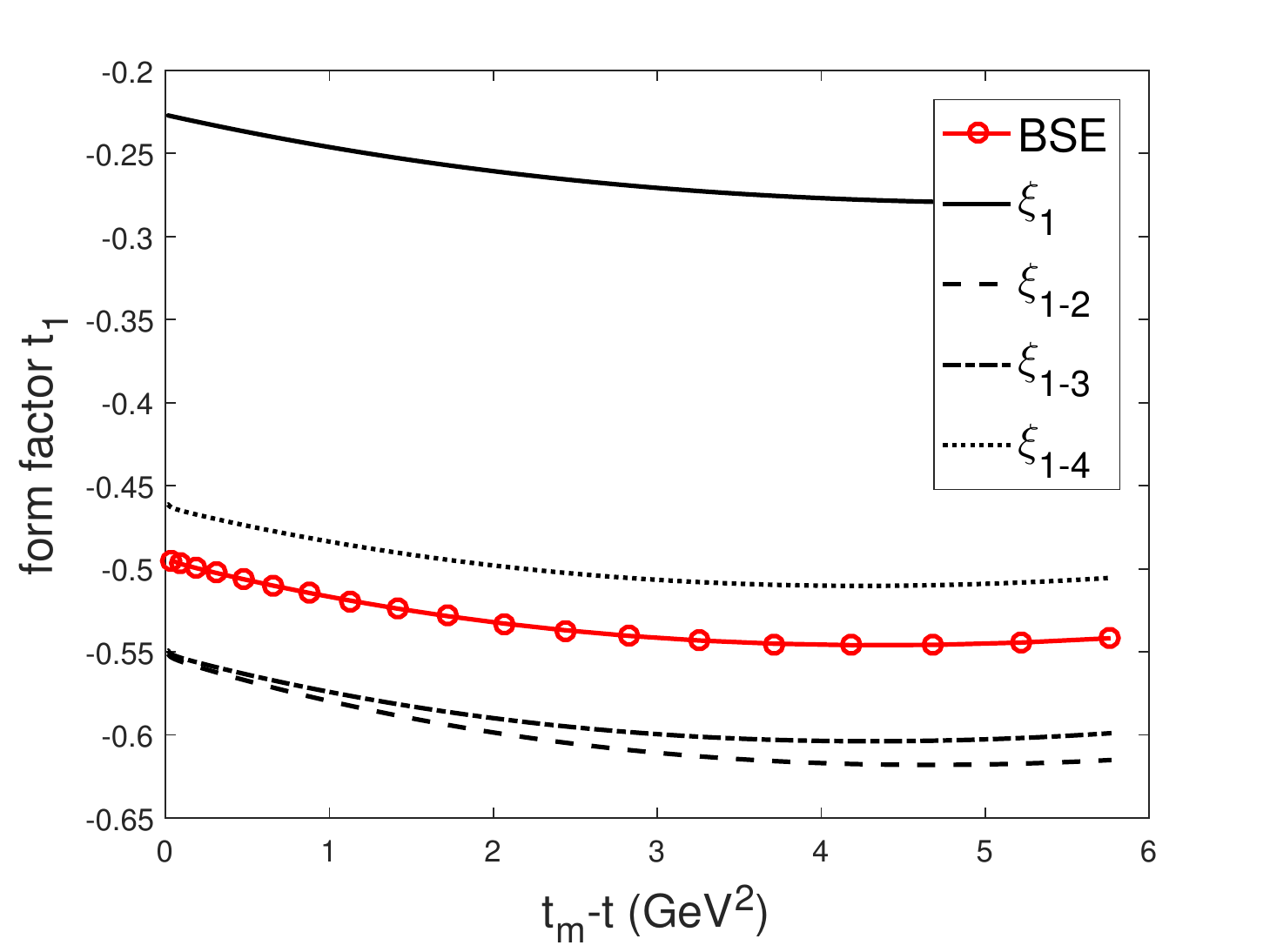}}
\subfigure[$B_c\to \chi_{c1}(2P):t_{2}$]{\label{fig:t2-bcxc12p}
			      \includegraphics[width=0.4\textwidth]{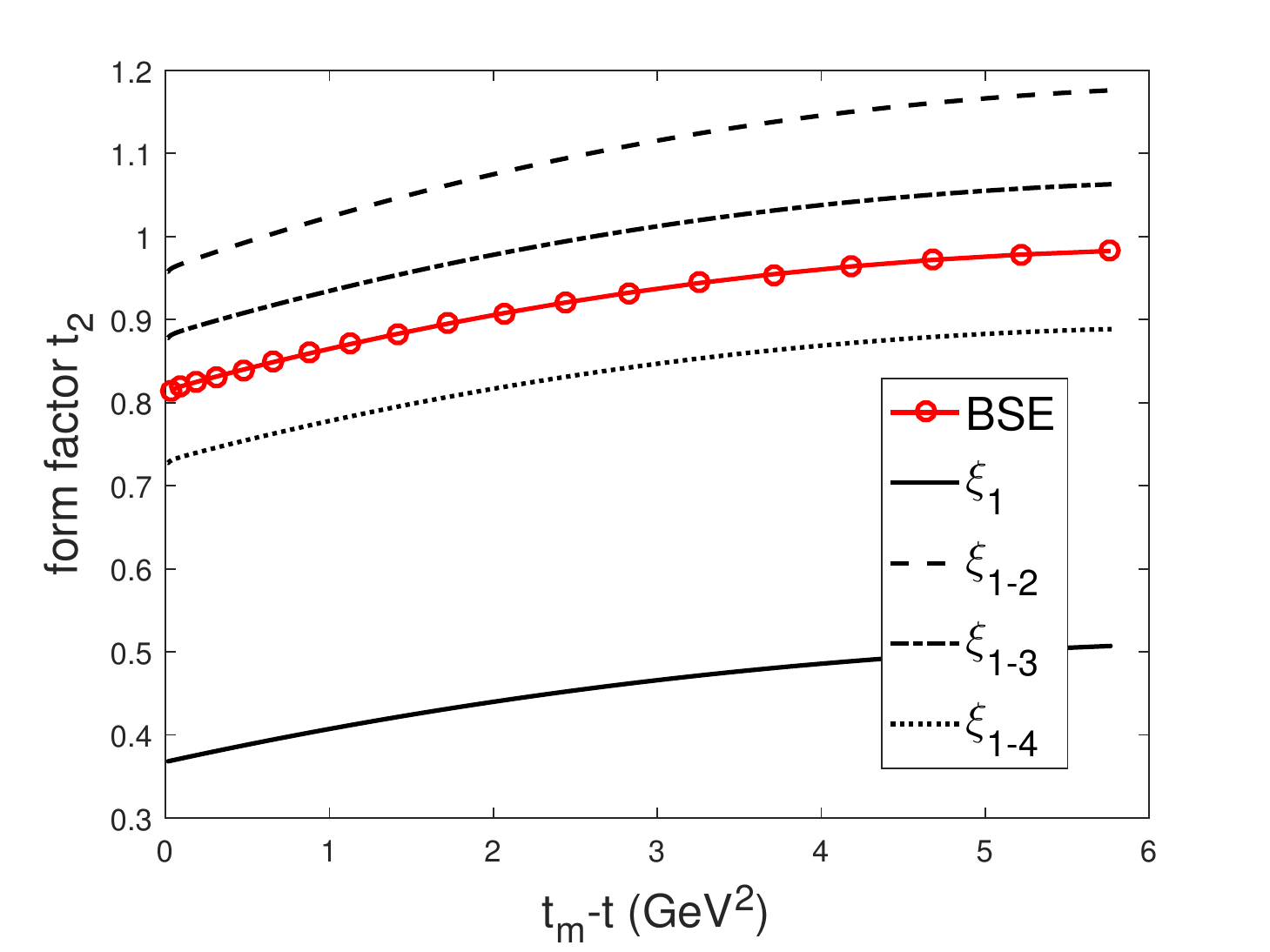}}
\subfigure[$B_c\to \chi_{c1}(2P):t_{3}$]{\label{fig:t3-bcxc12p}
			      \includegraphics[width=0.4\textwidth]{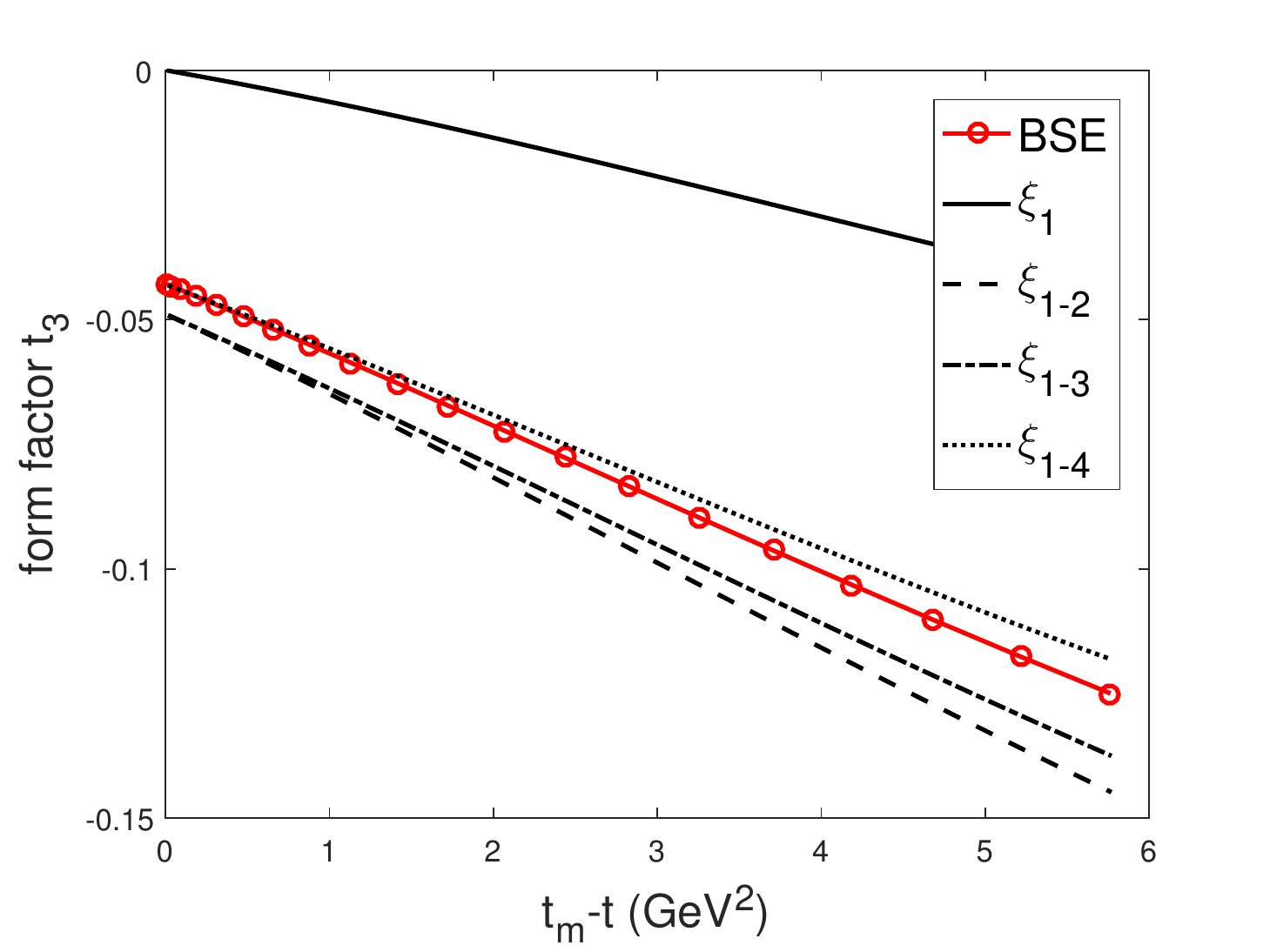}}
\subfigure[$B_c\to \chi_{c1}(2P):t_{4}$]{\label{fig:t4-bcxc12p}
			      \includegraphics[width=0.4\textwidth]{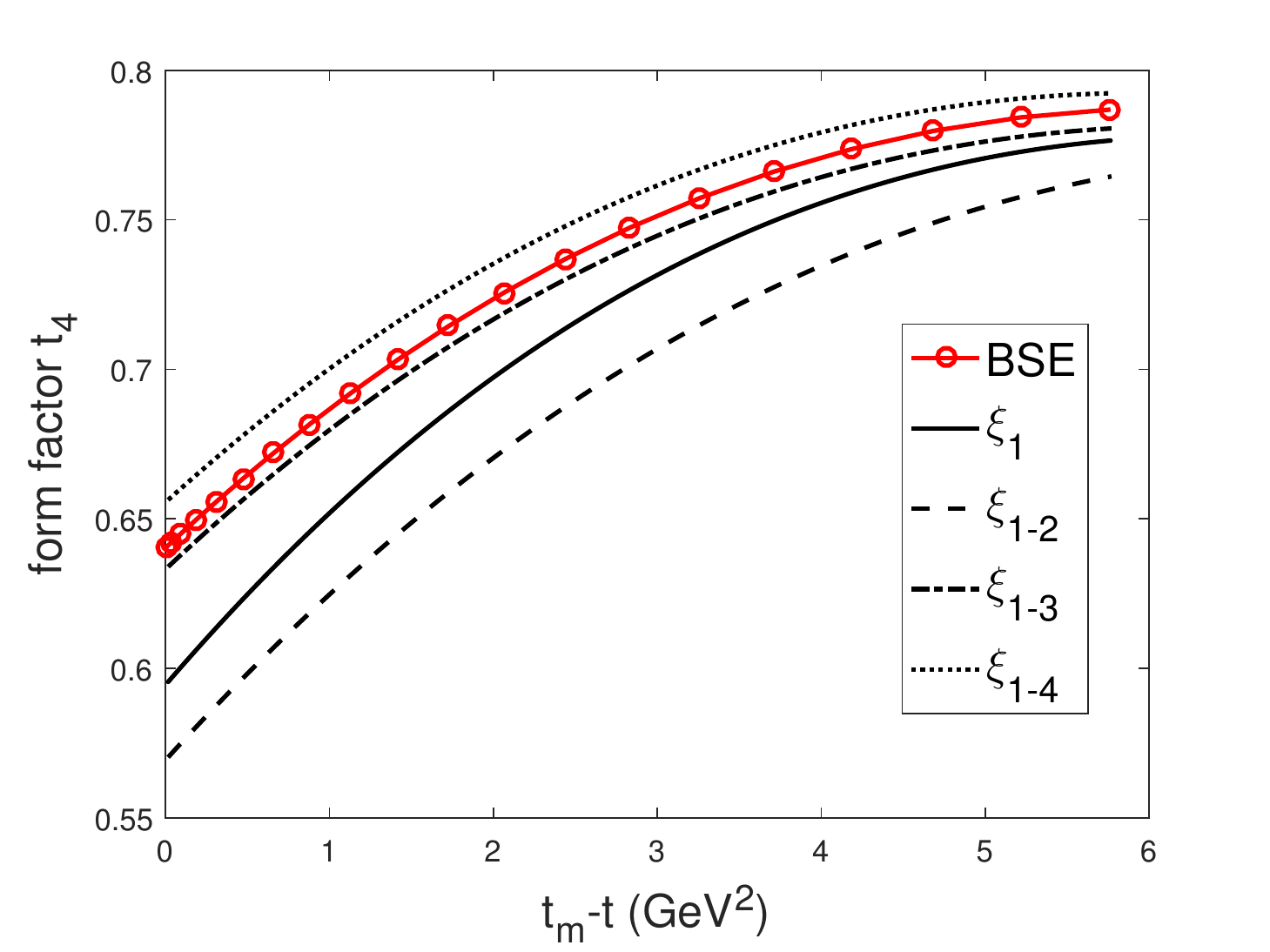}}
\caption{The form factors of $B_c\to \chi_{c1}(2P)$  calculated by IWFs and instantaneous Bethe-Salpeter method, where $t\equiv(P-P_f)^2$ is the momentum transfer, and $t_m-t=2MM_f(v\cdot v'-1)$. The meaning of each type line is the same as that in Fig.~\ref{fig:ffs1}.}\label{fig:ffs6}
\end{figure}

\begin{figure}[!hbp]
	\centering
	\subfigure[$B_c\to \chi_{c2}(2P):t_{1}$]{\label{fig:t1-bcxc22p}
		\includegraphics[width=0.4\textwidth]{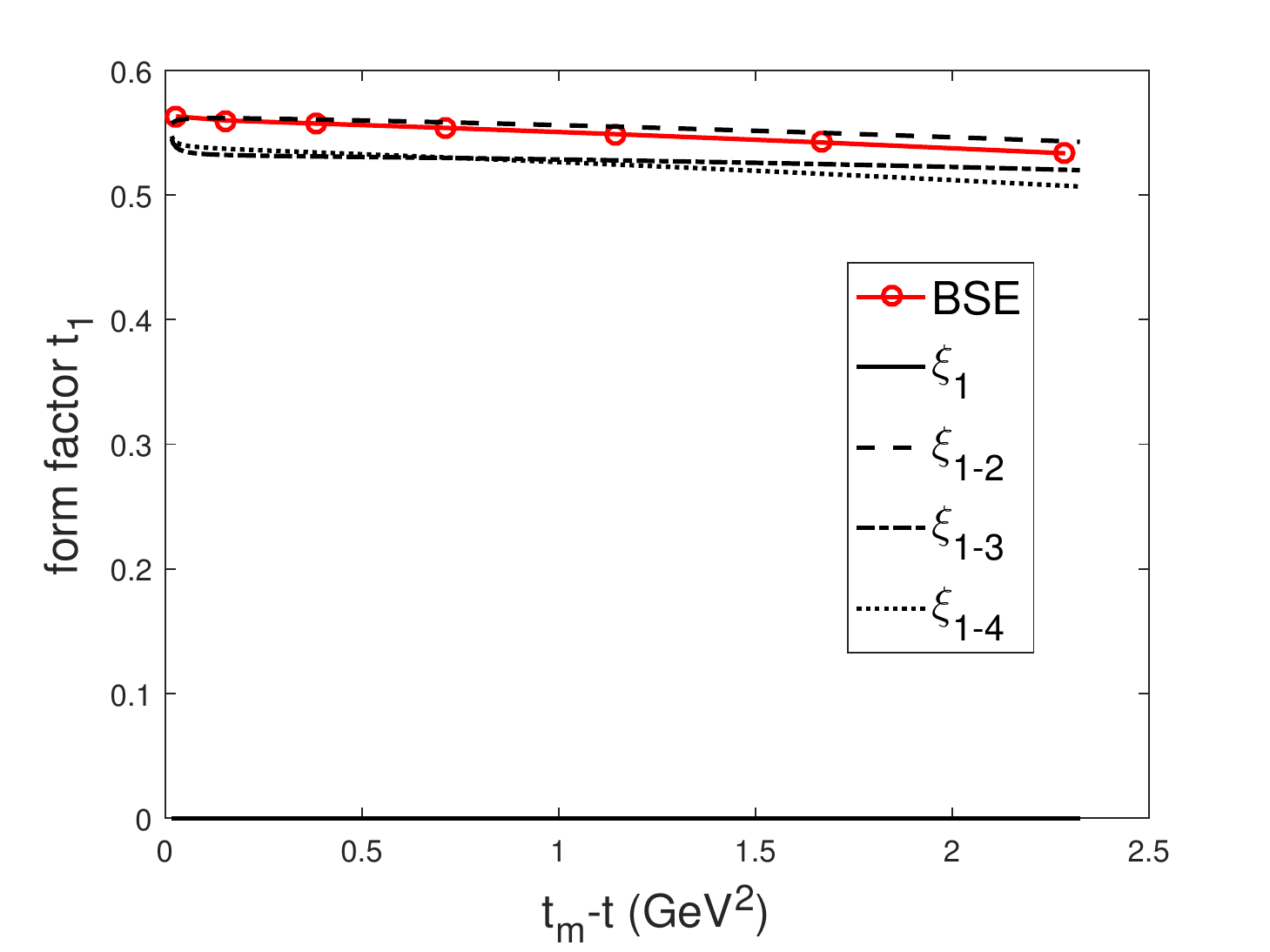}}
	\subfigure[$B_c\to \chi_{c2}(2P):t_{2}$]{\label{fig:t2-bcxc22p}
		\includegraphics[width=0.4\textwidth]{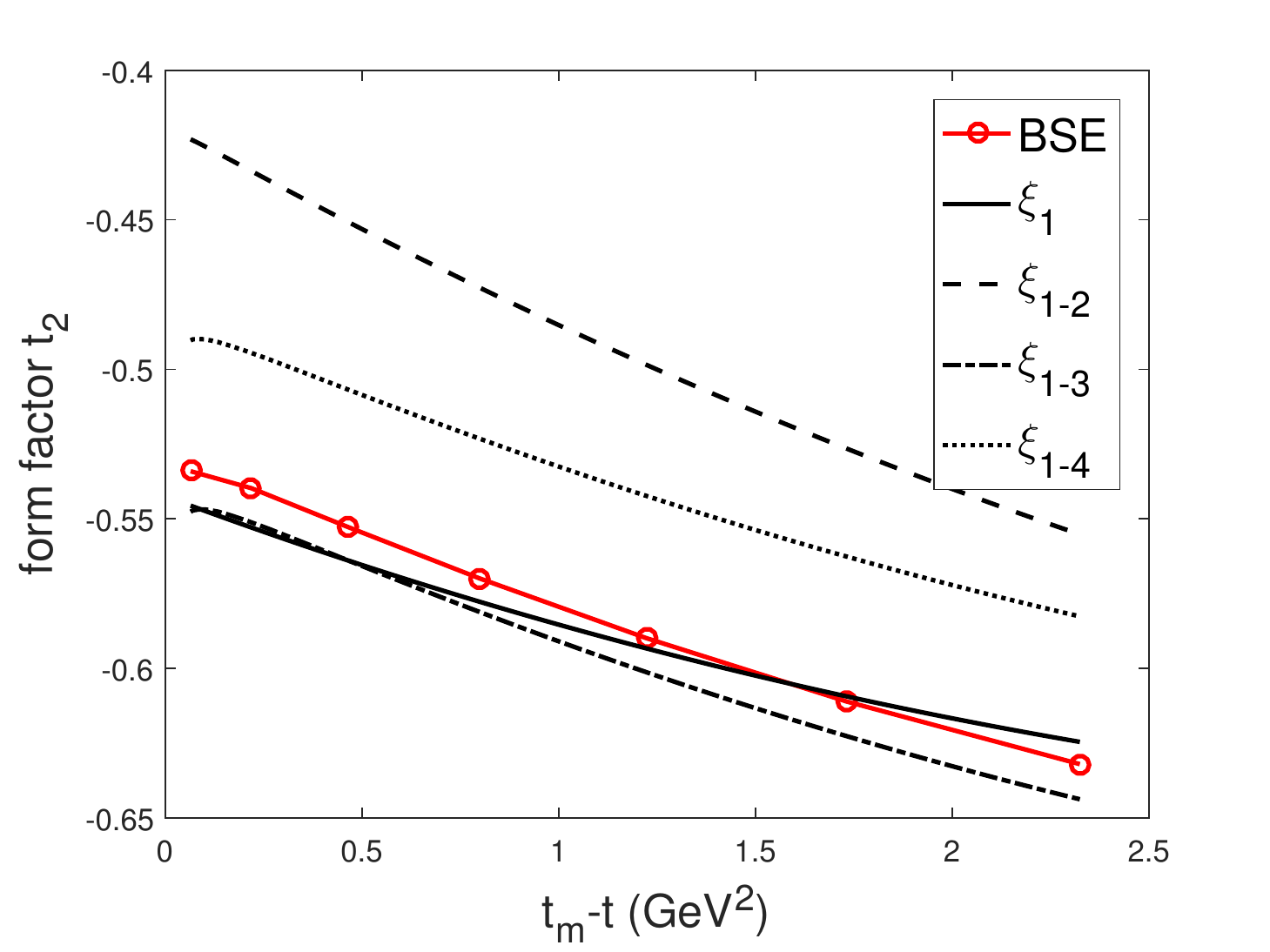}}
	\subfigure[$B_c\to \chi_{c2}(2P):t_{3}$]{\label{fig:t3-bcxc22p}
		\includegraphics[width=0.4\textwidth]{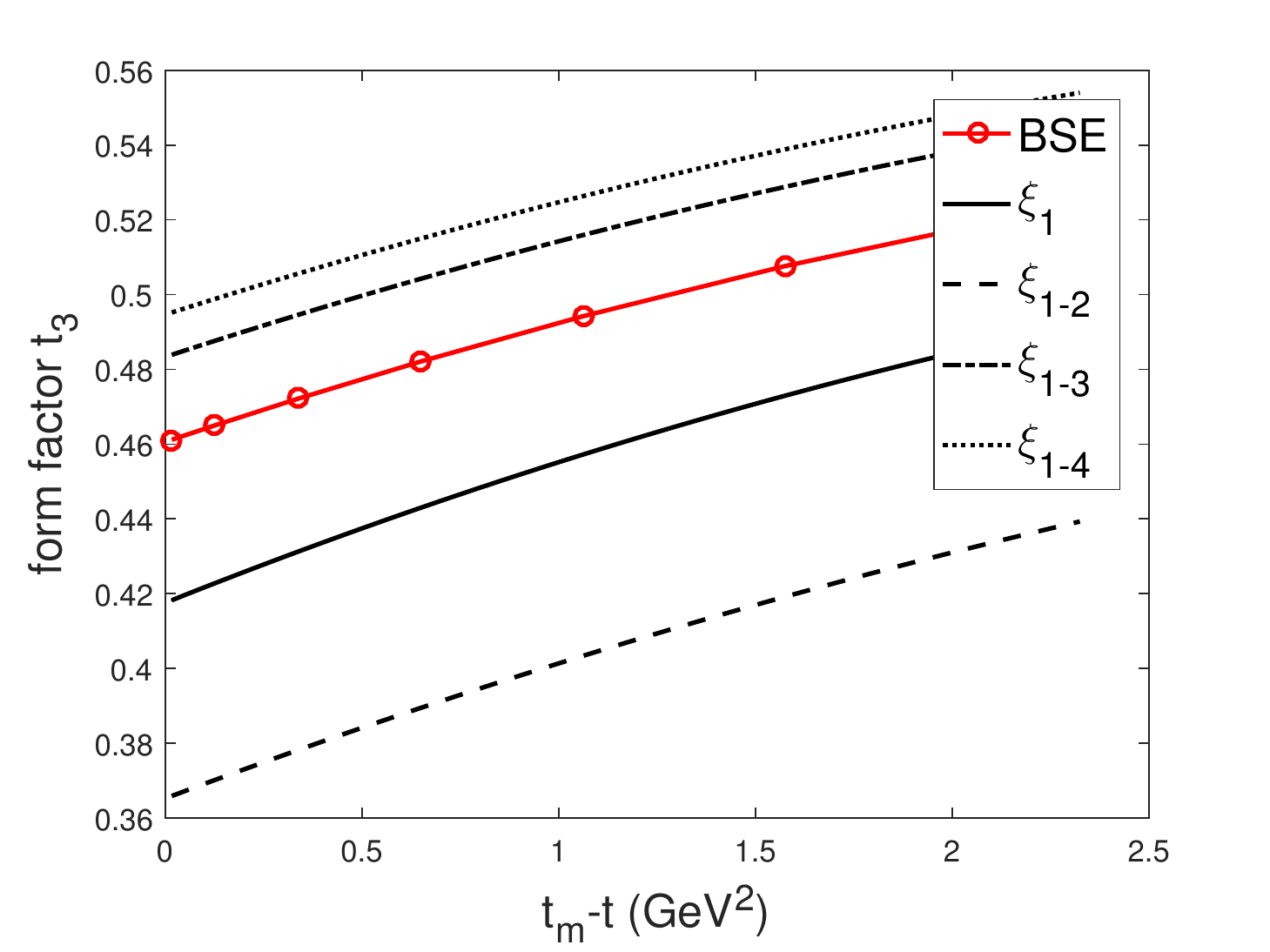}}
	\subfigure[$B_c\to \chi_{c2}(2P):t_{4}$]{\label{fig:t4-bcxc22p}
		\includegraphics[width=0.4\textwidth]{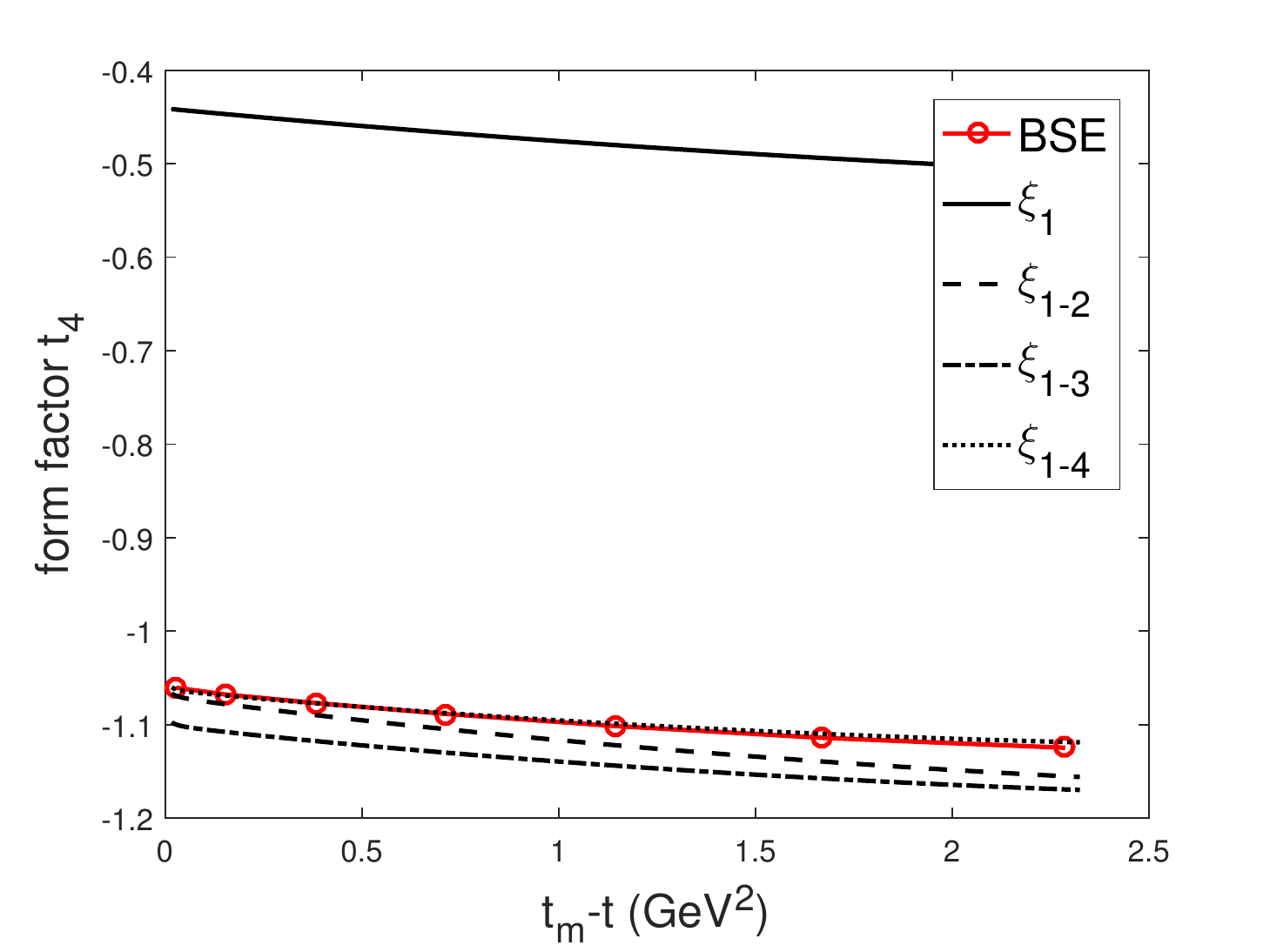}}
	\caption{The form factors of $B_c\to \chi_{c2}(2P)$  calculated by IWFs and instantaneous Bethe-Salpeter method, where $t\equiv(P-P_f)^2$ is the momentum transfer, and $t_m-t=2MM_f(v\cdot v'-1)$. The meaning of each type line is the same as that in Fig.~\ref{fig:ffs1}.}\label{fig:ffs6}
\end{figure}

The above analysis of relativistic corrections is qualitative, because the kinematic factors multiplied by IWFs are different and complex. In order to discuss these relativistic corrections precisely, the form factors in these processes are calculated by different order corrections in turn. Their numerical results are compared with those calculated by instantaneous Bethe-Salpeter method directly, as Figs.~\ref{fig:ffs1}-\ref{fig:ffs6} shows. In these Figs, $t\equiv(P-P_f)^2$ is the momentum transfer, and $t_m$ is the maximum of $t$, so $t_m-t=2MM_f(v\cdot v'-1)$. The circle-solid line (BSE) denotes the form factor calculated by instantaneous Bethe-Salpeter method directly, and we regard it as the relatively precise result because this method is almost covariant; the solid line denotes the leading order (LO) of form factor calculated only by IWF; the dash line denotes the result with IWF and first order (1st) correction; the dot-dash line denotes the result with IWF, the first and second order (2nd) corrections; the dot line denotes the result with IWF, the first, second and third order (3rd) corrections.

For the process $B_c\to\eta_c$, as Figs.~\ref{fig:s+-bcetac}--\ref{fig:s--bcetac} shows, there is some gap between the Leading order $S_+$ and the result from BSE. The $S_+$ with 1st correction is close to BSE. When the 3rd correction is taken into account, the result becomes very accurate. The difference between the LO $S_-$ and BSE is slightly larger, but due to the small contribution of $S_-$ to the decay width, the nonrelativistic results may be approximate. Though the high order corrections do not make $S_-$ and BSE exactly the same, they become closer. For the process $B_c\to J/\psi$, as Figs.~\ref{fig:t1-bcjpsi}--\ref{fig:t4-bcjpsi} shows, the form factor $t_3$ makes main contribution to the decay width. The LO $t_3$ has a little gap with BSE, and the result with high order corrections is more accurate. The LO $t_1$ is zero that is agree with HQET, see Eq.~(\ref{eq:limit1--}), but far from BSE. The LO $t_2$ and $t_4$ are different from BSE similarly. The high order corrections bring them closer to BSE, of which 1st correction is the most important one. In the mode $1S\to 1S$, the nonrelativistic results may be approximate, but high order corrections can make the result more precise. Note that, the accurate result of $t_1$ cannot be obtained by correcting $\xi_{00}$, as HQET did, so we need to introduce new high-order correction functions.

For the process $B_c\to h_c$, as Figs.~\ref{fig:t1-bchc}--\ref{fig:t4-bchc} shows, the form factor $t_2$ makes the main contribution to the decay width. The LO $t_2$ is slightly different from BSE. The 1st, 2nd and 3rd corrections are not small but almost cancel each other out. These corrections make $t_2$ closer to BSE. The LO $t_3$ and $t_4$ is zero which can be used to examine our method. The $t_1$, $t_2$ and $t_4$ with 1st correction are still a lot different from BSE, and therefore the higher order corrections are necessary. For the process $B_c\to \chi_{c0}$, as Figs.~\ref{fig:s+-bcxc0}--\ref{fig:s--bcxc0} shows, the difference between the LO $S_+$ and BSE is large. At least the 1st correction needs to be considered in order to reach an approximate result, though the $S_-$ with 1st correction is not accurate enough. The higher order corrections can make the results more accurate. For the process $B_c\to \chi_{c1}$, as Figs.~\ref{fig:t1-bcxc1}--\ref{fig:t4-bcxc1} shows, the form factor $t_2$ makes main contribution to the decay width. The corrections from 1st IWF are great except $t_4$. The higher order corrections are small but still important for accurate calculation. For the process $B_c\to \chi_{c2}$, as Figs.~\ref{fig:t1-bcxc2}--\ref{fig:t4-bcxc2} shows, the form factor $t_3$ makes main contribution to the decay width. The LO $t_3$ is close to BSE, and the high order corrections almost cancel each other out. It leads to the unusually small result of the total relativistic correction. The 1st correction makes $t_1$, $t_2$ and $t_4$ closer to BSE, but makes the main form factor $t_3$ farther from BSE. The result may be more imprecise if only the IWF and 1st correction are considered, so the higher-order corrections are very important. At zero recoil, the IWF $\xi_{11}$ is zero and the kinematical factor $(v\cdot v_f)/|\vec v_f|$ will lead to the divergence, see Eq.~(\ref{eq:heavylimit}). However most LO form factors are limited values at zero recoil. In general, the relativistic corrections are large, and the 1st corrections can only abtain the approximate results in the mode $1S\to 1P$.

In the modes $1S\to 2S$ and $1S\to 2P$, the form factors are no longer kinematically depressed but a little kinematically enhanced. The relativistic corrections are similar to those discussed above, but are greater and more complicated, as Figs.~\ref{fig:ffs4}-\ref{fig:ffs6} shows. Generally, there are big gaps between the leading order form factors and those from BSE directly. The newly introduced high-order correction functions make significant contributions in these relativistic corrections.


\section{\label{sec:conclusion}Conclusion}
In this work, we extract the Isgur-Wise functions in the framework of instantaneous Bethe-Salpeter equation. The Isgur-Wise function is the overlapping integrals of the wave functions for the initial state and the final state. The overlapping integrals which are with the relative momentum $\vec q\:'$ being inserted are the relativistic corrections ($1/m_q$ corrections) to the Isgur-Wise function, and the number of $\vec q\:'$ contained in the function corresponds to the order of the correction. We choose the semileptonic $B_c$ decays to charmonium to calculate the numerical results of Isgur-Wise functions and form factors, where the final states include $1S$, $1P$, $2S$ and $2P$. In the mode $1S\to 1S$, the Isgur-Wise function is coincident with the heavy quark effective theory, but it is not strict unity at zero recoil due to the relativistic correction. Another part of relativistic correction comes from the higher order correction functions, which make the results more accurate. The Isgur-Wise function can be generalized to other modes, including but not limited to $1S\to 1P$, $1S\to 2S$ and $1S\to 2P$ which this paper studies. The behavior of Isgur-Wise function almost exclusively depends on the configurations $nL$ of initial and final states, so they can still simplify the calculations of form factors. Some newly introduced higher order corrections provide a great relativistic corrections, even though the initial and final states are both the double-heavy mesons. These corrections can not be obtained by correcting the Isgur-Wise function, as HQET would do, so the higher order correction functions are necessary for accurate calculations.

\section{Acknowledgments}
This work was supported in part by the National Natural Science
Foundation of China (NSFC) under Grant No.~11405037, No.~11575048 and No.~11505039, and also in part by PIRS of HIT No.B201506.




\appendix

\section{\label{appendix} Equation and solution for heavy mesons}
BS equation for a quark-antiquark bound state generally is written as \cite{Chang:2014jca}
\begin{equation}
(\slashed p_1-m_1)\chi_P(q)(\slashed p_2+m_2)=\mathrm i\int\frac{\mathrm d^4k}{(2\pi)^4}V(P,k,q)\chi_P(k),
\end{equation}
where $p_1,p_2;m_1,m_2$ are the momenta and masses of the quark and antiquark, respectively; $\chi_P(q)$ is the BS wave function with the total momentum $P$ and relative momentum $q$; $V(P,k,q)$ is the kernel between the quark-antiquark in the bound state. $P$ and $q$ are defined as
\begin{equation}
\begin{aligned}
&\vec p_1=\alpha_1\vec P+\vec q,\quad\alpha_1=\frac{m_1}{m_1+m_2},\\
&\vec p_2=\alpha_2\vec P-\vec q,\quad\alpha_2=\frac{m_2}{m_1+m_2}.
\end{aligned}
\end{equation}

We divide the relative momentum $q$ into two parts, $q_{P_{||}}$ and $q_{P_\perp}$, a parallel part and an orthogonal one to $P$, respectively
\begin{equation}
q^\mu=q_{P_{||}}^\mu+q_{P_\perp}^\mu,
\end{equation}
where $q_{P_{||}}^\mu\equiv(P\cdot q/M^2)P^\mu,~q_{P_\perp}^\mu\equiv q^\mu-q_{P_{||}}^\mu$, and $M$ is the mass of the relevant meson. Correspondingly, we have two Lorentz-invariant variables
\begin{equation}
q_P=\frac{P\cdot q}{M},~q_{P_T}=\sqrt{q_P^2-q^2}=\sqrt{-q_{P_\perp}^2}.
\end{equation}
If we introduce two notations as below
\begin{equation}
\begin{aligned}
&\eta(q_{P_\perp}^\mu)\equiv \int\frac{k_{P_T}^2\ud k_{P_T}\ud s}{(2\pi)^2}V(k_{P_\perp},s,q_{P_\perp})\varphi(k_{p_\perp}^\mu),\\
&\varphi(q_{p_\perp}^\mu)\equiv\mathrm i\int\frac{\mathrm dq_P}{2\pi}\chi_P(q_{P_{||}}^\mu,q_{P_\perp}^\mu).
\end{aligned}
\end{equation}
Then the BS equation can take the form as follow
\begin{equation}
\chi_P(q_{P_{||}}^\mu,q_{P_\perp}^\mu)=S_1(p_1^\mu)\eta(q_{P_\perp}^\mu)S_2(p_2^\mu).
\end{equation}
The propagators of the relevant particles with masses $m_1$ and $m_2$ can be decomposed as
\begin{equation}
S_i(p_i^\mu)=\frac{\Lambda_{i_P}^+(q_{P_\perp}^\mu)}{J(i)q_P+\alpha_iM-\omega_{i_P}+\mathrm i\varepsilon}+\frac{\Lambda_{i_P}^-(q_{P_\perp}^\mu)}{J(i)q_P+\alpha_iM+\omega_{i_P}-\mathrm i\varepsilon},
\end{equation}
with
\begin{equation}
\begin{aligned}
\omega_{i_P}&=\sqrt{m_i^2+q_{P_T}^2},\\
\Lambda_{i_P}^{\pm}(q_{P_\perp}^\mu)&=\frac{1}{2\omega_{i_P}}\left[\frac{\slashed P}{M}\omega_{i_P}\pm J(i)(\slashed q_{P_\perp}+m_i)\right],
\end{aligned}
\end{equation}
where $i=1,2$ for quark and antiquark, respectively, and $J(i)=(-1)^{i+1}$.

Then the instantaneous Bethe-Salpeter equation can be decomposed into the coupled equations
\begin{equation}
\begin{aligned}
(M-\omega_{1p}-\omega_{2p})\varphi^{++}(q_{P_\perp})&=\Lambda_1^+(P_{1p_\perp})\eta(q_{P_\perp})\Lambda_2^+(P_{2p_\perp}),\\
(M+\omega_{1p}+\omega_{2p})\varphi^{--}(q_{P_\perp})&=-\Lambda_1^-(P_{1p_\perp})\eta(q_{P_\perp})\Lambda_2^-(P_{2p_\perp}),\\
\varphi^{+-}(q_{P_\perp})=0,\quad&\qquad\varphi^{-+}(q_{P_\perp})=0.
\end{aligned}\label{eq:phi}
\end{equation}

The instantaneous kernel has the following form
\begin{equation}
V(P,k,q)\sim V(|k-q|),
\end{equation}
especially when the two constituents of meson are very heavy. The kernel we used contains a linear scalar interaction for color-confinement, a vector interaction for one-gluon exchange and a constant $V_0$ which as a `zero-point', i.e. 
\begin{equation}
I(r)=\lambda r+V_0-\gamma_0\otimes\gamma^0\frac{4}{3}\frac{\alpha_s(r)}{r}\tag{A.4},
\end{equation}
where $\lambda$ is the `string constant', $\alpha_s(r)$ is the running coupling constant. In order to avoid the infrared divergence, a factor $e^{-\alpha r}$ is introduced, i.e.
\begin{equation}
\begin{split}
V_s(r)&=\frac{\lambda}{\alpha}(1-e^{-\alpha r}),\\
V_v(r)&=-\frac{4}{3}\frac{\alpha_s(r)}{r}e^{-\alpha r}.
\end{split}\tag{A.5}
\end{equation}
In momentum space the kernel reads:
\begin{equation}
I(\vec q\,)=V_s(\vec q\,)+\gamma_0\otimes\gamma^0V_v(\vec q\,)\tag{A.6},
\end{equation}
where
\begin{equation}
\begin{split}
V_s(\vec q\,)&=-\left(\frac{\lambda}{\alpha}+V_0\right)\delta^3(\vec q\,)+\frac{\lambda}{\pi^2}\frac{1}{(\vec q\,^2+\alpha^2)^2},\\
V_v(\vec q\,)&=-\frac{2}{3\pi^2}\frac{\alpha_s(\vec q\,)}{\vec q\,^2+\alpha^2},\\
\alpha_s(\vec q\,)&=\frac{12\pi}{27}\frac{1}{\mathrm{In}(a+\vec q\,^2/\Lambda_{QCD}^2)}.
\end{split}\tag{A.7}
\end{equation}
The fitted parameters are $a=e=2.7183$, $\alpha=0.06$ GeV, $\lambda=0.21$ ${\rm GeV}^2$, $\Lambda_{QCD}=0.27$ GeV; $V_0$ is fixed by fitting the mass of the ground state. With these parameters, the mass spectrums, decay constants and some branching fractions of the double heavy mesons, including $B_c$, charmonium and bottomium, can be obtained. These results are in good accord with the experimental data. More details can be referred to in the literatures \cite{Chang:2010kj,Fu:2011zzo}.

The instantaneous Bethe-Salpeter wave function for $2^{++}$ states mesons have the general form\cite{Wang:2009er}
\begin{equation}
\begin{aligned}
\varphi_{2^{++}}(q_\perp)&=\epsilon_{\mu\nu}q_\perp^\mu q_\perp^\nu\left[\zeta_1(q_\perp)+\frac{\slashed{P}}{M}\zeta_2(q_\perp)
+\frac{\slashed{q}_\perp}{M}\zeta_3(q_\perp)+\frac{\slashed{P}\slashed{q}_\perp}{M^2}\zeta_4(q_\perp)\right]\\
&~~~~+M\epsilon_{\mu\nu}\gamma^\mu q_\perp^\nu\left[\zeta_5(q_\perp)+\frac{\slashed{P}}{M}\zeta_6(q_\perp)
+\frac{\slashed{q}_\perp}{M}\zeta_7(q_\perp)+\frac{\slashed{P}\slashed{q}_\perp}{M^2}\zeta_8(q_\perp)\right]
\end{aligned}
\end{equation}
with
\begin{equation}
\begin{split}
\zeta_1(q_\perp)&=\frac{q_\perp^2\zeta_3(\omega_1+\omega_2)+2M^2\zeta_5\omega_2}{M(m_1\omega_2+m_2\omega_1)}\\
\zeta_2(q_\perp)&=\frac{q_\perp^2\zeta_4(\omega_1-\omega_2)+2M^2\zeta_6\omega_2}{M(m_1\omega_2+m_2\omega_1)}\\
\zeta_7(q_\perp)&=\frac{M(\omega_1-\omega_2)}{m_1\omega_2+m_2\omega_1}\zeta_5\\
\zeta_8(q_\perp)&=\frac{M(\omega_1+\omega_2)}{m_1\omega_2+m_2\omega_1}\zeta_6\label{eq:1-}
\end{split}
\end{equation}
The wave function corresponding to the positive projection has the form
\begin{equation}
\begin{split}
\varphi_{2^{++}}^{++}(q_\perp)&=\epsilon_{\mu\nu}q_\perp^\mu q_\perp^\nu\left[B_1(q_\perp)+\frac{\cancel P}{M}B_2(q_\perp)+\frac{\cancel q_\perp}{M}B_3(q_\perp)+\frac{\cancel P\cancel q_\perp}{M^2}B_4(q_\perp)\right]\\
&+M\epsilon_{\mu\nu}\gamma^\mu q_\perp^\nu\left[B_5(q_\perp)+\frac{\cancel P}{M}B_6(q_\perp)+\frac{\cancel q_\perp}{M}B_7(q_\perp)+\frac{\cancel P\cancel q_\perp}{M^2}B_8(q_\perp)\right]
\end{split}
\end{equation}
where
\begin{equation}
\begin{split}
B_1&=\frac{1}{2M(m_1\omega_2+m_2\omega_1)}[(\omega_1+\omega_2)q_{\perp}^2\zeta_3+(m_1+m_2)q_{\perp}^2\zeta_4+2M^2\omega_2\zeta_5-2M^2m_2\zeta_6]\\
B_2&=\frac{1}{2M(m_1\omega_2+m_2\omega_1)}[(m_1-m_2)q_{\perp}^2\zeta_3+(\omega_1-\omega_2)q_{\perp}^2\zeta_4+2M^2\omega_2\zeta_6-2M^2m_2\zeta_5]\\
B_3&=\frac{1}{2}\left[\zeta_3+\frac{m_1+m_2}{\omega_1+\omega_2}\zeta_4-\frac{2M^2}{m_1\omega_2+m_2\omega_1}\zeta_6\right]\\
B_4&=\frac{1}{2}\left[\frac{\omega_1+\omega_2}{m_1+m_2}\zeta_3+\zeta_4-\frac{2M^2}{m_1\omega_2+m_2\omega_1}\zeta_5\right]\\
B_5&=\frac{1}{2}\left[\zeta_5-\frac{\omega_1+\omega_2}{m_1+m_2}\zeta_6\right],\qquad A_6=\frac{1}{2}\left[-\frac{m_1+m_2}{\omega_1+\omega_2}\zeta_5+\zeta_6\right]\\
B_7&=\frac{M}{2}\frac{\omega_1-\omega_2}{m_1\omega_2+m_2\omega_1}\left[\zeta_5-\frac{\omega_1+\omega_2}{m_1+m_2}\zeta_6\right]\\
B_8&=\frac{M}{2}\frac{m_1+m_2}{m_1\omega_2+m_2\omega_1}\left[-\zeta_5+\frac{\omega_1+\omega_2}{m_1+m_2}\zeta_6\right]
\end{split}
\end{equation}
If the masses of the quark and antiquark are equal, the normalization condition reads as
\begin{equation}
\int\frac{\ud\vec q}{(2\pi)^3}\frac{8\omega_1\vec q\:^2}{15m_1}\left[5\zeta_5\zeta_6M^2+2\zeta_4\zeta_5\vec q\:^2-2\vec q\:^2\zeta_3\left(\zeta_4\frac{\vec q\:^2}{M^2}+\zeta_6\right)\right]=2M.
\end{equation}

\bibliography{reference}

\end{document}